\documentclass[prx,twocolumn,english,superscriptaddress,floatfix,aps,10pt,nolongbibliography]{revtex4-2} 
\usepackage{amsmath, amssymb, amscd, amsthm, amsfonts, amscd, dsfont, mathtools, wasysym, amsthm}
\usepackage{graphicx}%
\usepackage{dcolumn}%
\usepackage{bm}
\usepackage{standalone}
\usepackage[caption=false]{subfig}
\usepackage{enumitem}
\usepackage[utf8]{inputenc}
\usepackage{microtype}
\usepackage[plain]{algorithm}
\usepackage{algpseudocode}

\usepackage{bbm}
\usepackage{braket}
\usepackage{comment}
\usepackage{booktabs} %
\usepackage{multirow}
\usepackage{diagbox}

\usepackage{pifont}
\newcommand{\cmark}{\ding{51}}%
\newcommand{\xmark}{\ding{55}}
\usepackage{refcount}
\newcommand{\tabref}[1]{%
  \hyperref[#1]{Table~\MakeUppercase{\romannumeralstring{\getrefnumber{#1}}}}%
}
\newcommand{\romannumeralstring}[1]{%
  \expandafter\romannumeral\expandafter#1\relax
}

\usepackage{silence}
\newcommand{\ketbra}[2]{| #1 \rangle \langle #2 | }

\DeclareMathOperator{\Tr}{Tr}

\newcommand{\XL}{\bar{X}}
\newcommand{\YL}{\bar{Y}}
\newcommand{\ZL}{\bar{Z}}
\newcommand{\sL}{\bar{\sigma}}
\newcommand{\kpsi}{\ket{\psi}}
\newcommand{\id}{\mathds{1}}
\newcommand{\sre}{\tilde{\cal M}_2}
\newcommand{\rom}{{\cal R}}

\newtheorem{lemma}{Lemma}
\newtheorem{theorem}{Theorem}

\usepackage[dvipsnames]{xcolor}
\usepackage{hyperref}
\hypersetup{
    colorlinks = true,
    linkcolor = Blue,%
    citecolor = MidnightBlue,%
    urlcolor = Blue}%
\begin{document}

\title{Spreading of Magic Resource under Unitary Clifford Dynamics}

\author{Mircea Bejan}
\email{mircea.bejan@kcl.ac.uk}
 \affiliation{Department of Physics, King’s College London, Strand WC2R 2LS, UK\looseness=-1}
 \affiliation{T.C.M. Group, Cavendish Laboratory, University of Cambridge, J.J. Thomson Avenue, Cambridge, CB3 0HE, UK\looseness=-1}
\author{Pieter W. Claeys}
\affiliation{School of Physics, Trinity College Dublin, Dublin 2, Ireland}
\affiliation{Max Planck Institute for the Physics of Complex Systems, 01187 Dresden, Germany\looseness=-1}
\author{Jiangtian Yao}
\email{jyao@pks.mpg.de}
\affiliation{Max Planck Institute for the Physics of Complex Systems, 01187 Dresden, Germany\looseness=-1}

\date{October 2026}

\begin{abstract}

Nonstabilizerness, or quantum magic resource, presents a valuable resource in quantum error correction and computation. 
We study the dynamics of locally injected nonstabilizerness in unitary Clifford circuits, where the total nonstabilizerness is conserved.
However, the absence of physical observables quantifying nonstabilizerness precludes a direct microscopic or hydrodynamic description of its local distribution and dynamics. Using insights from stabilizer quantum error correcting codes, we rigorously show that the spatial distribution of nonstabilizerness can be inferred from a canonical representation of low-magic states, dubbed the bipartite magic gauge.
Moreover, we propose two operationally relevant magic length scales. 
We numerically establish that, at early times, both length scales grow ballistically at distinct velocities set by the entanglement velocity, after which nonstabilizerness delocalizes.
Our work sheds light on the spatiotemporal structure of quantum resources and complexity in many-body dynamics, opening up avenues for investigating their transport properties and further connections with quantum error correction.
\end{abstract}
\maketitle

Building scalable quantum computers hinges on the ability to protect logical information from being corrupted. To protect this information, one encodes it via a quantum error correcting (QEC) code~\cite{shor, CalderbankShor, steane, gottesman1997stabilizer, surface, gkp, qLDPC}; however, using certain QEC codes limits the logical operations implementable transversally, i.e., with physically local operations~\cite{EastinKnill,zeng2007transversalityversusuniversalityadditive,BravyiKoenig}.
To perform universal quantum computation, one needs to augment QEC codes with special states. For example, magic states are required for stabilizer codes~\cite{bravyi2005universal}. These states contain a quantum resource, dubbed nonstabilizerness (NS) or ``magic'', that can make the classical simulation of quantum systems hard~\cite{feynman1982simulating, Bravyi_PBC2016,bravyi2016improved, magic3}.

Magic and other quantum resources such as entanglement, non-Gaussianity, etc., not only play a crucial role in building practical quantum devices~\cite{Bravyi_PBC2016,bravyi2016improved, magic3, schuch2008entropy, vidal2003efficient, vidal2004efficient, Markov_2008, boson_sampling, ferm_sampling, DiasFermionic, DiasBosonic}, but also have profound implications on quantum many-body dynamics.
Signatures of quantum chaos emerge when injecting NS into Clifford dynamics~\cite{single_T_gate, Leone2021quantumchaosis}, which alters the entanglement spectral distribution from Poisson to Wigner-Dyson (commonly 
associated with integrable and chaotic dynamics, respectively)~\cite{single_T_gate}.
Generalized notions of quantum thermalization require NS~\cite{dt1,dt2,dt9,dt3,dt4,dt5,dt6,designs_via_free_probability,dt9,dt10,dt7,dt8} because ensembles of magic-free (stabilizer) states form quantum state $k$-designs  only with $k\leq 3$~\cite{cliff3design,cliffordgroupfailsgracefully,dt7,dt8}.
Magic resource has also been used in studying a variety of aspects of quantum many-body systems, including (symmetry-protected) topological order~\cite{magicMB1,magicMB8,wei2025longrangenonstabilizernessquantumcodes}, ground and critical states~\cite{magicMB2, magicMB4, magicMB6, LRMandPM,santra2025quantumresourcesnonabelianlattice,falcão2025magicdynamicsmanybodylocalized,PhysRevA.106.042426}, conformal field theories~\cite{magicMB3, cao2025gravitationalbackreactionmagical, magicMB10, hoshino2025stabilizerrenyientropyconformal}, quantum scars~\cite{magicMB7,magicMB5,sil2025quantumcomplexityconstrainedmanybody}, monitored and noisy dynamics~\cite{bejan2024prxq,FuxPRR,tirrito2025magicphasetransitionsmonitored, trigueros2025nonstabilizernesserrorresiliencenoisy, scocco2025risefallnonstabilizernessrandom, TarabungaTirrito2025,GuPRXQ, wang2025magictransitionmonitoredfree, zhang2024quantummagicdynamicsrandom, zhang2025designsmagicaugmentedcliffordcircuits, hou2025stabilizerentanglementenhancesmagic}, and quantum chaos~\cite{njgn-fksh, Passarelli2025chaosmagicin, santra2025complexitytransitionschaoticquantum,PhysRevB.100.125115,varikuti2025impactcliffordoperationsnonstabilizing}.

Unlike the accumulation of entanglement~\cite{Mezei_2017,nahumPRX17,cvkPRX18,bertiniPRX} or of NS~\cite{bejan2024prxq,Turkeshi_2025,tirrito2025anticoncentrationnonstabilizernessspreadingergodic,tirrito2025universalspreadingnonstabilizernessquantum} in ergodic quantum systems, the \emph{spreading} of NS under dynamics that conserves it remains largely unexplored, raising an important question: How does locally injected magic resource spread?
Answering this may be relevant practically for the optimal allocation of classical and quantum simulation resources, and physically for unveiling the local mechanisms behind quantum phenomena such as generalized thermalization.

\begin{figure}[tp]
    \centering
    \includegraphics[width=6.5cm]{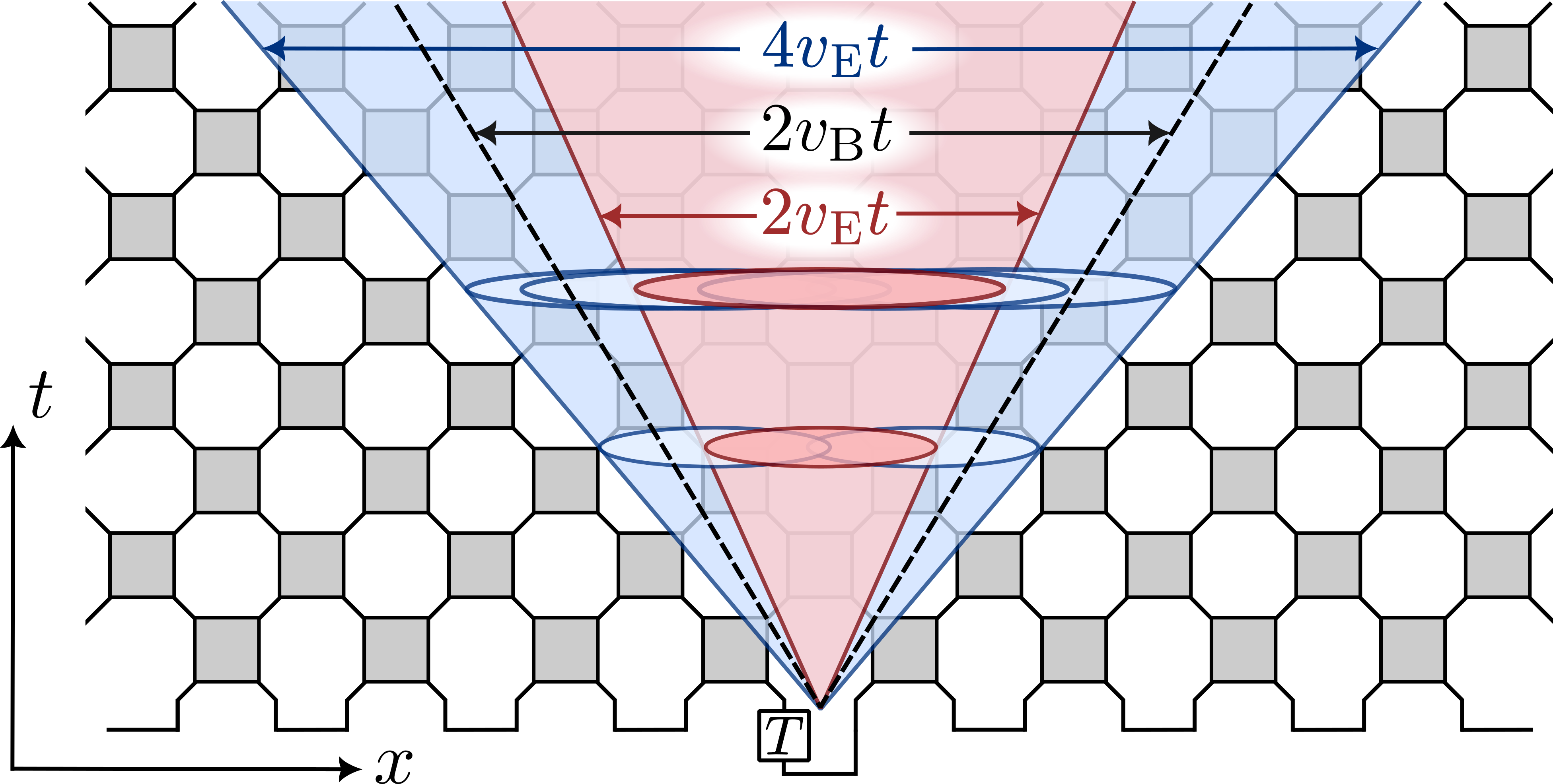}
    \caption{Magic resource spreading in a $T$-doped random Clifford circuit with $L$ qubits, time $t$, and a brickwall of random 2-qubit Clifford gates (grey). Magic resource is locally injected by a $T$ gate acting on a short-range entangled stabilizer state with the two central qubits in a Bell pair. The typical size of the smallest contiguous region where nonstabilizerness can be extracted (red ellipse) spreads at the entanglement velocity $v_{\rm E}$ (red cone). The union of the smallest mutually independent such intervals (blue and red ellipses) indicates the region outside of which operations cannot reduce the nonstabilizerness (blue cone), which spreads at $2v_{\rm E}$. For comparison, operators spread at the butterfly velocity $v_{\rm B}$ (black dashed cone).}
    \label{fig:circ_spreading}
    \vspace{-\baselineskip}
\end{figure}%

Here we study how magic resource, injected via a single local unitary in a short-range entangled state, spreads under random local Clifford unitary evolution---a model of scrambling quantum many-body dynamics. While Clifford dynamics conserves the total amount of NS, it creates entanglement and spreads operators, thereby spreading the initially localized magic resource. Our focus is precisely on the spatiotemporal distribution of this resource. 

We identify two magic length scales (MLS), which are the sizes of (\textit{i}) the smallest contiguous region from which magic resource can be extracted or destroyed, and (\textit{ii}) the region outside of which operations cannot reduce the NS [Fig.~\ref{fig:circ_spreading}]. A key perspective in identifying these length scales is that the time-evolved state encodes a single logical qubit through a dynamically generated stabilizer QEC code. This enables a mapping of the spatial structure of magic resource to that of the code's logical operators.

We rigorously prove in \autoref{thm:bip} the existence of a gauge of logical operators and stabilizers which reveals the spatial structure of magic resource, for any bipartition of the system. This provides an efficient classical algorithm to \textit{exactly} compute the NS in any subsystem, regardless of the amount of entanglement present. Conversely, exactly computing  NS in typical (especially volume-law entangled) states %whereas a naive calculation 
would be exponentially costly in (sub)system size.
Using our approach, we numerically find that, at early times, the two MLS spread ballistically at rates given by the entanglement velocity ($v_{\rm E}$) and twice thereof ($2v_{\rm E}$), respectively. 
Once these MLS reach their maximal values, magic resource delocalizes, i.e., it becomes extractable from any $\approx L/2$ qubits, while the QEC code approaches a random $[[L,1]]$ stabilizer~code.

\textbf{\textit{Quantum circuit setup.}}---We consider the dynamics of magic resource in unitary quantum circuits with $L$ qubits and depth $t$, cf. Fig.~\ref{fig:circ_spreading}. 
Each Clifford gate is chosen either as the identity gate or from a uniform distribution over the 2-qubit Clifford group~\cite{Nielsen2010} with probability $p \in [0,1]$ and $1-p$, respectively. 
Such identity-doped circuits admit, upon averaging, parametrically tunable butterfly $v_{\rm B}(p)$ and entanglement $v_{\rm E}(p)$ velocities~\cite{JYPC_PRR, SM}
\begin{align}
v_{\rm B}(p)=\frac{3}{5} \frac{1-p}{1+p},
v_{\rm E}(p)\approx \frac{\log(1-v_{\rm B})+\log(1+v_{\rm B})}{\log(1-v_{\rm B})-\log(1+v_{\rm B})}. \label{eq:vPdep}
\end{align}

Applying a magic-injecting non-Clifford gate on a stabilizer state dynamically generates a stabilizer quantum error correcting code~\cite{bejan2024prxq}.
For simplicity, we consider a pure short-range entangled stabilizer state $\rho_0 = \ketbra{\psi_0}{\psi_0}$ and apply a $T_j$ gate on qubit $j$, localizing the NS at this site. Because $T_j= \alpha \id + \beta Z_j$, this adds NS if $Z_j$ anticommutes with at least one stabilizer of $\rho_0$. We can always choose the generators of the stabilizer group $\mathcal{S}(\ket{\psi_0})$ such that $\mathcal{S}(\ket{\psi_0}) = \langle s_1, \mathcal{G}\rangle$, where $\mathcal{G} = \langle g_2, \dots, g_L\rangle$, with $[T,g_i]=0\ \forall i$ and $\{Z_j, s_1\} = 0$~\cite{Nielsen2010}. Thus, $\rho = T \rho_0 T^\dagger$ is
\begin{align}
    \rho = \ketbra{\psi}{\psi} = \frac{1}{2^L} \left( \mathds{1} + \frac{\ZL - \YL}{\sqrt{2}} \right) \prod_{i=2}^L \left( \mathds{1} +g_i \right),
    \label{eq:rho}
\end{align}where we used $T s_1 T^\dagger = (s_1 + i s_1 Z_j)/\sqrt{2}$ and defined $\XL \coloneqq Z_j, \ZL \coloneqq s_1$ and $\YL \coloneqq i \XL \ZL$. $\ket{\psi}$ encodes one logical qubit in $L$ physical qubits via a stabilizer QEC code~\cite{gottesman1997stabilizer, gottesman1998heisenberg} with stabilizer group ${\cal G}$ and logical operators $\XL, \YL, {\rm and}\ \ZL$~\cite{bejan2024prxq}.
For concreteness, we here take $\ket{\psi_0} = \bigotimes_{i=1}^{L/2}\ket{\Phi^+}_{2i-1,2i}$ defined from Bell states $\ket{\Phi^+}_{ij} = (\ket{0_i0_j}+\ket{1_i1_j})/\sqrt{2}$, locally inject NS by applying a non-Clifford gate $T\sim e^{-i\pi Z/8}$ on qubit $L/2$ and evolve the state using an identity-doped Clifford circuit.%

\textit{\textbf{Quantifying magic resource.}}---There exist multiple quantifiers of nonstabilizerness in a state $\rho$. In our setup, the presence of magic resource is essentially encoded in the structure of Eq.~\eqref{eq:rho} [see also Eq.~\eqref{eq:rho_bip}]. For illustration purposes, we focus on two commonly used quantifiers: the second stabilizer R\'enyi entropy (SRE)~\cite{sre} and the robustness of magic (ROM)~\cite{RoM}. While SRE is not a faithful magic measure for mixed states, since it can be nonzero for classical mixture of stabilizer states [e.g., $\rho_\lambda = \lambda \ketbra{0}{0} + (1-\lambda) \ketbra{1}{1}$ for any $\lambda \in (0,1), \lambda \neq 1/2$], we include it for comparison convenience. 
Conversely, ROM is a faithful magic measure even for mixed states.
Since ROM is also a magic monotone, it is particularly well-suited to quantifying NS in subsystems in our setup or, more generally, in setups with mixed initial states.

The second SRE for pure states is $\mathcal{M}_2 (\rho) \coloneqq S_2 (\Xi(\rho))-L$, with Pauli spectrum $\Xi (\rho)\coloneqq \{\Tr[\rho P]^2/2^L\}$ for a basis of $L$-qubit Pauli operators $\{P\}$~\footnote{The Pauli spectrum is defined in terms of a basis of $L$-qubit Pauli operators. Such a basis is formed by all Hermitian Pauli strings $P$ with phase $+1$.} and R\'enyi-2 entropy~$S_2$. For mixed states, the Pauli spectrum needs to be normalized~\cite{sre}, corresponding to subtracting the R\'enyi-2 entropy of~$\rho$, which leads to the mixed-state SRE
\begin{align}
   \tilde{\mathcal{M}}_2(\rho) \coloneqq \mathcal{M}_2(\rho) -S_2(\rho)=-\log_2 
\frac{\sum_P \Tr \left[P\rho \right]^4}{\sum_P \Tr\left[ P\rho \right]^2}. 
\label{eq:SRE}
\end{align}
The ROM of an $L$-qubit state $\rho$ is~\cite{RoM}
\begin{align}
    {\cal R}(\rho) \coloneqq \min_{\{x_i\}} \Big\{  \sum_i \vert x_i \vert \ \Big{\vert} \  \rho = \sum_i x_i \sigma_i, \ \sum_i x_i =1 \Big\}, \label{eq:rom}
\end{align}
where the decomposition is over \textit{all} pure $L$-qubit stabilizer states $\sigma_i$. For pure, mixed, and convex combinations (i.e., $x_i >0$ in the optimal decomposition) of stabilizer states, the ROM is ${\cal R}(\rho) = 1$, whereas for nonstabilizer states it satisfies ${\cal R}(\rho) > 1$~\cite{RoM}. Conversely, while $\tilde{\mathcal{M}}_2(\rho)=0$ for \textit{pure} stabilizer states, $\tilde{\mathcal{M}}_2(\rho) >0$ does not imply that $\rho$ is a nonstabilizer state if $\rho$ is \textit{mixed}. 

To quantify how magic resource is distributed across the system, we compute the SRE and ROM for various subsystems by tracing out their respective complements. If such a subsystem has the same amount of magic resource as the full system, we say that these are ``contained" within the extent of the subsystem; otherwise, the size of the subsystem needs to be increased until it holds the full amount of nonstabilizerness.

\textbf{\textit{Operational definitions of magic resource.}}---%
We next define two operational notions of nonstabilizerness linked to locality. We focus on a subsystem~$A \subset [L] \coloneqq \{1,\dots,L\}$ and pure states with a single global unit of nonstabilizerness; see~\cite{SM} for generalizations.

Our first operational definition is \textit{unitarily-extractable magic resource}.
Considering an ancilla qubit $C$ initialized as $\ket{0}_C$, we say magic resource is unitarily-extractable from $A$ if there exists a Clifford unitary $U = \id_B \otimes U_{AC}$ such that $U\kpsi \otimes \ket{0}_C = \ket{\psi'}_{AB} \otimes \ket{\psi'}_C$, with magic state $\rom(\ket{\psi'}_C) > 1$. In our setup, it suffices to focus on $\ket{\psi'}_C = \ket{T}_C$ with magic state $\ket{T}=(\ket{0}+e^{\frac{i\pi}{4}}\ket{1})/\sqrt{2}$.
As we show in~\cite{SM}, NS can be extracted from $A$ iff $\rho_A$ has a full magic unit, i.e., ${\cal R}(\rho_A) > 1$.
Moreover, by \autoref{thm:bip}, this is equivalent to all logicals being reducible to $A$~\cite{SM}.
Altogether, these equivalent notions of extractable magic resource correspond to NS (and logical information) being concentratable in $A$.
Here we excluded extracting NS to $C$ via measurements since these can teleport it---e.g., by measuring Pauli operators, which are not logicals, in a region $B$ that contains NS, we can move this NS to its complement $A$, even if $A$ contained no~NS.

Instead, measurements lead to our second operational definition, dubbed \textit{projectively-destroyable magic resource}.
We say magic resource is destroyable in a subsystem $A$ if there exists a Pauli operator fully supported in $A$ which, if measured, collapses $\kpsi$ to a stabilizer state~\footnote{This notion of magic is relevant to active quantum error correction and monitored quantum dynamics, where a key question related to the complexity of classically simulating the dynamics is which measurements reduce the amount of magic in a state~\cite{bejan2024prxq}.}. All possible Paulis that can destroy these are the logicals, and one can efficiently [in ${\cal O}(L^3)$ time] determine if there is any logical whose support can be reduced to $A$~\cite{GF2}. Nonetheless, in this work, we mainly focus on unitary extraction to avoid subtleties stemming from the teleportation of magic resource via measurements.

Computing the precise spatial structure of magic resource requires analyzing all possible subsystems $A$. This is generically hard because there are $2^L$ choices, requiring $\exp(L)$ runs. 
This challenge arises regardless of how hard it is to determine if a single subsystem contains nonstabilizerness.
To quantify the key features of this spatial structure, we hence focus on subsystems that are contiguous intervals\vphantom{\hphantom{\cite{BravyiTerhal2009, gullansPRXpurif}}}~\footnote{Focusing on contiguous systems also naturally leads to interesting connections between magic length scales for projectively-destroyable magic resource and the linear and contiguous code distances of QECC~\cite{BravyiTerhal2009, gullansPRXpurif}.}. This choice enables us to resolve the magic structure efficiently and points towards two relevant magic length scales, sharpening our view on the spatiotemporal structure of magic resource besides the broad structure already captured by the causal set and ``causal cone of magic'' proposed in Ref.~\cite{bejan2024prxq}.

\begin{table*}[htp]
    \centering
    \caption{
    Reducibility of logicals $\ZL, \YL$ to a subsystem and the subsystem magic resource. 
    All possible cases ($i$--$v$) are discussed in the proof of \autoref{thm:bip}.
    The columns labelled ``$X$ red.'' indicate whether the support of a logical $\sL$ can be reduced to be fully contained in subsystem $X$; \cmark~(\xmark) indicates yes (no). $\sre(\rho_A)$ and ${\cal R} (\rho_A)$ are the stabilizer 2-R\'enyi entropy and robustness of magic of $\rho_A$ [Eqs.~\eqref{eq:SRE} and~\eqref{eq:rom}], respectively.}
    \label{tab:bip}
    \resizebox{\textwidth}{!}{%
    \begin{tabular}{c
                    @{\hskip 1em} cc 
                    @{\hskip 1em} cc 
                    @{\hskip 1em} cc 
                    @{\hskip 1em} cc 
                    @{\hskip 1em} cc}
    \toprule
    \toprule
    \multirow{2}{*}{\diagbox[width=14em]{Quantity}{Case}} 
        & \multicolumn{2}{c}{($i$)} 
        & \multicolumn{2}{c}{($ii$)} 
        & \multicolumn{2}{c}{($iii$)} 
        & \multicolumn{2}{c}{($iv$)} 
        & \multicolumn{2}{c}{($v$)} \\
    \cmidrule(lr){2-3} \cmidrule(lr){4-5} \cmidrule(lr){6-7} \cmidrule(lr){8-9} \cmidrule(lr){10-11}
    & \multicolumn{1}{c}{$A$ red.} & \multicolumn{1}{c}{$B$ red.} 
    & \multicolumn{1}{c}{$A$ red.} & \multicolumn{1}{c}{$B$ red.} 
    & \multicolumn{1}{c}{$A$ red.} & \multicolumn{1}{c}{$B$ red.} 
    & \multicolumn{1}{c}{$A$ red.} & \multicolumn{1}{c}{$B$ red.} 
    & \multicolumn{1}{c}{$A$ red.} & \multicolumn{1}{c}{$B$ red.} \\
    \midrule
    Support of $\ZL$ 
    & \multicolumn{1}{c}{\cmark} & \multicolumn{1}{c}{\xmark} 
    & \multicolumn{1}{c}{\xmark} & \multicolumn{1}{c}{\cmark} 
    & \multicolumn{1}{c}{\xmark} & \multicolumn{1}{c}{\xmark} 
    & \multicolumn{1}{c}{\cmark} & \multicolumn{1}{c}{\cmark} 
    & \multicolumn{1}{c}{\xmark} & \multicolumn{1}{c}{\xmark} \\
    Support of $\YL$ 
    & \multicolumn{1}{c}{\cmark} & \multicolumn{1}{c}{\xmark} 
    & \multicolumn{1}{c}{\xmark} & \multicolumn{1}{c}{\cmark} 
    & \multicolumn{1}{c}{\xmark} & \multicolumn{1}{c}{\xmark} 
    & \multicolumn{1}{c}{\xmark} & \multicolumn{1}{c}{\xmark} 
    & \multicolumn{1}{c}{\cmark} & \multicolumn{1}{c}{\cmark} \\
    SRE $\sre (\rho_A)$ 
      & \multicolumn{2}{c}{$\log_2(4/3)$} 
      & \multicolumn{2}{c}{0} 
      & \multicolumn{2}{c}{0} 
      & \multicolumn{2}{c}{$\log_2(6/5)$} 
      & \multicolumn{2}{c}{$\log_2(6/5)$} \\
    ROM ${\cal R} (\rho_A)$ 
      & \multicolumn{2}{c}{$\sqrt{2}$} 
      & \multicolumn{2}{c}{1} 
      & \multicolumn{2}{c}{1} 
      & \multicolumn{2}{c}{$1$} 
      & \multicolumn{2}{c}{$1$} \\
    \bottomrule
    \bottomrule
    \end{tabular}
    }
    \end{table*}
    
\textbf{\textit{Magic length scales.}}---%
The spatial structure of magic resource is captured by the set of mutually-irreducible intervals where this can be extracted. Multiple intervals may arise since the logical information of a QECC may be altered by physically acting on distinct qubit sets, as reflected in the freedom of choosing the generators~\footnote{Here, by generators, we mean both the logical and stabilizer generators of, for example, $\rho$ from Eq.~\eqref{eq:rho}.}. 
Consider two magic-extractable subsystems $A_1$ and $A_1'$ with $A_1$ contained in $A_1'$ (i.e., $A_1 \subset A_1'$). Since NS is extractable from $A_1$ alone, $A_1'$ is redundant. Let another magic-extractable subsystem $A_2$ be contained in $A_1'$ ($A_2 \subset A_1'$) but independent from $A_1$ ($A_1 \not\subset A_2$ and $A_2 \not \subset A_1$):
By the same logic, $A_2$ is more precise than $A_1'$, however, it is equally relevant as $A_1$ as they are mutually irreducible. 

\begin{figure}[h]
    \centering
    \includegraphics[width=0.6\linewidth]{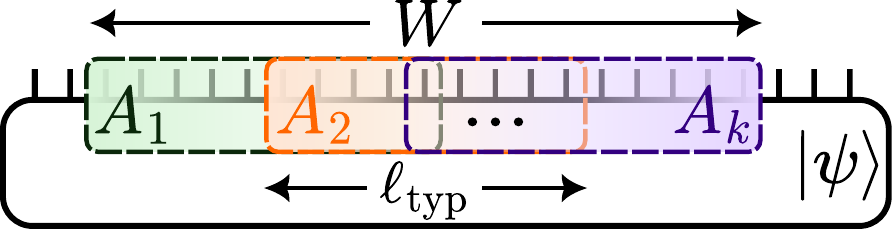}
    \caption{Magic length scales: typical magic length $\ell_{\rm typ}$ and full linear extent of magic $W$, defined in terms of mutually-irreducible and minimal-length contiguous subsystems $\{A_i\}_{i=1}^k$ where nonstabilizerness is extractable from~$\kpsi$.}     %
    \label{fig:MLMI}
    \vspace{-\baselineskip}
\end{figure}

We collect these \textit{minimal linear magic intervals} (MLMI) in the set ${\cal I}(\rho) = \{ A_1, \dots, A_k \}$ [Fig.~\ref{fig:MLMI}]. MLMIs must mutually overlap: Suppose there exist two non-overlapping MLMI $A_1$ and $A_2$, where NS is extractable. By definition, one can extract the nonstabilizerness with local operations on $A_2$ only; however, by the same definition, any operations on the complement of $A_1$, which includes $A_2$, cannot reduce the magic resource in $A_1$. Hence, we arrive at a contradiction. To obtain these subsystems in practice, one first finds all magic-extractable contiguous intervals, then applies the above logic recursively.

The first magic length scale is the typical length $\ell_{\rm typ}$ of the MLMIs [Fig.~\ref{fig:MLMI}]. 
To quantify the spreading of magic resource, we consider how $\ell_{\rm typ}$ grows. For illustration purposes, we introduce a related quantity: We define the \textit{linear magic length} (LML) $\ell(t)$ as the length of the shortest left-right symmetric and centered (on the injection site) interval where NS is extractable. The LML grows within the geometric lightcone of the non-Clifford gate, unlike $\ell_{\rm typ}$, which can inherit subtleties from entanglement already present in the state~\cite{SM}. Moreover, the growth rate of $\ell(t)$ is consistent with that of $\ell_{\rm typ}(t)$~\cite{SM}.
The second length scale is the width of the union of the MLMIs [Fig.~\ref{fig:MLMI}], which we call the \textit{full linear extent of magic} (FLEOM) and denote by $W \coloneqq  {\rm width}\left( \cup_{A \in {\cal I}} A \right)$. Physically, local operations such as measurements, unitaries, decoherence, etc., applied outside of $\cup_{A\in \mathcal{I}} A$ cannot reduce the total NS in the system.

We can compute these two MLS efficiently, i.e., with poly$(L)$ classical resources.
For contiguous subsystems, the computational hardness comes from assessing a subsystem's nonstabilizerness. In the Supplemental Material~\cite{SM}, we provide efficient algorithms for computing both the subsystem magic $\sre(\rho_A), {\cal R}(\rho_A)$, and Pauli spectrum $\Xi(\rho_A)$ exactly. While the latter might naively have an exponential $\sim 4^{|A|}$ cost~\cite{sre}, the following construction significantly reduces the cost by exposing the local structure of operators in Eq.~\eqref{eq:rho}.

\textit{\textbf{Bipartite magic gauge.}}---We introduce a gauge of stabilizer generators and logical operators of $\rho$ for any bipartition, which enables the classical computation of the ROM and Pauli spectrum, and thus SRE, in $\mathcal{O}(L^3)$~time. We dub this the \textit{bipartite magic gauge} (BMG). We rigorously show its existence and provide an algorithm for its construction in~\cite{SM}, assuming one knows the stabilizers of $\rho$. 
This is our first key analytical result. 

\begin{theorem} \label{thm:bip}
    Let $\kpsi$ be a pure $L$-qubit state with one unit of nonstabilizerness (i.e., one unitarily-extractable magic state), and encoding one logical qubit, where $\rho = \kpsi\bra{\psi} = \frac{1}{2^{L}} ( \id + \frac{\ZL - \YL}{\sqrt{2}} ) \prod_{j=2}^L (\id + g_j)$ with logical operators $\sL$ and stabilizer group ${\cal G}=\langle g_2, \dots, g_L\rangle$.
    For any bipartition into subsystem $A$ and its complement $B$, there exists a \textbf{bipartite magic gauge}, i.e., a choice of logical operators and stabilizer generators such that
    \begin{align}
        \rho = %
        \left(\frac{\id}{2} + \frac{\ZL - \YL}{2\sqrt{2}} \right)
        \prod_{i=1}^{n_A} \frac{\id + a_i}{2} \prod_{j=1}^{n_B}\frac{\id +  b_j}{2} \prod_{k=1}^{m} \frac{\id + h_k}{2}
        \label{eq:rho_bip}
    \end{align}
    with $L -1 = n_A + n_B + m$, where the support of stabilizers $a_i, b_j$, and $h_k$ can(not) be reduced to $A$($B$), $B$($A$), and neither $A$ nor $B$, respectively. 
    Furthermore, using the reducibility of the logicals $\ZL$ and $\YL$, we can compute the magic resource of subsystem $A$, as summarized in \autoref{tab:bip}.
\end{theorem}

To guide our intuition for the various cases in \autoref{tab:bip}, it is useful to consider our second rigorous result [\autoref{lemma:bip} (proven in the End Matter)] on the spatial structure of logicals. The magic resource of a subsystem~$A$ depends on whether logical operators, which enter in $\rho$, survive the tracing $\Tr_B[\cdot]$; how many stabilizers survive is unimportant since their contribution to $\sre$ from ${\cal M}_2$ is canceled by that of the entanglement 2-R\'enyi entropy $S_2$, cf. Eq.~\eqref{eq:SRE}, and also for ${\cal R}(\rho_A)$. 
It is crucial to determine if these logicals can be chosen to be fully supported on~$A$, and this is where \autoref{lemma:bip} comes into play.

\begin{lemma} \label{lemma:bip}
    For any stabilizer code encoding one logical qubit in the pure state of $L$ qubits, and a bipartition into subsystems $A$ and $B= [L] \setminus A$, it is true that either 
    (i) all three logicals $\bar{X},\bar{Y},\bar{Z}$ are reducible to $A$, or (ii) all three are reducible to~$B$, or (iii) only one logical can be supported only on $A$ or $B$, while the other two are supported on both $A$ and $B$.
\end{lemma}

A few comments are in order. 
First, the states $\kpsi$ considered in \autoref{thm:bip} extend beyond unitary circuits [Fig.~\ref{fig:circ_spreading}] to monitored Clifford+$T$ circuits, provided measurements do not remove the magic resource.
Second, the logicals $\ZL, \YL$ and prefactors $a_\pm = \pm1/\sqrt{2}$ are generic\footnote{We can relabel the logicals $\XL, \YL,$ and $\ZL$ as needed, provided they follow the \(\mathfrak{su}(2)\) anti-commutation rules; the prefactors $a_\pm$ depend on the non-Clifford gate used to produce $\kpsi$, but for suitably chosen $\ZL$ and $\YL$, they satisfy $|a_\pm| \neq 0$, ensuring that $\sre(\rho_A) > 0$ and ${\cal R}(\rho_A)>1$ for a subsystem \(A\) where at least one logical can be reduced. The specific values of $|a_\pm|$ set the values of ``full'' [case ($i$)] and ``partial'' [cases ($iv, v$)] NS. See also~\cite{SM}.}; the phenomenology holds for other local non-Clifford gates, with $\rom (\rho_A) > 1$ indicating the presence of nonstabilizerness. Third,
while \autoref{thm:bip} considers $\kpsi$ with a single magic unit, or $[[L,1]]$ stabilizer codes, we anticipate that it can be extended to states with multiple ($k\geq 1$) magic units, corresponding to $[[L,k]]$ stabilizer codes.

\textbf{\textit{Phenomenology of magic resource~spreading.}}---Here we report numerical results on the growth of LML and FLEOM and heuristically argue their early time growth rates. 

\begin{figure}[tp]
    \centering
    \includegraphics[width=8.6cm]{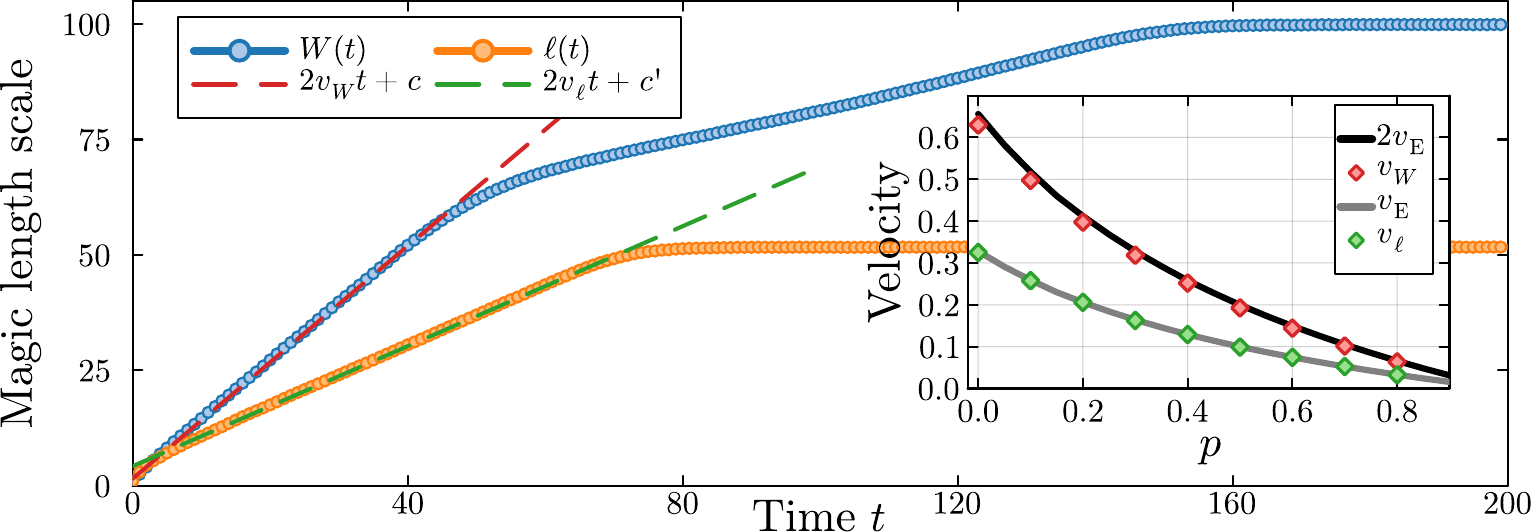}
    \caption{
    Growth of magic length scales over time in random Clifford circuits for $L=102$ and identity doping rate~$p=0$. Blue (orange) circles show FLEOM (LML).  
    At early times, both MLS grow ballistically at fitted rates $2v_W$ and $2v_{\rm \ell}$, respectively (dashed lines). 
    Inset: $p$-dependence of $v_W$ and $v_{\rm \ell}$ compared to the entanglement velocity $v_{\rm E}(p)$.
    Data averaged over 2$\times 10^3$ circuit realizations, negligible error bars not shown.
    }
    \vspace{-\baselineskip}
    \vspace{-\baselineskip}
    \label{fig:width_rand_cliff}
\end{figure}

Focusing on random Clifford circuits with open boundary conditions (BC) at doping rate $p$ and extractable NS for FLEOM and LML, we find that, for early times, both $W(t)$ and $\ell(t)$ grow ballistically at rates $v_W$ and $v_{\rm \ell}$, respectively [Fig.~\ref{fig:width_rand_cliff}].
While LML saturates to $\ell(t\rightarrow \infty)\simeq L/2$, the ballistic growth of FLEOM slows down at intermediate times~\footnote{For intermediate times, the data from Fig.~\ref{fig:width_rand_cliff} is consistent with a growth rate of FLEOM given by $\partial_t W \sim 4v_{\rm E}/3$. We attribute the distinct intermediate-time growth of $W(t)$ to a combination of effects due to boundary conditions and shrinking of large logicals, as detailed in~\cite{SM}. }, only to later saturate at $W(t\rightarrow \infty)=L$. 
Throughout the entire $p$ range, the $p$-dependence of the early-time velocities agrees remarkably well with that of the entanglement velocity $v_{\rm E}(p)$ [Eq.~\eqref{eq:vPdep}], suggesting that $v_W(p) \simeq 2 v_{\rm E}(p)$ and $v_{\ell} (p) \simeq v_{\rm E}(p)$ [Fig.~\ref{fig:width_rand_cliff} inset].
This agreement is further established in~\cite{SM} using results on doped dual-unitary circuits and periodic BC.

Magic resource spreading at $\partial_t W/2 \simeq 2v_{\rm E}$ can be faster than the geometric lightcone with $v_{\rm LC}=1$, but does this violate causality? No. Magic resource spreading is influenced by both entanglement growth and operator spreading, giving rise to an upper bound $[W(t)-W(0)]/2t \leq 3v_{\rm LC}$~\cite{SM}. To shed light on this interplay (see~\cite{SM} for details), it is instructive to consider  
    $UT \kpsi = (UTU^\dagger) U \kpsi,$
for Clifford unitary $U$ and short-range entangled $\kpsi$. While $UTU^{\dagger}$ spreads at the butterfly velocity $v_{\rm B}$, entanglement builds up in $U\kpsi$ at rate $v_{\rm E}$, and both contribute to nonstabilizerness spreading. What is nontrivial is how the structure of the operator $UTU^\dagger$ interacts with that of the entanglement in $U\kpsi$. 

That $v_{\ell} \approx v_{\rm E}$ can be understood through the dynamics of purity. To link purity to magic resource, a key observation is that if $\rho_A = \frac{\id_A}{d_A}$ is maximally mixed, then subsystem $A$ cannot host NS.
For random circuits where entanglement grows ballistically, $v_{\rm E}$ corresponds to the decay rate of the circuit-averaged subsystem purity
$\overline{{\cal P}(\rho_A(t))} = \overline{\Tr[\rho^2_A(t)]}$~\cite{nahumPRX17,cvkPRX18}.
The decay $\overline{{\cal P}(\rho_A(t))}\propto e^{-v_{\rm E}t}$ implies that, starting from a product state, at time $t$ any $\rho_A$ with $A$ contiguous and $|A| \lesssim 2v_{\rm E}t$ ($|A| \gtrsim 2 v_{\rm E}t$) is (not) maximally mixed with high probability; hence, $\ell(t) \gtrsim 2v_{\rm E}t$ with high probability.
Conversely, for $A$ to host NS, by counting the logicals in $\Xi(\rho_A)$, $A$ must be in the causal cone set by entanglement and operator spreading and satisfy $|A| \gtrsim 2v_{\rm E} t$~\cite{SM}. Focusing on symmetric-centered intervals, there hence exists with high probability a subsystem $A_\epsilon$ supporting at least one logical, with $|A_\epsilon| = (1+\epsilon)2v_{\rm E}t$ for some $0<\epsilon \ll 1$, and thus, $\ell(t) = |A_\epsilon|
 \implies \partial_t \ell = 2v_{\rm E} (1+\epsilon) \simeq 2v_{\rm E}$. 
Beyond the scope of this heuristic argument, an interesting open problem is to find the scaling of $\epsilon$ with system size $L$. We lastly remark on a similarity between the growth of LML and that of entanglement in a stabilizer state. There, $v_{\rm E}$ characterizes the rate at which each stabilizer generator grows when imposing a collective constraint---the ``clipped gauge''~\cite{nahumPRX17}. In our setup, $v_{\ell}$ characterizes the rate at which both logical generators grow 
under the constraint that the interval containing them is both \textit{minimal} and \textit{centered}. Remarkably, despite these
different constraints, the growth rates $v_\ell$ and $v_{\rm E}$ agree exceptionally well.

A way to understand the growth rate $v_W \simeq 2v_{\rm E}$ at early times is as follows. Consider the left- and rightmost MLMIs. Assuming these two MLMIs have typical sizes $\ell_{\rm typ}(t)$, then $W(t)\lesssim2\ell_{\rm typ}(t)$ by the mutual-overlap condition. 
Because the circuit dynamics is otherwise structureless, it is reasonable to expect that there is no other mechanism constraining the spreading of MLMIs. One might naively anticipate an ${\cal O} (1)$ overlap between the left- and rightmost MLMIs; however, due to operator fronts broadening (see~\cite{SM} for a depiction)~\cite{nahumPRX18, cvkPRX18}, we instead expect an ${\cal O} (\sqrt{t})$ overlap. We numerically confirm, as detailed in~\cite{SM}, that the left- and rightmost MLMIs indeed overlap on $~\sim t^{0.45}$ qubits at early times. 
 Altogether, since $\ell_{\rm typ} (t)\simeq  2v_{\rm E}t$, we thus expect to leading order $W(t) \simeq 2 \ell_{\rm typ}(t)$, which implies $W(t)\simeq 4v_{\rm E}t$.

\textbf{\textit{Magic resource delocalizes at late times.}}---After a saturation time $t_{\rm sat}\sim \mathcal{O}(L)$, the FLEOM reaches the full system size, indicating that nonstabilizerness delocalizes [Fig.~\ref{fig:width_rand_cliff}]. Physically, in this regime, the full amount of nonstabilizerness can be unitarily extracted from \textit{any} (i.e., not necessarily contiguous) subsystem with $\gtrsim L/2$ qubits [see Fig.~\ref{fig:channel_cap}(a)]. In other words, nonstabilizerness no longer has a notion of locality since the spatial proximity of qubits becomes irrelevant, and an extensive number of qubits $\gtrsim L/2$ is needed to distill a magic state. In contrast, to unitarily extract a magic state at earlier times, subsystems with at least $\approx \ell_{\rm typ}(t)$ qubits are needed, and these subsystems typically overlap with or are contained within the FLEOM's lightcone. Note that the LML saturates before the FLEOM and approaches $\ell(t\to \infty) \simeq L/2$ [Fig.~\ref{fig:width_rand_cliff}].
In the End Matter, by considering the dynamical QECC, we argue that, at late times, this code approaches a random $[[L,1]]$ stabilizer code, which is known to saturate the so-called quantum Singleton bound~\cite{qSingleton, gottesman1997stabilizer}, thereby explaining the saturation value $\ell(t\to \infty) \simeq L/2$. By using a proxy for the quantum channel capacity of a channel that erases a finite fraction of randomly chosen qubits~\cite{schumacher_quantum_1996,lloyd_capacity_1997,  bennett_capacities_1997}, we also numerically show that nonstabilizerness can be unitarily extracted from any $\gtrsim L/2$ qubits.

\textbf{\textit{Discussion and outlook.}}---%
We have shown, for pure states with one magic unit and using insights from QECC, how to find the spatiotemporal structure of magic resource. Specifically, we proved the existence of a bipartite magic gauge and, based on it, developed algorithms to exactly compute subsystem nonstabilizerness efficiently.
We proposed two length scales that are operationally relevant to magic distillation protocols. We reported numerical evidence for a ballistic growth of these length scales in random Clifford circuits, followed by magic delocalization at late times.

Moving forward, while we provided analytical bounds and heuristic arguments for the typical magic velocities, it would be desirable to understand more rigorously the underlying physical mechanisms, e.g., through coarse-grained models and hydrodynamical descriptions of nonstabilizerness spreading.
It would also be interesting to generalize \autoref{thm:bip} to states with multiple units of magic, and extend the associated algorithm (see also~\cite{aziz2025classicalsimulationslowmagic,cao2025efficientalgorithmcomputeentanglement}). To this end, we note that our operational definitions extend to such states, cf.~\cite{SM}. 
Using these to examine ``interaction effects" between spatially separated magic sources in monitored and open quantum dynamics would be appealing~\cite{fisher2023rqc,bejan2024prxq,FuxPRR}. More generally, exploring how quantum resources (including NS, non-Gaussianity, and coherence) spread may clarify how complexity builds up in quantum many-body systems~\cite{Monta_L_pez_2024,sierant2025fermionicmagicresourcesquantum, aditya2025mpembaeffectsquantumcomplexity}, how these systems deeply thermalize~\cite{dt1,dt2,dt9,dt3,dt4,dt5,dt6,dt7,dt8,dt9,dt10}, and how logical information is causally organized in the spacetime of dynamical QEC codes~\cite{Cotler_2019}.

\textit{\textbf{Acknowledgments.}}---M.B. thanks Benjamin B\'eri and Campbell McLauchlan for insightful discussions and collaboration on related work. J.Y. acknowledges helpful discussions with Philippe Suchsland, Xhek Turkeshi, and Sagar Vijay. 
This work was supported by EPSRC PhD Studentship 2606484 and UKRI FLF grant MR/Z000297/1.
P.W.C. and J.Y. acknowledge support from the Max Planck Society.

\textit{Note added.}---While finalizing this manuscript, we became aware of an independent work by Maity and Hamazaki~\cite{maity2025localspreadingstabilizerrenyi}, which studies the spreading of local magic resource under Clifford dynamics. While they use a different approach and focus on distinct spatiotemporal structures, our results are consistent where they overlap: Certain magic length scales grow ballistically until nonstabilizerness delocalizes at late times.

\newpage
\begin{onecolumngrid}
\begin{center}
    \textbf{{\fontsize{11}{11}\selectfont End Matter}}
\end{center}
\medskip
\end{onecolumngrid}
\appendix
\begin{twocolumngrid}

\textbf{\textit{Appendix A: Proof of \autoref{lemma:bip}.}}---
Here we prove one of our key technical results, \autoref{lemma:bip}, which exposes the spatial structure of both logicals and stabilizer generators of $\kpsi$. The proof of our other key technical result, \autoref{thm:bip}, follows from this Lemma and is found in~\cite{SM}.

Let the stabilizer group of our code be $\mathcal{G} = \langle g_2, \dots, g_L\rangle$, where $g_i$ are the stabilizer generators. The logical operators of this code are nontrivial Hermitian Pauli operators $\bar{\sigma}$ that commute with all generators $[\bar{\sigma}, g]=0\  \forall g \in \mathcal{G}$ and are not elements of $\pm\mathcal{G}$. Thus, a logical operator $\bar{\sigma}$ is in one of the three cosets $\bar{X}\mathcal{G}, \bar{Y}\mathcal{G},$ and $\bar{Z}\mathcal{G}$ (up to a sign).

It is useful to represent the code and its logical operators through a pure stabilizer state $\ket{\Phi_{ABR}}$ on an enlarged system with an additional reference qubit $R$~\cite{lee2024randomlymonitoredquantumcodes}. 
We depict this in Fig.~\ref{fig:encoding}, where for illustration purposes we take subsystems $A$ and $B$ as contiguous regions; however, our proof holds for any bipartition with $B = [L] \setminus A$.
The stabilizer group of $\ket{\Phi_{ABR}}$ is defined as $\mathcal{S} \coloneqq \langle\bar{X}X_R, \bar{Z}Z_R, g_2, \dots, g_L\rangle$, which also contains $-\bar{Y}Y_R$.

Using this mapping, we shall determine whether logical operators $\bar{\sigma}$ of the code can be fully supported on a subsystem ($A$ or $B$). To illustrate this, note that any logical $\bar{\sigma}$ corresponds to a stabilizer of $\ket{\Phi_{ABR}}$ which has nontrivial support on the reference qubit~$R$, whereas any stabilizer $g \in {\cal G}$ of $\kpsi$ is represented by a stabilizer of $\ket{\Phi_{ABR}}$ with trivial support on $R$. Hence, to determine the spatial structure of the logicals, it suffices to work with the corresponding stabilizers of $\ket{\Phi_{ABR}}$.

\begin{figure}[h]
    \centering
    \includegraphics[width=7cm]{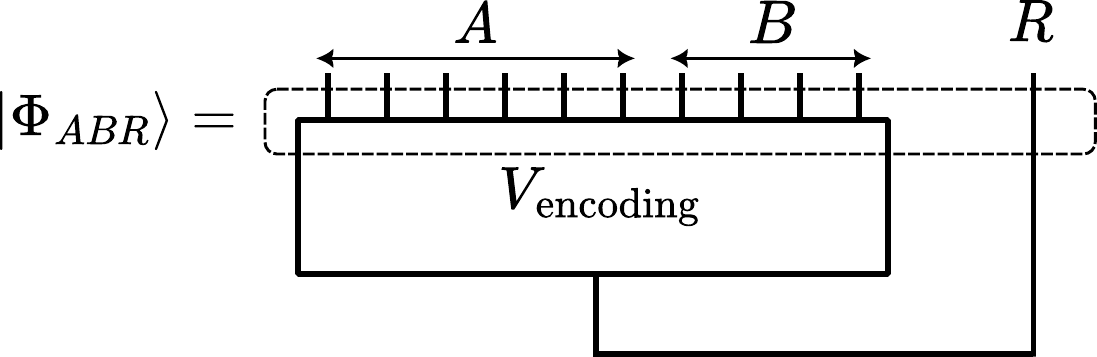}
    \caption{Representation of a $[[n,k]]$ stabilizer code with $n$ qubits and $k$ logical qubits as a pure stabilizer state $\ket{\Phi_{ABR}}$ on an enlarged system $ABR$, with $n=|AB|$ and $k=|R|$.
    $\ket{\Phi_{ABR}}$ is the Choi state of the isometry $V_{\rm encoding}$.
    The spatial structure of the logical operators, for any subsystem $A$ and its complement $B$, can be inferred from the spatial structure of the stabilizers of $\ket{\Phi_{ABR}}$, as discussed in the proof of \autoref{lemma:bip}.
    }
    \label{fig:encoding}
\end{figure}

$\ket{\Phi_{ABR}}$ is locally Clifford unitarily (LCU) equivalent, i.e., with Clifford $U \equiv U_A \otimes U_B \otimes U_R$, to a tensor product of tripartite GHZ states, EPR (Bell) pairs, and single-qubit states~\cite{GHZ_trio}.
Here, since $k=|R|=1$, there can be at most one GHZ state $\ket{{\rm GHZ}_{abR}} = \frac{1}{\sqrt{2}}(\ket{0_a0_b0_R} + \ket{1_a 1_b 1_R})$ for some qubits $a \in A, b \in B$; $\ket{{\rm GHZ}_{abR}}$ has stabilizers in $\langle X_a X_b X_R, Z_a Z_b, Z_b Z_R \rangle$. The EPR pairs can be between any two subsystems out of $A,B$ and $R$, and are defined as $\ket{\Phi^+_{ij}} = \frac{1}{\sqrt{2}}(\ket{0_i 0_j} + \ket{1_i 1_j})$ for qubits $i$ and $j$, with stabilizers in $\langle X_i X_j, Z_i Z_j \rangle$. Since the reference qubit $R$ is maximally entangled with the subsystem $AB$ for $\ket{\Phi_{ABR}}$ and $U$ is local, these remain entangled, and thus, after the LCU, subsystem $R$ must be involved in either a GHZ state or an EPR pair. 
Hence, we can choose exactly two generators of $U\ket{\Phi_{ABR}}$ that are supported on $R$ with the operators $X_R$ and $Z_R$; these are not necessarily $U\bar{X}X_R U^\dagger$ and $U\bar{Z}Z_R U^\dagger$ since, when choosing these two generators, we might need to multiply $U\bar{\sigma}\sigma_R U^\dagger$ by some $UgU^\dagger$ for $g\in\mathcal{G}$ and we can also choose $\sigma = Y$. 

Next, focusing on $U\ket{\Phi_{ABR}}$, we proceed into three cases depending on which entangled state $R$ belongs to: ($i$) $\ket{\Phi^+_{aR}}$ for qubit $a\in A$, ($ii$) $\ket{\Phi^+_{bR}}$ for $b\in B$, or ($iii$) $\ket{{\rm GHZ}_{abR}}$ for $a \in A$ and $b \in B$. We take $U_R = \id_R$ for simplicity, but our conclusions hold for any Clifford gate $U_R$ (this can only permute how we label our logicals).

Case ($i$). Since the stabilizers of $\ket{\Phi^+_{aR}}$ are in $\langle X_a X_R, Z_a Z_R\rangle$, we can choose the representatives of the logical operators (encoded in $U\ket{\Phi_{ABR}}$) to be $\bar{X}' = X_a$ and $\bar{Z}' = Z_a$, thus, $\bar{Y}' = Y_a$. Reverting the LCU transformation $U\ket{\Phi_{ABR}} \mapsto U^\dagger U\ket{\Phi_{ABR}} = \ket{\Phi_{ABR}}$, we find that the representatives of the logical operators of our initial stabilizer code (encoded in $\ket{\Phi_{ABR}}$) can be taken $\bar{\sigma} = U^\dagger \bar{\sigma}' U$ for $\bar{\sigma}' \in \{ X_a, Y_a ,Z_a \}$. Hence, since $U$ is local, for each coset $\bar{\sigma} \mathcal{G}$, there exist representative logical operators $\bar{\sigma}$ that are simultaneously fully supported on subsystem $A$, i.e., $\bar{\sigma}\vert_B = \id_B$.

Case ($ii$). This is analogous to case ($i$), with $A$ replaced by $B$, and we find that there exist representative logical operators $\bar{\sigma}$ that are fully supported on subsystem~$B$.

Case ($iii$). Here, the two logical generators representatives $\bar{X}'X_R$ and $\bar{Z}'Z_R$, i.e., the generators of $U\ket{\Phi_{ABR}}$ supported on $R$, must be chosen from the stabilizer group of the GHZ state $\langle X_a X_b X_R, Z_a Z_b, Z_b Z_R\rangle$. Thus, we can take without loss of generality $\bar{X}' = X_a X_b$ and $\bar{Z}' = Z_a$ (or $\bar{Z}' = Z_b$), leading to $\bar{Y}' = Y_a X_b$ (or $\bar{Y}' = X_a Y_b$). Then, because $U$ is local, reverting the LCU does not change whether a logical $\bar{\sigma} = U^\dagger \bar{\sigma}' U$ is supported on subsystem $A$, or $B$, or both. Hence, two logicals (here, $\bar{X}$ and $\bar{Y}$) are supported on both $A$ and $B$, while the third (here, $\bar{Z}$) can be chosen to be supported only on either $A$ or $B$; moving the third logical from $A$ to $B$ and vice versa can be done by multiplying it by a stabilizer $g \in {\cal G}$ (here, $Z_aZ_b$).   
This concludes our proof.
\hfill $\square$

\begin{figure*}[htp]
    \centering
    \includegraphics[width=17.2cm]{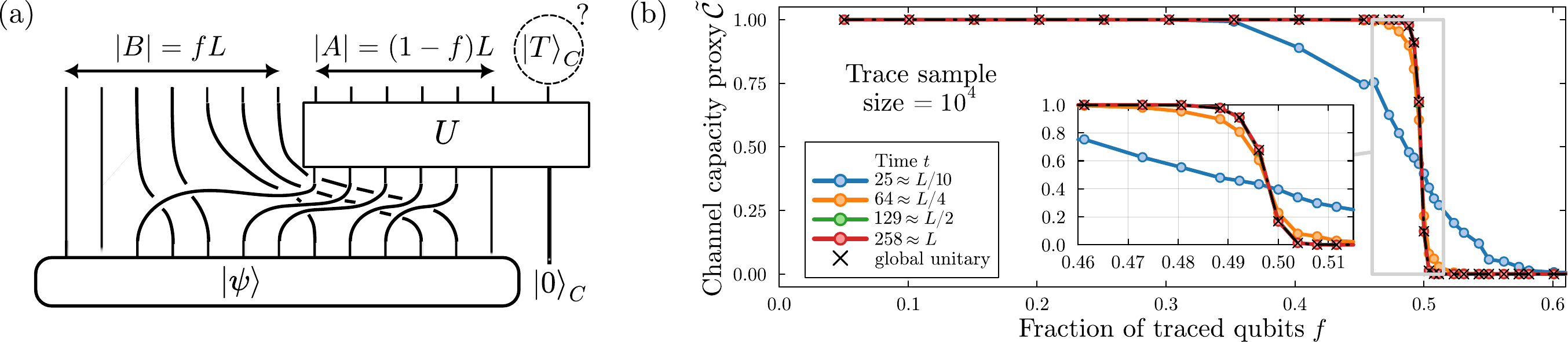}
    \caption{(a): Quantum circuit setup for unitarily extracting the magic resource from subsystem $A$ of state $\kpsi$ onto an ancilla qubit $C$ into a magic state $\ket{T}_C$ using a Clifford unitary $U$. The complementary subsystem $B$ contains a finite fraction $f$ of randomly chosen qubits. At late times, the extraction succeeds with high probability for any $A$ with $|A|\gtrsim L/2$.
    (b): Channel capacity proxy $\tilde{\cal C}$ of the channel that traces out $fL=|B|$ randomly chosen qubits after evolution up to time $t$ under a random brickwork Clifford circuit with identity doping $p=0.1$ (circles) and a global random Clifford (crosses). For $t\gtrsim t_{\rm sat}$, $\tilde{\cal C} = {\cal O}(1)$ for $f \leq 0.5$ indicating the preservation of magic resource; for $f \gtrsim 0.5$, $\tilde{\cal C}$ quickly drops to zero and magic resource is lost (see inset). Data averaged over $10^4$ realizations of $B$ for a single random circuit instance, negligible error bars not shown.}
    \label{fig:channel_cap}
\end{figure*}

\textbf{\textit{Appendix B: Details on magic resource delocalization at late times.}}---By focusing on QEC aspects of the dynamically generated stabilizer code, we here show that this code approaches a random $[[L,1]]$ stabilizer code at late times, which in turn implies that one can unitarily distill a magic state from \textit{any} subsystem with $\gtrsim L/2$ qubits, and thus that magic resource delocalizes.

To show this, we consider a subsystem $B$ with a finite fraction $f = |B|/L$ of qubits chosen at random, and ask if magic can be unitarily extracted from its complement~$A \coloneqq [L] \setminus B$, as depicted in Fig.~\ref{fig:channel_cap}(a). By \autoref{lemma:bip}, this question is equivalent to asking if the logical information of the code can be transmitted through a quantum channel ${\cal N}_B = {\cal E}_B \circ U$, with a Clifford unitary $U$ and an error channel ${\cal E}_B$, chosen as either tracing out or measuring in a complete stabilizer basis of $B$. To answer the latter question, we use a proxy $\tilde{\cal C}({\cal N}_B[\rho])$ for the single-copy channel capacity~\cite{schumacher_quantum_1996,lloyd_capacity_1997,  bennett_capacities_1997}, with $\tilde{\cal C} = 1\ (0)$ if the logical qubit is (not) preserved.

We can evaluate the single-copy channel capacity proxy $\tilde{\cal C}$ through the coherent information $I_c(\Psi_{AR})$ by using the construction from the proof of~\autoref{lemma:bip} [Fig.~\ref{fig:encoding}]. 
Let $\Phi$ be the stabilizer state on ${ABR}$ that encodes $U\rho U^\dagger$. Acting with the error channel ${\cal E}_B$ yields $\Psi = {\cal E}_B \otimes \id_{AR} [\Phi]$.
The logical information (or equivalently, magic resource) is preserved if the mutual information between subsytem $A$ and the ancilla $R$ satisfies $I_\Psi(A,R) = 2$~\cite{lee2024randomlymonitoredquantumcodes}, where $I_\Psi(X,Y) = S(\Psi_X) + S(\Psi_Y)-S(\Psi_{XY})$ with von Neumann entanglement entropy $S$.
By a simple calculation, we can recast this condition in terms of the coherent information $I_c(\Psi_{AR}) = S(\Psi_{A}) - S(\Psi_{AR})$ as $I_c(\Psi_{AR})=1$; conversely, $I_c(\Psi_{AR})=0$ if the logical information is lost.\\

Numerically, we focus on a single circuit (code) realization with fixed $f$, sample $N$ sets $B$ of $fL$ qubits, and thus estimate $\tilde{\cal C}({\cal N}_B)$ at different evolution times. (We checked multiple code realizations, finding similar results.)
Random stabilizer codes are known to saturate the quantum Singleton bound~\cite{gottesman1997stabilizer, qSingleton}, which is a bound on the code distance: for $f \lesssim0.5$, their logical information is preserved asymptotically ($L\to \infty$), whereas for $f>0.5$, it is highly likely lost due to the No Cloning theorem~\cite{gottesman1997stabilizer}. 
We find that, at $t>t_{\rm sat}$, the channel capacity proxy $\tilde{\cal C}({\cal N}_B) = 1$ for $f\lesssim 0.48$, finite up to $f \simeq 0.5$, and then drops to zero for $f > 0.5$, as expected from the quantum Singleton bound [Fig.~\ref{fig:channel_cap}(b)]. Comparing our results with global random Clifford gates, which generate random stabilizer codes, we find excellent agreement, further supporting the approach of the dynamical code to a random $[[L,1]]$ stabilizer code and the delocalization of magic resource at late times.

\end{twocolumngrid}

\newpage
\appendix
\begin{onecolumngrid}
    
\newpage
\begin{center}
    {\fontsize{11}{11}\selectfont
        \textbf{Supplemental Material: Spreading of Magic Resource under Unitary Clifford Dynamics\\[5mm]}}
    {\normalsize Mircea Bejan, Pieter W. Claeys, Jiangtian Yao \\[1mm]}
    
\end{center}
\normalsize\

\setcounter{equation}{0}
\setcounter{figure}{0}
\setcounter{table}{0}
\setcounter{theorem}{0}
\setcounter{lemma}{0}
\setcounter{page}{1}

\renewcommand{\theequation}{S\arabic{equation}}
\renewcommand{\thefigure}{S\arabic{figure}}
\renewcommand{\thesection}{S\arabic{section}}
\makeatletter
\@removefromreset{equation}{section}
\makeatother

In Appendix~\ref{app:BMGproofs}, we present details on the bipartite magic gauge by proving \autoref{thm:bip}. In Appendix~\ref{app:algos}, we report three efficient algorithms for computing subsystem nonstabilizerness and constructing the BMG. Appendix~\ref{app:equiv_def} discusses equivalent definitions of unitarily-extractable magic. Appendix~\ref{app:extensions} presents extensions of our results to other initial states. In Appendix~\ref{app:causal}, we present bounds on the linear magic length from the locality of the unitary dynamics, and in Appendix~\ref{app:interplay}, we discuss the interplay between entanglement growth and operator spreading and its effect on magic resource spreading. Appendix~\ref{app:rand_cliff_velocity_scales} presents a derivation of the entanglement and butterfly velocity in identity-doped Clifford circuits. The connection between the shape of the Pauli spectrum and the stabilizer R\'enyi entropy is detailed in Appendix~\ref{app:PS_SRE}. Additional numerical results are presented in Appendices~\ref{app:distribution} and~\ref{app:du} on the distribution of the length of typical MLMIs and on magic resource spreading in identity-doped dual-unitary Clifford circuits, respectively. In Appendix~\ref{app:overlap}, we provide numerics on the overlap of the left- and rightmost MLMIs. In Appendix~\ref{app:L_scaling}, we provide additional numerical data supporting the system-size independence of the spreading velocities.

\section{Details on the bipartite magic gauge. Proof of \autoref{thm:bip}}
\label{app:BMGproofs}

In this appendix, we first prove the existence and discuss the structure of the bipartite magic gauge (BMG) for a pure state $\ket{\psi}$ with one unit of magic resource.
To prove \autoref{thm:bip}, we shall use \autoref{lemma:bip}, where the structure of both logicals and stabilizer generators of $\kpsi$ is already apparent. Building on the proof of \autoref{lemma:bip}, we can break down the proof of \autoref{thm:bip} into three main steps.

For concreteness and without loss of generality, we focus on contiguous subsystems $A$ and $B$, and use the mapping $\kpsi \mapsto \ket{\Phi_{ABR}}$ shown in Fig.~\ref{fig:encoding}; we shall also use the local Clifford unitary (LCU) $U = U_A \otimes U_B \otimes U_R$ that disentangles qubits within each subsystem, cf. proof of \autoref{lemma:bip}.
First, by considering the spatial structure of the generators of $U\ket{\Phi_{ABR}}$, and using the locality of $U^\dagger$, we argue how Eq.~\eqref{eq:rho_bip} and cases ($i$--$v$) from \autoref{tab:bip} arise. We then show that the subsystem magic resource is LCU-invariant, i.e., $\sre (\rho_A) = \sre (\rho'_A)$ and ${\cal R}(\rho_A) = {\cal R}(\rho'_A)$ with $\rho' = (U_A \otimes U_B) \rho (U^\dagger_A \otimes U^\dagger_B)$. Finally, we explicitly compute $\sre (\rho'_A)$ and ${\cal R}(\rho'_A)$. We treat the SRE and ROM in turn.

\subsection{Spatial structure of generators}
Here, we show that one can choose logicals $\ZL,\YL$ and stabilizer generators of $\kpsi$ as given by Eq.~\eqref{eq:rho_bip} and the five cases from \autoref{tab:bip}. Consider the stabilizers of $U\ket{\Phi_{ABR}}$: As before, there exists a LCU $U$ such that $U\ket{\Phi_{ABR}}$ is a tensor product of tripartite GHZ states, Bell pairs between distinct subsystems ($A,B$ and $R$), and single-qubit states~\cite{GHZ_trio}; thus, if we assume for simplicity that $U_R = \id_R$, the stabilizer group of $U\ket{\Phi_{ABR}}$ is 
\begin{align}
    {\cal S}(U\ket{\Phi_{ABR}}) = \langle {\cal G}^U_A, {\cal G}^U_B, {\cal G}^U_{AB}, \XL^U X_R, \ZL^U Z_R\rangle,
\end{align}where without loss of generality, we have ${\cal G}^U_A = \langle \{ Z_{i} \}_{i=1}^{n_A}\rangle$, ${\cal G}^U_B = \langle \{ Z_{L-j} \}_{j=1}^{n_B}\rangle$, and 
\begin{align}
    {\cal G}^U_{AB} = \langle \{h^U_k\}\rangle =  \begin{cases}
                        \langle \{ X_{|A|-k+1}X_{|A|+k}, Z_{|A|-k+1}Z_{|A|+k}\}_{k=1}^{n_{AB}}\rangle, \qquad &\text{if $\nexists$ GHZ state}, \\
                        \langle \{ X_{|A|-k+1}X_{|A|+k}, Z_{|A|-k+1}Z_{|A|+k}\}_{k=1}^{n_{AB}}, Z_{n_A + 1} Z_{L-n_B-1}\rangle, \qquad &\text{if $\exists$ GHZ state,}
                      \end{cases}
\end{align}with $L-1 = n_A + n_B + m$, where the number of generators of ${\cal G}^U_{AB}$ is $m = \log_2|{\cal G}^U_{AB}|$. As in the proof of \autoref{lemma:bip}, because, for the state $\ket{\Phi_{ABR}}$, $R$ is maximally entangled with $AB$ and $U$ is local, subsystems $R$ and $AB$ remain entangled; thus, $U\ket{\Phi_{ABR}}$ must contain either a Bell pair $\ket{\Phi^+_{aR}}$ or $\ket{\Phi^+_{bR}}$, or a GHZ state $\ket{{\rm GHZ}_{abR}}$ for $a \in A, b\in B$. If we have $\ket{\Phi^+_{aR}}$ or $\ket{\Phi^+_{bR}}$, then we can pick $\XL^U = X_q, \ZL^U=Z_q \implies \YL^U = Y_q$ with $q = n_A+1$ or $q = L-n_B-1$, respectively; whereas for $\ket{{\rm GHZ}_{abR}}$, we have $\{ \XL^U, \YL^U, \ZL^U \} = \{ Z_a, X_a X_b, \pm Y_aX_b \}$ with $a=n_A+1, b= L-n_B-1$ and the $\pm$ sign is chosen to satisfy the $\mathfrak{su}(2)$ (anti-)commutation relations. Note that, for the GHZ case, $Z_b$ is a logical operator equivalent to $Z_a$ since $Z_aZ_b \in {\cal G}^U_{AB}$.

The spatial structure of the generators of $U\ket{\Phi_{ABR}}$ is the one from Eq.~\eqref{eq:rho_bip} and \autoref{tab:bip}, and this is preserved by inverting the LCU. Hence, it is also the structure of the generators of $\rho$.
We start with the stabilizer generators of $\rho$:
It is clear that the generators $\{ Z_i\}_{i=1}^{n_A}$ of ${\cal G}^U_A$ cannot be reduced to subsystem $B$ by multiplying them by any stabilizer from $\langle{\cal G}^U_B, {\cal G}^U_{AB}\rangle$; the same holds for the generators of ${\cal G}^U_B$ (${\cal G}^U_{AB}$) with respect to reducing them to $A$ ($A$ or $B$). By reverting the LCU, i.e., $g^U = U g U^\dagger \mapsto U^\dagger g^U U = g$, we obtain the stabilizer generators of $\rho$; however, this cannot modify the reducibility relations we just established because $U$ is local. Hence, we recover the generators of $\rho$ from Eq.~\eqref{eq:rho_bip} by using $a_i = U_A^\dagger Z_i U_A$ for $i\in [n_A]$, $b_j = U^\dagger_B Z_{L-j} U_B$ for $j \in [n_B]$, and $h_k = (U^\dagger_A \otimes U^\dagger_B) h^U_k (U_A \otimes U_B)$ for $k \in [m]$. 
We now turn to the logicals, which correspond to generators of $U\ket{\Phi_{ABR}}$ with nontrivial support on qubit $R$, and consider each case of \autoref{lemma:bip}.
A key feature in obtaining the logicals of $\kpsi$ (or equivalently $\ket{\Phi_{ABR}}$) is that their locality and reducibility are the same as that of the logicals of $U\ket{\Phi_{ABR}}$ because $U^\dagger$ is local. Unfolding cases ($i$-$iii$) of \autoref{lemma:bip}, we have: ($i$) all three logicals $\sL^U$ are fully reducible to $A$ and cannot be reduced (by multiplying them by a stabilizer) to $B$, and by locality the same holds $\sL = U^\dagger \sL^U U$, leading to case ($i$) in \autoref{tab:bip}; ($ii$) leads analogously to case ($ii$) in \autoref{tab:bip}; when there is a GHZ state, it depends which of the three logicals $\sL^U$ (or $\sL$) can be reduced to either $A$ or $B$---choosing this to be $\XL$, $\ZL$, and $\YL$ leads to cases ($iii$), ($iv$), and ($v$), respectively. 

\subsection{Subsystem SRE}
We now focus on the SRE before considering the ROM. To show that the subsystem SRE is LCU-invariant, it suffices to perform the following calculation. Recalling that $\sre(\rho) = {\cal M}_2(\rho) - S_2(\rho)$ [Eq.~\eqref{eq:SRE}] and denoting $\rho' = (U_A \otimes U_B) \rho (U^\dagger_A \otimes U^\dagger_B)$, we have
\begin{align}
    \sre(\rho'_A) &= {\cal M}_2\left( \Tr_B [(U_A \otimes U_B) \rho (U^\dagger_A \otimes U^\dagger_B)]\right) - S_2(\rho_A) \\
    &= {\cal M}_2\left( U_A \Tr_B [ U_B \rho U^\dagger_B] U^\dagger_A\right) - S_2(\rho_A) \\
    &= {\cal M}_2 \left( \Tr_B[U^\dagger_B U_B \rho] \right) - S_2(\rho_A) \\
    &= {\cal M}_2(\rho_A) - S_2(\rho_A) = \sre(\rho_A), 
\end{align}where we used the local unitary invariance of the entanglement R\'enyi entropy in the first line, the partial trace properties in the second line, and the fact that $U_A$ is a Clifford operation in the third line.

We explicitly calculate $\sre(\rho_A)$ by computing $ \sre(\rho'_A)$ in case ($i$) from \autoref{tab:bip}. The other cases can be similarly obtained; see also App.~\ref{app:PS_SRE} for a general recipe for obtaining the SRE from the Pauli spectrum. Focusing on case ($i$), 
\begin{align}
    \rho'_A = \frac{1}{d_A} \left( \id_A + \frac{Z_{q} - Y_q}{\sqrt{2}}\right) \prod_{i=1}^{q-1}(\id_A + Z_i), \qquad d_A = 2^{|A|},\ q = n_A +1.
\end{align}Thus, the nonzero entries of the Pauli spectrum 
$\Xi(\rho'_A)$ are $\frac{1}{d_A}$ and $\frac{1}{2d_A}$ repeated $2^{q-1}$ and $2^q$ times, respectively. Plugging this into Eq.~\eqref{eq:SRE} yields $\sre(\rho'_A) = \log_2(4/3)$.

\subsection{Subsystem ROM}
We now turn to the ROM, introduced in Ref.~\cite{RoM}, where the properties of ROM essential to our proof are derived. First, we prove the LCU-invariance of the ROM. Consider a (possibly mixed) state $\rho$ on a bipartite system $AB$. Let $\rho' =(U_A \otimes U_B) \rho (U^\dagger_A \otimes U^\dagger_B)$ for some Clifford unitaries $U_A$ and $U_B$. Let the optimal decomposition of $\rho_A$ be $\rho_A = \sum_i x_i \sigma_i$ [cf. Eq.~\eqref{eq:rom}]. A direct calculation reveals that $\rho'_A = U_A \rho_A U^\dagger_A = \sum_i x_i (U_A \sigma_i U^\dagger_A) \equiv \sum_i x'_i \sigma_i$, where the entries of $x$ are reshuffled into $x'$. Hence, $\rom(\rho'_A) \leq \sum_i |x'_i| = \sum_i |x_i| = \rom (\rho_A).$ The decomposition $\rho'_A = \sum_i x'_i \sigma_i$ is in fact optimal, which we prove by contradiction. Suppose there exists another decomposition $\rho'_A = \sum_i y'_i \sigma_i$ with $\sum_i |y'_i| < \sum_i |x'_i|$, in which case we could revert the LCU to find $\rho_A = \sum_i y'_i (U^\dagger_A \sigma_i U_A) \equiv \sum_i y_i \sigma_i$ with $\sum_i |y_i| = \sum_i |y'_i| < \sum_i |x_i| = \rom(\rho_A)$. Hence, we find $\rom(\rho_A) = \rom(\rho'_A)$.

We next show that $\rom (\rho \otimes \sigma) = \rom(\rho)$ for any (possibly mixed) stabilizer state $\sigma$. Consider a (possibly mixed) state $\rho$ and a (possibly mixed) stabilizer state $\sigma$. Let the ROM-optimal decomposition of $\rho$ be $\rho = \sum_i x_i \sigma_i$, thus $\rom(\rho) = \sum_i |x_i|.$ Conversely, $\sigma$ is decomposable as $\sigma = \sum_i y_i \sigma_i$ with $\sum_i |y_i| = 1$ and $y_i \geq 0, \ \forall i$. Consider now the ROM $\rom(\rho\otimes \sigma)$. By the submultiplicativity of ROM~\cite{RoM}, $\rom (\rho \otimes \sigma) \leq \rom(\sigma) \rom(\sigma) = \rom (\rho).$ However, by the monotonicity of ROM~\cite{RoM}, applying the stabilizer channel $\cal E$ that traces out the subsystem of $\sigma$, we have $\rom({\cal E}(\rho\otimes \sigma)) = \rom (\rho) \leq \rom (\rho \otimes \sigma)$. By using both inequalities, we find $\rom(\rho \otimes \sigma) = \rom (\rho)$.

Finally, combining the two observations above, we explicitly compute $\rom(\rho_A)$ for the various cases presented in \autoref{tab:bip}. Case (\textit{i}): We know that there exists a local Clifford unitary $U = U_A \otimes U_B$ such that $\rho'_A = \ketbra{T}{T} \otimes \sigma$ for some stabilizer state $\sigma$. Hence, using $\rom(\ketbra{T}{T})=\sqrt{2}$~\cite{RoM} and the above observations, we find $\rom(\rho_A) = \sqrt{2}$. Cases (\textit{ii}) and (\textit{iii}): Using the same type of LCU transformation, we can find a $\rho'_A = \sigma$ for some stabilizer state $\sigma$; thus, $\rom(\rho_A) =1$, as expected from the SRE calculation. Cases (\textit{iv}) and (\textit{v}): Again, by applying an LCU, we can find
\begin{align}
    \rho'_A = \left( \frac{\id}{2} + \frac{Z_a}{2\sqrt{2}}\right) \otimes \sigma
\end{align}for a stabilizer state $\sigma$ and some qubit $a \in A$, where $U \ZL U^\dagger = Z_a$ and $-U \YL U^\dagger = Z_a$ in cases (\textit{iv}) and (\textit{v}), respectively. Using the methods of Ref.~\cite{RoM}, one finds that $\rom \left( \frac{\id}{2} + \frac{Z_a}{2\sqrt{2}}\right) =1$; hence, $\rom (\rho_A) = 1$, signaling that a full magic unit \textit{cannot} be extracted from $\rho_A$, as expected from the reducibility structure of logicals. To see this explicitly, note that $\frac{1}{2}\left(\id+\frac{Z_a}{\sqrt{2}}\right)=\left(\frac{\sqrt{2}+1}{2\sqrt{2}}\right)\left(\frac{\id+Z_a}{2}\right)+\left(\frac{\sqrt{2}-1}{2\sqrt{2}}\right)\left(\frac{\id-Z_a}{2}\right)$.  
We emphasize that this feature is characteristic of ROM, but not SRE which is $\log _2(6/5)$ rather than $\log_2 (4/3)$, the latter of which further requires knowing the angle of the non-Clifford gate to determine whether $\rho_A$ has a full or half magic unit.\hfill $\square$

\section{Subsystem nonstabilizerness algorithms} \label{app:algos}

Building on the bipartite magic gauge from \autoref{thm:bip}, we here present three algorithms for computing subsystem magic via the reducibility of logical operators. While Algorithm 1 (App.~\ref{app:alg1}) is the simplest and most efficient one, the latter two algorithms offer more insight into the structure of the state. Algorithm 2 (App.~\ref{app:alg2}) is suited for a certain class of operators and might be of independent interest, whereas Algorithm 3 (App.~\ref{app:alg3}) computes the full BMG decomposition.
Each algorithm is based on subroutines that perform stabilizer simulations~\cite{gottesman1998heisenberg,aaronsonPRA2004,Audenaert_2005}, each requiring a poly$(L,t)$ classical resources. However, the number of times these routines need to be repeated grows exponentially with the amount of magic in the system: For states with stabilizer nullity $\nu$~\cite{Beverland_2020}, we expect the classical runtime to scale as poly$(L,t) \exp(\nu)$. Here $\nu =1$, thus making our classical simulation efficient. 

\subsection{Algorithm 1: Subsystem nonstabilizerness via logicals only} \label{app:alg1}
The first algorithm computes the magic resource of a subsystem $A$, as quantified by SRE $\sre(\rho_A)$ and ROM $\rom(\rho_A)$, for a pure state $\rho$ with one unit of magic, without accessing the Pauli spectrum. Instead, by using \autoref{thm:bip} (see also~\autoref{tab:bip}), one can compute these by assessing if the logical operators $\ZL$ and $\YL$ are reducible to subsystem $A$ upon multiplication with certain stabilizer generators of $\rho$, corresponding to $\sL \to g_{i_1} \cdots g_{i_k} \sL$ for some $i_1 < i_2 < \dots < i_k$ with $k \leq L-1$. This can be done efficiently by performing Gaussian elimination over binary fields~\cite{GF2}. One then reads off the subsystem magic $\sre(\rho_A)$ and $\rom(\rho_A)$ from \autoref{tab:bip} based on the reducibility of both $\ZL$ and $\YL$.

\subsection{Algorithm 2: Extended normal form for subsystem nonstabilizerness}\label{app:alg2}
We write $\rho = \rho^{(Z)} + \rho^{(-Y)}$ from Eq.~\eqref{eq:rho} with
\begin{align}
    \rho^{(\sigma)}\coloneqq \frac{1}{2^{L+1}}\left( \id + \sqrt{2} \sL \right) \prod_{i=2}^L (\id + g_i) \label{eq:rhoSigma}
\end{align}for any logical $\sL$. Since these operators are reminiscent of the stabilizer states $\frac{1}{2^{L}}\left( \id \pm \sL \right) \prod_{i=2}^L (\id + g_i)$, we call $\rho^{(\sigma)}$ \textit{pseudostabilizer operators}. Such operators are fully specified by a set of generators $\{\sL, g_2, \dots, g_L\}$ and a prefactor $\beta>0, \beta\neq1$ of the logical generator (LG); e.g., above $\beta = \sqrt{2}$. The key difference, however, stems from this $\beta$ factor, as we will shortly explain.

We here show that each of $\rho^{(Z)}$ and $\rho^{(-Y)}$ can contribute at most half a unit of magic to the magic resource of $\rho$, and their Pauli spectra can be used to infer that of $\rho$. We then discuss how to bring the pseudostabilizer operators $\rho^{(\sigma)}_A$ to a form similar to the bipartite magic gauge, by exploiting their similarities with density matrices of stabilizer states; henceforth, we call the subsystem of interest $A$ and its complement $B$. 
At the core of this algorithm, we combine insights from \autoref{thm:bip} with the Clifford normal form for a single party (CNFP) algorithm from Ref.~\cite{Audenaert_2005}. In brief, we apply a modified version of CNFP to update both $\rho^{(Z)}$ and $\rho^{(-Y)}$, and then, based on \autoref{thm:bip}, argue that the subsystem magic can be read off from these. We next summarize the standard CNFP and our extensions. 

CNFP takes in generators of a possibly mixed stabilizer state and a bipartition while outputting the canonical form of the generators, which enables the computation of bipartite entanglement measures. To do so, CNFP applies to the generators Clifford gates restricted to each side of the bipartition, as well as multiplies generators with each other $(g_i, g_j) \mapsto (g_i, g_i g_j)$ or $(g_i, g_j) \mapsto(g_i g_j, g_j)$. 

\begin{figure}[tp]
    \centering
    \includegraphics[width=7cm]{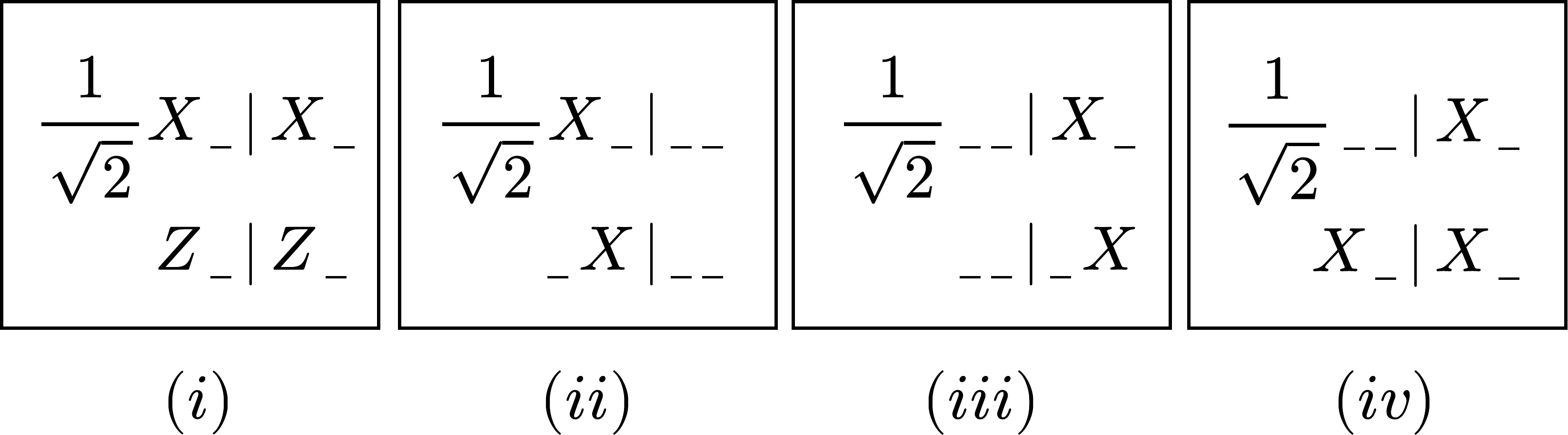}
    \caption{The four possible outcomes after applying the extended CNFP algorithm to both sides of the tableau. The vertical lines separate qubits in subsystems $A$ and $B$. %
    }
    \vspace{-\baselineskip}
    \label{fig:4_cases}
\end{figure}

However, a key difference in updating the generators of pseudostabilizer operators $\rho^{(\sigma)}$ is that, due to the $\beta \neq 1$ prefactor of $\sL$, the only allowed multiplication between a stabilizer $g$ and a logical $\sL$ generator is $(\sL, g) \mapsto (g \sL, g)$; $(\sL, g) \mapsto (\sL, g \sL)$ is forbidden because this would not faithfully represent $\rho^{(\sigma)}$. Therefore, we modify the CNFP algorithm by imposing that, if a logical and a stabilizer generator have to be multiplied into each other, we always multiply the stabilizer into the logical one. This modified procedure ends in poly$(L)$ time~\cite{Audenaert_2005}. 

At the end of this part of the algorithm, the stabilizer-like tableau~\cite{aaronsonPRA2004, Audenaert_2005}, with rows representing generators, has an upper part with pairs of $(X_{i_k} X_{j_k}, Z_{i_k} Z_{j_k})$ generators on distinct qubits (i.e., $i_k \in A, j_k \in B$ and $i_k \neq i_l, j_k \neq j_l$ for $k\neq l$), and a lower part with $X$ strings supported only on the other qubits. 
If the row corresponding to $\sL$ is in the upper part [Fig.~\ref{fig:4_cases} case (\textit{i})], $\rho^{(\sigma)}_A$ does not contribute any magic to $\rho_A$ since this generator will not survive tracing out over $B$, and we stop the program. 

Conversely, if $\sL$'s row is in the lower part of the tableau, we process this as follows.
First, by performing a Gaussian elimination over the lower rows corresponding to stabilizer generators, we update these such that at most one generator is a $X_a X_b$ operator with $a \in A, b \in B$ while the others are single-qubit $X$ operators; then, we reduce the support size of $\sL$ by multiplying such $X$ generators into it. The final support size of the LG will be 1. 

Hence, three more outcomes for the logical generator arise (see Fig.~\ref{fig:4_cases}).
(\textit{ii}) [(\textit{iii})] It is a single-qubit $X_i$ operator supported on $A$ ($B$) and no other generator is supported on this qubit; thus, it does (not) survive under tracing $B$ out, and $\rho^{(\sigma)}_A$ has a nonzero (zero) contribution to the magic of $\rho_A$.
(\textit{iv}) It is a single-qubit $X_a$ or $X_b$ operator supported on $A\ni a$ or $B\ni b$ and there is another generator $X_a X_b$, while all other generators are supported on other qubits. Thus, this logical generator can be redefined to be supported in either $A$ or $B$ [cf. cases ($iii,iv,v$) in \autoref{tab:bip}], and $\rho^{(\sigma)}_A$ adds nonzero magic to $\rho_A$. We emphasize 
that ($iv$) does not occur for regular CNFP, since there one may reduce the support of the $X_a X_b$ generator via $(X_a, X_a X_b) \mapsto (X_a, X_b)$. 

To conclude, using Algorithm 2, one can infer the magic of $\rho_A$ by reading off the magic of $\rho^{(Z)}_A$ and $\rho^{(-Y)}_A$, where each can contribute at most half a unit of magic [\autoref{tab:bip}]. Similarly, one can also directly deduce the Pauli spectrum of $\rho_A$ from that of $\rho^{(Z)}_A$ and $\rho^{(-Y)}_A$ (see also App.~\ref{app:PS_SRE}). For example, any Pauli operator $P = \ZL g$ with a nonzero coefficient in the PS of $\rho^{(Z)}_A$ cannot exist in the PS of $\rho^{(-Y)}_A$ since $\ZL$ and $\YL$ are independent. Conversely, any Pauli $P = g$ with a nonzero coefficient must be shared by $\rho^{(Z)}_A$ and $\rho^{(-Y)}_A$ [cf. Eq.~\eqref{eq:rhoSigma}]. 

\subsection{Algorithm 3: Pauli spectrum via the bipartite magic gauge} \label{app:alg3}

While Algorithm 1 extracts the subsystem magic, one might also be interested in accessing the entire Pauli spectrum (PS) of $\rho_A$, for example, to compute certain entanglement measures. While Algorithm 2 already enables the computation of the PS, this third algorithm presented here enables us to access the Pauli spectrum by explicitly finding the BMG. Note that although the stabilizer generators of $\rho^{(Z)}_A$ and $\rho^{(-Y)}_A$ can be chosen to be the same, Algorithm 2 does not guarantee that the (updated) stabilizer generators coincide.

We focus on a pure state $\rho$ with one unit of magic, given without loss of generality by Eq.~\eqref{eq:rho}. It is useful to decompose $\rho$ in terms of a logical and a stabilizer part
\begin{align}
    \rho = \left( \mathds{1} + \frac{\ZL - \YL}{\sqrt{2}} \right) \phi, \quad \phi = 2^{-L}\prod_{i=2}^L \left( \mathds{1} +g_i \right),
\end{align}where $\phi$ is a mixed stabilizer state. We shall bring $\rho$ to the form in Eq.~\eqref{eq:rho_bip} and find the Pauli spectrum in three steps.

\textit{Step 1.} By applying an algorithm from Ref.~\cite{Audenaert_2005}, we first find a local Clifford unitary $C = C_A \otimes C_B$ that brings $\phi \mapsto \phi' = C\phi C^\dagger$, where $\phi'$ is a tensor product of Bell pairs between $A$ and its complement $B$, and a separable state. More precisely, we apply for $\phi$ the Clifford normal form for a single party (CNFP) algorithm from Ref.~\cite{Audenaert_2005} twice, once for $A$ and once for $B$, while keeping track, at each step, of the Clifford unitaries that will make up $C$ upon multiplication.
Similar to Algorithm 2, the stabilizer tableau of $\phi'$ contains an upper part with pairs of $(X_{i_k} X_{j_k}, Z_{i_k} Z_{j_k})$ generators on distinct qubits (i.e., $i_k \in A, j_k \in B$ and $i_k \neq i_l, j_k \neq j_l$ for $k\neq l$), and a lower part with $X$ strings supported only on the other qubits.
While the bottom rows, containing only $X$ strings, can be argued to give rise to a separable state~\cite{Audenaert_2005}, we still need to massage them further to recover the structure of the BMG, particularly, the irreducibility to subsystems $A$ and/or $B$.

\textit{Step 2.}
We next (almost) diagonalize the lower blocks of the stabilizer tableau of $\phi'$ using standard Gaussian elimination, which corresponds to reducing the support of the $X$-string generators in the lower blocks to contain at most two qubits (see also Algorithm 2). In fact, since $\phi$ has $L-1$ stabilizers, the lower part of $\phi'$'s new tableau contains single-qubit $X$ operators and at most a single two-qubit $X_a X_b$ operator with $a \in A, b \in B$. We call these generators $\{ g'_i\}$.

The BMG is readily available at this stage. To find the generators $\{a_i\}, \{ b_j\},$ and $\{ h_k\}$ from Eq.~\eqref{eq:rho_bip}, one takes the generators $\{g'_i\}$ and reverts the local Clifford unitary $g'_i \mapsto C^\dagger g'_i  C$. Under this transformation, $X_i$ with $i \in A$ map to $a_i$, $X_j$ with $j \in B$ map to $b_j$, while $X_aX_b$ and $Z_a Z_b$ with $a\in A, b\in B$ each map to a distinct $h_k$.

\textit{Step 3.}
To find the Pauli spectrum, using Algorithm 1 (App.~\ref{app:alg1}), we check the reducibility of the logical operators $\ZL, \YL$ with respect to the generators $\{a_i\}, \{b_j\}$, and $\{h_k\}$ found at the previous step.
From this, as outlined in the proof of \autoref{thm:bip} from the previous subsection, one can easily infer the number and values of the nonzero entries in the Pauli spectrum.
We note that one can also infer the Pauli spectrum by checking the reducibility of the updated logical $\sL \mapsto \sL' = C \sL C^\dagger$, for $\sL = \ZL, \YL$, with respect to the generators $\{ g'_i\}$.

\section{Equivalent definitions of unitarily-extractable magic resource}
\label{app:equiv_def}
In this appendix, we show that unitarily-extractable magic can be interpreted in at least three different ways: It can be extracted to an external system through local Clifford gates, all logical information can be concentrated on the extraction region, and this region also contains a full unit of magic.
In particular, by focusing on pure states $\rho$ with a single unit of magic and denoting by $A$ a subsystem with one unit of magic, we first show that $\rom(\rho_A) > 1$ is equivalent to all three $\XL, \YL, \ZL$ being reducible to $A$, which corresponds to concentrating the logical information on $A$. We then show that being able to reduce $\XL, \YL$ and $\ZL$ to $A$ is equivalent to being able to extract magic from $A$, which by transitivity will complete our proof.

First, if subsystem $A$ has one unit of magic [as in our setup, $\rom(\rho_A) = \sqrt{2}$ and $\sre(\rho_A) = \log_2(4/3)$], by~\autoref{thm:bip} case ($i$), the support of both logical operators $\YL$ and $\ZL$ can be reduced to $A$. Furthermore, by \autoref{lemma:bip}, if two logicals (here, $\YL$ and $\ZL$) are reducible to $A$, then the third (here, $\XL$) is also reducible to $A$; this completes the forward implication $\rom(\rho_A) =\sqrt{2} \implies$ all logical information is reducible to $A$. The backward implication ``$\impliedby$'' is trivial.

\begin{figure}[htp]
    \centering
    \includegraphics[width=0.3\linewidth]{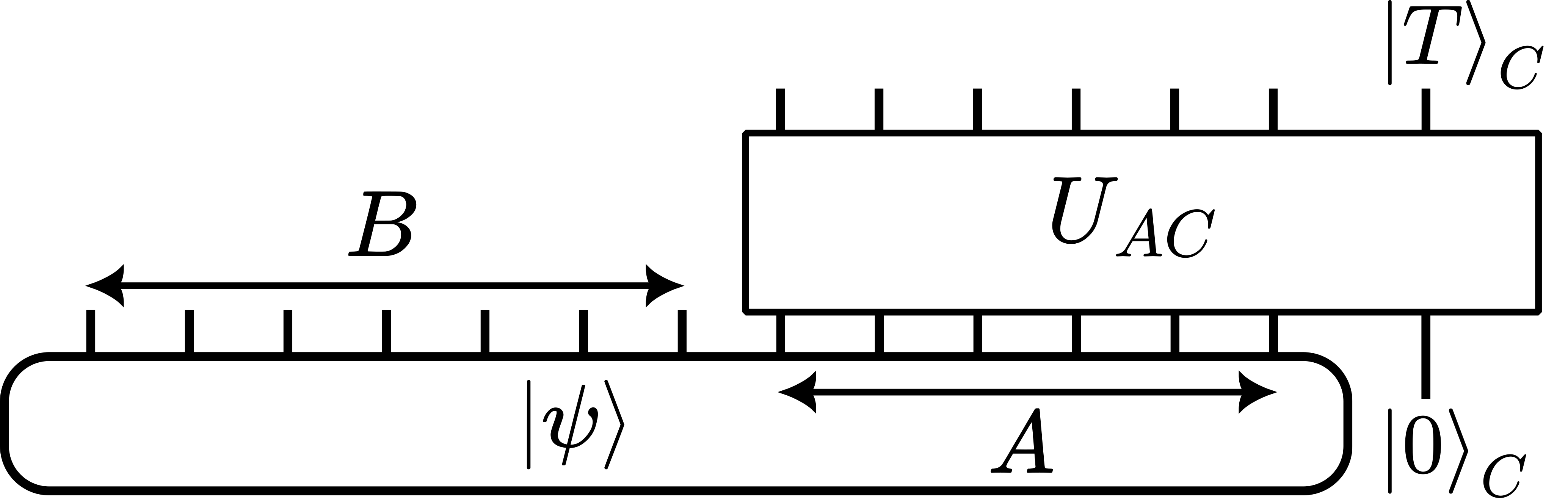}
    \caption{Procedure to unitarily extract magic from a pure state $\kpsi$ with one unit of magic. If magic can be concentrated to a subsystem $A$, then, by applying a Clifford unitary $U_{AC}$ supported only on subsystem $A$ and an external system $C$ (initialized in the stabilizer state $\ket{0}_C$), one can distil a magic state $\ket{T}_C$ on $C$.}
    \label{fig:uExtrM}
\end{figure}

Second, let $C$ be a single-qubit external system, initialized in state $\ket{0}_C$, to which we want to distil the magic (see Fig.~\ref{fig:uExtrM}). We shall prove that, by acting with a Clifford unitary acting only on $A$ and $C$, we can distil a magic state on $C$, i.e., $\ket{T}_C = (\ket{0} + e^{i\pi/4}\ket{1})/\sqrt{2}$, if and only if all logical information is reducible to $A$. The forward direction ``$\implies$'' is trivial: Starting from the post-distillation state, apply a SWAP between $C$ and a qubit in $A$, and then, reverse the $AC$ Clifford unitary. Now, we focus on the backwards ``$\impliedby$'' direction.
Since all three logicals $\XL, \YL, \ZL$ are reducible to $A$, by the proofs of \autoref{thm:bip} and \autoref{lemma:bip} (see App.~\ref{app:BMGproofs}), there exists local Clifford unitary $U_A \otimes U_B$ such that $(U_A\otimes U_B)\ket{\psi} = \ket{T}_a \otimes \ket{\psi'}_{\bar{a}}$, for some qubit $a \in A$ and where $\ket{\psi'}_{\bar{a}}$ is a stabilizer state on the qubits $AB \setminus a$. Note that further applying $U^\dagger_B$ cannot alter the state $\ket{T}_a$; thus, we extract the magic from $A$ to $C$ by applying $\mathrm{SWAP}_{aC} (U_A \otimes \id_B \otimes \id_C) \kpsi \otimes \ket{0}_C = \ket{\tilde{\psi}}_{AB} \otimes \ket{T}_C$, with some stabilizer state $\ket{\tilde{\psi}}_{AB}$. This completes our proof.

\section{Generalizations to other non-Clifford gates and mixed initial states, and of operational definitions} \label{app:extensions}

In this Appendix we discuss various extensions of our results, demonstrating the generality of our conclusions and the applicability of our operational definitions to broader classes of initial states. We first discuss the quantization conditions for pure initial states and briefly certain mixed initial states, and then the operational definitions.  

\subsection{Non-Clifford gates}

We expect our results to extend directly to any local non-Clifford gate acting on a short-range entangled pure stabilizer state, irrespective of the quantization given by SRE and/or ROM. Without loss of generality we can focus on single-qubit non-Clifford gates generated by rotations about the $Z$-axis  $R(\phi) = e^{i \phi Z}$; single-qubit rotations about the $X$- or $Y$-axis and multi-qubit local non-Clifford gates can be accommodated by a suitable choice of pure short-range entangled initial state. We also note that pure short-range entangled states, particularly product states, can be easily prepared experimentally by projectively measuring single qubits in a stabilizer basis. 

The maximum values of subsystem nonstabilizerness, as quantified by the SRE and ROM, depend on $\phi$; however, all of our results rely on the spatial structure of logicals in the dynamically generated code, and \textit{this structure} (i.e., reducibility of logicals to a subsystem) \textit{is independent of} $\phi$, provided $R(\phi)$ injects nonstabilizerness. This key observation demonstrates how our results can be generalized. 
To illustrate our reasoning, consider applying $R_j(\phi) = e^{i\phi Z_j}$ instead of the $T_j \propto R_j(-\frac{\pi}{8})$ gate presented in the main text on the same initial state $\rho_0$. Using $R_j(\phi) s_1 R_j^\dagger(\phi) = \cos(2\phi) s_1 + i\sin(2\phi) Z_j s_1 $ and defining $\XL \coloneqq Z_j, \ZL \coloneqq s_1$ and $\YL \coloneqq i \XL \ZL$, we find [provided $R_j(\phi)$ injects nonstabilizerness under the same assumptions from the main text, which imply $\cos (2\phi) \neq 0$ and $\sin (2\phi) \neq 0$]
\begin{align}
    \rho (\phi) = R_j(\phi) \rho_0 R^\dagger_j(\phi) = \frac{1}{2^L} \left[ \mathds{1} + \cos (2\phi) \ZL + \sin (2\phi) \YL \right] \prod_{i=2}^L \left( \mathds{1} +g_i \right).
    \label{eq:rho_phi}
\end{align} Eq.~\eqref{eq:rho_phi} recovers Eq.~\eqref{eq:rho} for $\phi=-\frac{\pi}{8}$. Since all the arguments based on \autoref{thm:bip} and \autoref{lemma:bip} rely on the reducibility of the $\phi$\textit{-independent} logicals $\XL, \YL,$ and $\ZL$ to a subsystem, and because $\cos(2\phi) \neq 0$ and $\sin (2\phi) \neq 0$, our conclusions regarding the spreading extend to any such $\phi$. 

Furthermore, unlike the SRE, using the ROM eliminates any potential ambiguity between the half and full magic unit cases: ${\cal R} (\rho_A(\phi))=1$ for subsystems with half a unit of nonstabilizerness or no nonstabilizerness, whereas ${\cal R} (\rho_A(\phi))= {\cal R}(\phi) > 1$ for subsystems from which a full unit of nonstabilizerness is unitarily extractable. This result follows from the faithfulness and monotonicity of ROM~\cite{RoM}. One can clearly distinguish between subsystems with and without a full unit of nonstabilizerness by measuring ${\cal R}(\rho_A(\phi)) > 1$ and ${\cal R}(\rho_A(\phi))=1$, respectively, even if the true angle in an experiment is not exactly $\phi_{\rm exp} = \phi$, provided nonstabilizerness is still injected. Such measurements could then be used to trace the spreading of magic resource in spacetime. 
We therefore expect that our conclusions hold provided that one can unitarily extract exactly one single-qubit magic state from the system's global state.

\subsection{Mixed initial states}

Our results directly extend to mixed initial states that are convex sums of pure stabilizer states. As an illustrative example, consider applying a $T_1$ gate on the state $\rho_{0}^{(\lambda)} = (1-\lambda) \rho_+  + \lambda \rho_- $ with $\rho_\pm = \ketbra{\pm}{\pm}\otimes \ketbra{0}{0}^{\otimes L-1}$ and $\lambda \in [0,1] \setminus \{ \frac{1}{2} \}$, leading to $\rho^{(\lambda)} = T_1 \rho_{0}^{(\lambda)} T_1^\dagger$. The spatial structure of the logical operators of $T_1 \rho_+ T^\dagger_1$ and $T_1 \rho_- T^\dagger_1$ is identical (the logicals differ by some unimportant sign), and this holds even after applying a Clifford unitary evolution $U$; hence, the spreading of magic resource in $U\rho^{(\lambda)}U^\dagger$ is the same for any $\lambda \neq \frac{1}{2}$, as characterized by ROM. In particular, one can determine the nonstabilizerness spreading dynamics by focusing on the pure state $\lambda=0$ or $\lambda=1$, which can be obtained using the formalism in the main text. While this demonstrates that our conclusions can be relevant to even more experimentally motivated setups with mixed initial states, the dynamics of magic resource in generic open system dynamics represents an interesting open question. 

\subsection{Operational definitions}

The definitions of unitarily-extractable and projectively-destroyable magic resource can be generalized to states with multiple units of nonstabilizerness. For unitarily-extractable magic resource, one may consider sequentially extracting the nonstabilizerness onto one ancilla at a time and ask how many times one can do this extraction procedure; this quantifies how many units of nonstabilizerness were present in the initial (reduced) density matrix. For projectively-destroyable magic resource, a natural extension could be dubbed \textit{projectively-reducible magic resource}, wherein one asks whether the total extractable units of nonstabilizerness from a subsystem can be reduced by measuring a single Pauli operator supported inside the subsystem, and also how many units can be removed by multiple such local measurements. We note that these extended operational definitions might be particularly useful for studying the dynamics of magic resource with multiple units of nonstabilizerness.

\section{Locality bounds on the linear magic length}
\label{app:causal}

In this appendix, we show that the time evolution of the linear magic length $\ell$ is bounded by the locality of interactions. %
For brevity, we focus on destroyable magic and contiguous subsystems; a similar proof holds for unitarily-extractable magic. Recall that $\ell$ is the length of the shortest symmetric, centered on the injection site, and contiguous interval where magic is destroyable; that is, there exists a contiguous subsystem $A$ centered on the injection site, with $|A| = \ell$, and a logical operator $M = M\vert_A \otimes \id_{\bar{A}}$, where $\bar{A}$ is the complement of $A$.

Let the LML at time $t$ be $\ell(t) = \ell$ and $M \in \arg \ell (t)$ be one (of possible multiple) logical operator determining $\ell$. We are interested in how much  $\ell$ can increase or decrease in one time step under the action of a $k$-local brickwork Clifford circuit. To show that the change in $\ell(t)$ is bounded by locality, we shall show that $|\ell(t+1)-\ell(t)|\leq 2(k-1)$ for all $k$.

It is straightforward to see that, under the action of a $k$-local Clifford circuit $C$, the width of a Pauli operator $P\mapsto C P C^\dagger$ can increase (decrease) by at most $2(k-1)$, with a $(k-1)$ maximal increase at either operator front. This is, of course, valid for the width of any logical operator $M \in \arg \ell(t)$, hence, $\ell(t+1) \leq \ell(t)+2(k-1)$. We are not interested in any logical operator that would lead to a larger $\ell(t+1)$ since, by definition, $\ell$ involves a minimum.
It remains to show that $\ell$ cannot decrease by more than $2(k-1)$. By the same argument as above, the width of the symmetric-centered subsystem containing any updated logical operator $CMC^\dagger$ with any $M \in \arg \ell(t)$ is at least $\ell(t) - 2(k-1)$. By the definition of the LML, any other logical operator $N \notin \arg \ell(t)$ must be supported at time $t$ on a symmetric-centered region of width $w(t)$ with $w(t) > \ell(t)+2$; since the ends of $CNC^\dagger$ cannot move towards the ``center'' by more than $(k-1)$, the updated $w(t+1) \geq w(t) - 2(k-1) > \ell(t) - 2(k-2)$. This concludes our proof that $|\ell(t+1) - \ell(t)| \leq 2(k-1)$, implying the locality bounds on the change in $\ell(t)$. 

\section{Interplay between entanglement growth and operator spreading}
\label{app:interplay}

In this appendix, we discuss a broader class of $T$-doped Clifford circuits which makes apparent the effects of both entanglement growth and operator spreading on magic spreading. We begin with the circuit setup and then discuss entanglement- and/or operator-dominated spreading regimes, which we substantiate with numerical results. We end by highlighting a potentially important effect at intermediate times, the shrinking of large logical operators.

\subsection{\texorpdfstring{$\bm{T}$}{}-doped Clifford circuit setup}\label{subsec:evo_T_gate}

As mentioned in the main text, an illuminating view of how the interplay between operator and entanglement spreading affects the spreading of magic in the state $UT\kpsi$ is to rewrite it as $UT\kpsi = (UTU^\dagger) U \kpsi$. In fact, we shall consider the more general setup shown in the left panel of Fig.~\ref{fig:operator_vs_ent}. Without loss of generality, we consider unitarily evolving a product $L$-qubit stabilizer initial state $\kpsi$ to $(U_t TU^\dagger_t) V_\tau \kpsi$, where $U_t$ and $V_\tau$ are Clifford unitaries made of a brickwork of 2-qubit Clifford gates of depth $t$ and $\tau$, respectively, and the $T$ gate is applied on some qubit $i$. We are interested in the spreading of magic under this Clifford dynamics. While entanglement builds up due to the evolution under $V_\tau$, the $T$ operator spreads under the $U_t$ dynamics leading to the operator $\tilde{T} = U_t TU^\dagger_t$. 

Before discussing this interplay in more detail, for completeness, we explain when the $T$ gate successfully injects magic; this also stands as the derivation of Eq.~\eqref{eq:rho} by setting $U_t = V_\tau$. Let the stabilizer group of $V_\tau \kpsi$ be ${\cal S}(V_\tau \kpsi) = \langle s_1, s_2, \dots , s_L\rangle$. We have $\tilde{T} = \alpha \id + \beta \tilde{Z}_i$ with $\alpha, \beta \in \mathds{C}$ and Pauli operator $\tilde{Z}_i = U_t Z_i U^\dagger_t$. Since $V_\tau \ketbra{\psi}{\psi}V^\dagger_\tau = 2^{-L}\prod_{i=1}^L (\id + s_i)$, magic is injected successfully by $\tilde{T}$ if $\tilde{Z}_i$ anticommutes with at least a generator $s_i$~\cite{bejan2024prxq}. Assuming this is the case, we can redefine our generators such that only one of them, say $s_1$, anticommutes with $\tilde{Z}_i$: we take the new stabilizer generators of $V_\tau \kpsi$ to be $s_1$ and $\{ g_j \}_{j=2}^L$ with $g_j = s_j$ if $[\tilde{Z}_i, s_j]=0$ and $g_j = s_1 s_j$ if $\{ \tilde{Z}_i, s_j \} =0$. A direct calculation thus yields
\begin{align}
    \tilde{T} V_\tau \ketbra{\psi}{\psi} V^\dagger_\tau \tilde{T}^\dagger = \frac{1}{2^L} \left( \mathds{1} + \frac{\ZL - \YL}{\sqrt{2}} \right) \prod_{j=2}^L \left( \mathds{1} +g_j \right),
\end{align}where $\XL \coloneqq \tilde{Z}_i$, $\ZL \coloneqq s_1$ and $\YL \coloneqq i \XL \ZL$. One can thus apply the algorithms given in the End Matter to infer the spatiotemporal structure of magic in this state. We now turn back to more general considerations on magic spreading.

\begin{figure}[tp]
    \centering
    \includegraphics[width=11cm]{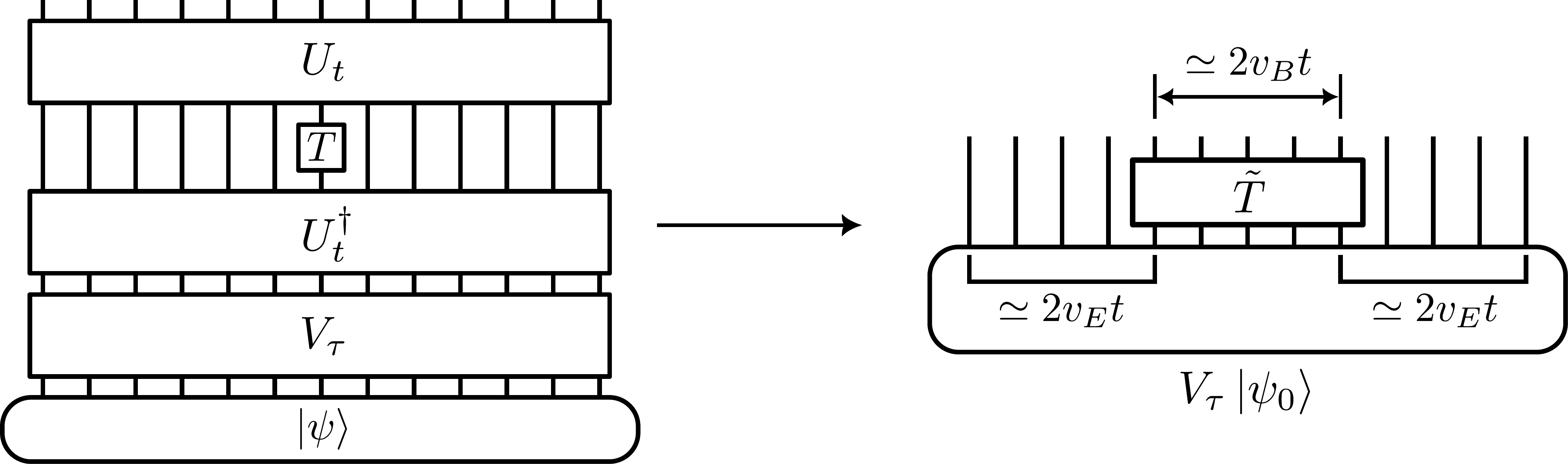}
    \caption{Interplay between operator and entanglement spreading, seen after inserting the resolution of identity. The non-Clifford operator ($T$ gate) spreads at $v_{\rm B}$ whereas entanglement builds up simultaneously at speed $v_{\rm E}$. Right panel: the extreme case where the effects of the non-Clifford operator propagate at the upper bound of ($v_{\rm B}+2v_{\rm E}$). The entanglement structure on qubits other than the two Bell pairs shown is omitted.
    }
    \label{fig:operator_vs_ent}
\end{figure}

\subsection{Spreading regimes}

The spread of magic is influenced by both the spreading of the operator $\tilde{T} = U_t T U^\dagger_t$ and the growth of entanglement in the state $\rho_\tau = V_\tau \ketbra{\psi}{\psi}V^\dagger_\tau$. More precisely, these together give an upper bound on the extent to which magic spreads as we next argue. We focus on contiguous subsystems for simplicity. For a subsystem $A$ of $\rho  = \tilde{T} \rho_\tau \tilde{T}^\dagger$ to contain a full unit of magic, it must have overlapping support with $\tilde{T}$, and ${\rm supp}(\tilde{T})$ typically contains qubits in a symmetric interval centered on the $T$-injection site with size $|{\rm supp}(\tilde{T})| \simeq 2 v_{\rm B}t$ [Fig.~\ref{fig:operator_vs_ent}], up to $\mathcal{O}(\sqrt{t})$ corrections~\cite{cvkPRX18, nahumPRX17}.

How far away such a subsystem $A$ can extend from ${\rm supp}(\tilde{T})$ is set by the entanglement in $\rho_\tau$ because $\tilde{T}$ must anticommute with some generator of $\rho^A_\tau$. It is thus useful to consider the shortest contiguous subsystem $A$ that starts at the rightmost (leftmost) qubit in ${\rm supp}(\tilde{T})$ and continues to the right (left), and crucially, can host a Bell pair within $\rho^A_\tau$ between the start and end qubits of $A$: assuming $\tilde{T}$ anticommutes with one generator of the Bell pair, it introduces magic, which is partially teleported to the qubit at the other end of $A$. Such subsystems are relevant because any subsystem for which $\rho^A_\tau$ is maximally mixed cannot fully absorb the magic; conversely, any longer subsystem with the same starting point can be reduced to the one above. 
Moreover, the size of $A$ is typically $\simeq 2 v_{\rm E} \tau$ (see Fig.~\ref{fig:operator_vs_ent}) since any smaller subsystem is maximally mixed with high probability~\cite{cvkPRX18, nahumPRX17}.
Putting together these observations, the full extent of magic is typically upper bounded by appending on each side of a $\simeq 2 v_{\rm B}t$-sized interval an interval of size $\simeq 2 v_{\rm E} \tau$ [cf. Fig.~\ref{fig:operator_vs_ent}]; hence, the FLEOM should typically satisfy $W(t, \tau) \leq 2(v_{\rm B} t + 2 v_{\rm E} \tau)$. 

However, whether this upper bound is saturated or not depends on the specific choices of $U_t$ and $V_\tau$. To this end, we next discuss a few key special cases, focusing on equal evolution times $t=\tau$ such that we can define a velocity $v_W = W(t)/2t$, and on $U_t$ and $V_t$ drawn from the same circuit family. We thus generally expect 
\begin{align} 
W(t) &\leq 2(v_{\rm B} + 2 v_{\rm E})t, \label{eq:Wupper}\\ 
v_W &\leq v_{\rm B} + 2 v_{\rm E} \eqqcolon v^{\max}_W,  \label{eq:vMupper}
\end{align}
the latter of which can be viewed as a general upper bound for a magic velocity. We now consider four cases in turn.

(\textit{i}) For $U = \id$ and $V \neq \id$, only entanglement can contribute to the magic spreading by creating Bell pairs between the qubit where the $T$ gate acts and other qubits; hence, it is reasonable to expect $v_W \simeq 2 v_{\rm E}$.

(\textit{ii}) For $U \neq \id$ and $V = \id$, only operator spreading captured in $\tilde{T}$ can contribute to the magic spreading, thus making it plausible to predict that $v_W \simeq v_{\rm B}$. 

(\textit{iii}) For both $U$ and $V$ nontrivial, and $U \neq V$, it is a priori unclear whether the upper bound Eq.~\eqref{eq:vMupper} is saturated, or if another velocity emerges. Indeed, we numerically find families of circuits that fall into both of these classes.

(\textit{iv}) For $U = V \neq \id$, we recover the circuits from the main text. Here, as in case (\textit{iii}), it is a priori unclear if the upper bound is saturated, but it is important to be aware of it for causality arguments. Moreover, despite the fact that ``resonant'' effects may appear due to $U^\dagger V = \id$, which might naively suggest that the magic spreading may be slowed down compared to (\textit{iii}), without further assumptions, these can in principle also boost the spreading of magic.

To scrutinize our upper bounds [Eqs.~\eqref{eq:Wupper} and~\eqref{eq:vMupper}] and better understand the various spreading regimes, we next numerically investigate cases ($i$-$iv$) from above.

\begin{figure}[tp]
    \centering
    \includegraphics[width=\textwidth]{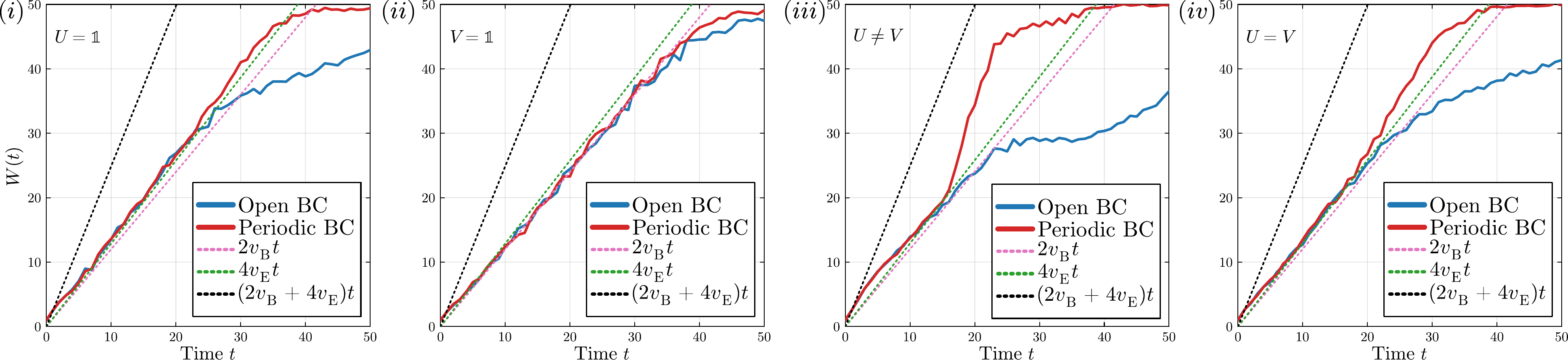}
    \caption{Interplay between entanglement growth and operators spreading in spreading magic: full linear extent of magic $W(t)$ vs evolution time $t$ under random brickwork Clifford circuits at no identity doping $p=0$ in the architecture from Fig.~\ref{fig:operator_vs_ent} for $L=50$. Initial states are random product stabilizer states. Data is averaged over $10^2$ circuit realizations where magic is successfully injected (imperceptible error bars: standard error of the mean). The blue and red solid curves represent $W(t)$ for open and periodic boundary conditions, respectively. Pink, green, and black dashed lines show spreading at the butterfly $v_{\rm B}$, twice entanglement $2v_{\rm E}$, and maximum $v^{\max}_W = v_{\rm B} + 2 v_{\rm E}$ velocity, respectively. Panels ($i$)-($iv$) depict various choices of the unitaries $U$ and $V$ from Fig.~\ref{fig:operator_vs_ent}. In all cases, the growth of $W(t)$ is consistent with the upper bound on magic growth [Eq.~\eqref{eq:Wupper}].}
    \label{fig:interplay}
\end{figure}

\subsubsection{Numerics}

Without loss of generality, we consider random (local) Clifford circuits without identity doping (i.e., $p=0$) and random product stabilizer initial states $\ket{\psi_0} = \otimes_i \ket{\pm \sigma_i}$ with $\ket{\pm \sigma_i}$ eigenstates of $\sigma_i \in \{X_i, Y_i, Z_i \}$, for both open and periodic boundary conditions (OBC and PBC). For circuits where both $U\neq \id$ and $V\neq \id$, we sample these independently from each other. In our simulations, we post-select circuit realizations where the $T$ gate adds magic; the fraction of discarded circuits becomes negligible for large enough $t$, thus, the simulations do not incur a large sample complexity. (We find that $\simeq 10^2$ circuit realizations yield sufficient magicful circuits.)

The numerical results, depicted in Fig.~\ref{fig:interplay}, can be summarized case by case as follows.
(\textit{i}) At early times, for both OBC and PBC, we observe $W(t) \simeq 4 v_{\rm E} t$ [cf. Fig.~\ref{fig:interplay}(\textit{i})], consistent with our expectations from the previous subsection. The effects of OBC eventually become apparent, leading to a slowdown in the growth of the FLEOM. (We discuss the effects of boundary conditions in more detail in the next subsection.)
(\textit{ii}) For both OBC and PBC, $W(t)$ grows as $\simeq 2 v_{\rm B}t$ until it saturates to its maximal value [Fig.~\ref{fig:interplay}(\textit{ii})]; this matches our expectations and further suggests that the effects of BCs are manifest in the entanglement structure of $\rho_\tau$, which is trivial here.
(\textit{iii}) At early times, for both OBC and PBC, the simulations show that the growth rate of $W(t)$ is consistent with a rate of the order of $4 v_{\rm E}$ and $2 v_{\rm B}$ [Fig.~\ref{fig:interplay}(\textit{iii})]. Then, for intermediate times, this rate decreases (increases) for OBC (PBC), after which $W(t)$ saturates. Notably, the value of $W(t)$ is consistent with our upper bound from Eq.~\eqref{eq:Wupper} at all times.
(\textit{iv}) Finally, here we observe a behaviour largely similar to case (\textit{i}): The only notable difference is that, for PBC at intermediate times (i.e.,  before saturation), $W(t)$ seems to grow slightly quicker [Fig.~\ref{fig:interplay}(\textit{iv})].

The numerical results from Figs.~\ref{fig:interplay} (\textit{i}), (\textit{iii}), and (\textit{iv}) suggest that the entanglement structure of $\rho_\tau$ may have effects on the spreading of magic at intermediate time, i.e., between an early time spreading regime and the late time saturation regime. We next suggest a potential mechanism underlying this behaviour.

\subsection{Effects of shrinking large logical operators}

Here, we discuss the shrinking of large logicals as a potential mechanism behind the intermediate-time spreading regimes from above. We start with another circuit example with $U=V$ and then conjecture why this mechanism might be important based on more general arguments.

\subsubsection{Example: Free fermionic self-dual kicked Ising circuits}
An illuminating example consists of the so-called self-dual kicked Ising (SDKI) circuit at the free-fermion, dual-unitary, and Clifford point (see e.g.~\cite{JYPC_PRR,sommersCrystallineQC}). In these circuits, also known as SDKI-f circuits, at $p=0$ doping, the logicals that give rise to MLMIs can be exactly tracked.
For SDKI-f circuits~\cite{JYPC_PRR}, the non-identity 2-qubit gates are $U_{\rm SDKI-f} = {\rm CZ} (H \otimes H) {\rm CZ}$, where ${\rm CZ}$ and $H$ are the control-$Z$ and Hadamard gate, respectively.
For concreteness, we fix $L=12k +6$ for some integer $k \geq 1$ with periodic boundary conditions, and choose the initial state as a tensor product of nearest neighbour Bell pairs $\ket{\psi_0} = \otimes_{i=1}^{L/2} (\ket{0_{2i-1} 0_{2i}} + \ket{1_{2i-1} 1_{2i}})/\sqrt{2}$, then act with a $T$ gate on qubit $L/2$. 
The logicals leading to the single MLMI at $t=0$ can be taken as $\YL(0) = Y_{L/2} X_{L/2+1}$ and $\ZL(0) = X_{L/2} X_{L/2+1}$. 
Focusing on $t \leq L/2$ since the dynamics has recurrences after $t=L/2$, the FLEOM satisfies $W(t) = 2t$ for $t < t_*$, jumps to $W(t_*) = L$, and stays $W(t) = L - \mathcal{O}(1)$ for $t_* < t \leq L/2$, where $t_* = L/2 v^{\max}_W = L/6$---this can be straightforwardly checked numerically. 
In terms of MLMIs, there is a single MLMI $(\frac{L}{2} - t, \frac{L}{2}+t+1)$ for $t < t_*$, followed by three MLMIs 
\begin{align}
    \left\{ \left( \frac{L}{2}-t_*, \frac{L}{2}+t_*+1 \right), \left( 1, \frac{L}{2}+t_*\right), \left( \frac{L}{2}+t_*, L\right) \right\} \qquad {\rm for} \ t=t_*,
\end{align}and multiple MLMIs for $t_* < t \leq L/2$, which we omit for brevity.
At $t= t_*$, the logicals which give rise to MLMIs are
\begin{align}
	\left( \frac{L}{2}-t_*, \frac{L}{2}+t_*+1 \right):\  \YL(0) &= Y_{\frac{L}{2}} X_{\frac{L}{2}+1} \textrm{ and} \ \ZL(0) = X_{\frac{L}{2}} X_{\frac{L}{2}+1},\\
	\left( 1, \frac{L}{2}+t_*\right):\ \YL'(0) &= \left( \prod_{i=1}^{6k} X_i\right) \id \id YX \left( \prod_{i=6k+5}^{10k+4} X_i \right) YY \left( \prod_{i=10k+7}^{L} X_i\right)
    \textrm{ and}\\  \ZL'(0) &=  \left( \prod_{i=1}^{10k+4} X_i \right) YY \left( \prod_{i=10k+7}^{L} X_i\right),
\end{align}
see also Fig.~\ref{fig:SDKIf_shrink}, while those of $\left( \frac{L}{2}+t_*, L\right)$ can be found by doing a reflection about the center of the system.

\begin{figure}[tp]
    \centering
    \includegraphics[width=\textwidth]{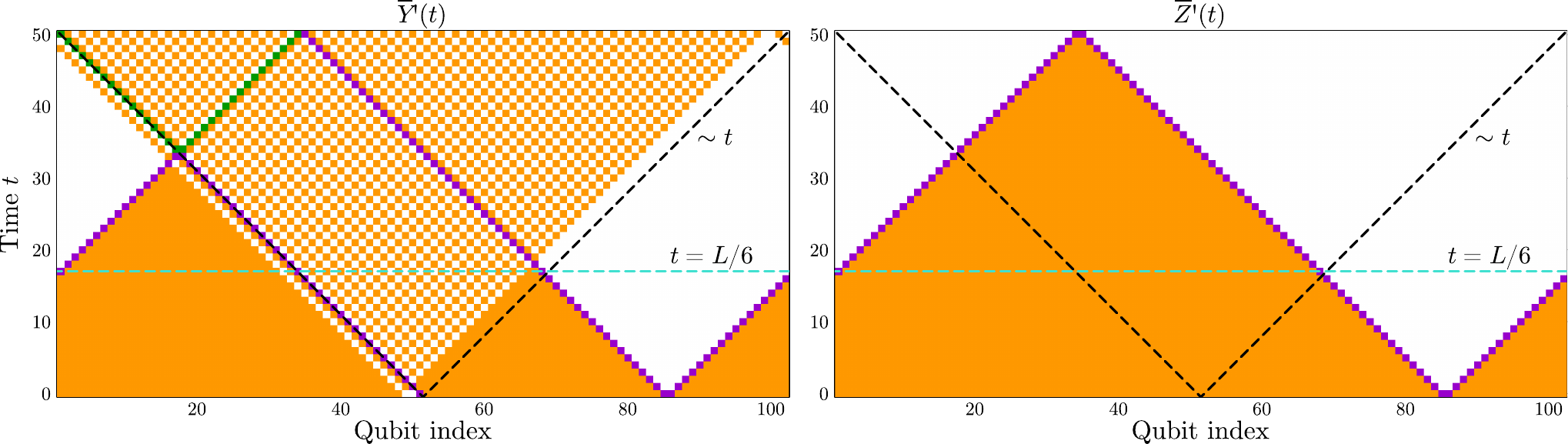}
    \caption{Time evolution of shrinking logical operators $\YL'(t)$ and $\ZL'(t)$ under an SDKI-f circuit with $L=102$ and up to $t=50$. White, orange, purple, and green pixels depict $\id_i$, $X_i$, $Y_i$, and $Z_i$ Pauli operators, respectively. $\YL'(t)$ and $\ZL'(t)$ shrink due to an early ($t=1$) nucleation of an island of identity operators and, at $t= L/2v^{\max}_W = L/6$ (horizontal turquoise dashed line), lead to a minimal linear magic interval that is mutually-irreducible with respect to the lightcone growing MLMI (black dashed lines).}
    \label{fig:SDKIf_shrink}
\end{figure}

To gain some intuition into the appearance of these new pairs of logicals, it is instructive to monitor the evolution of $\YL'(t)$ and $\ZL'(t)$: this is depicted in Fig.~\ref{fig:SDKIf_shrink}. It thus becomes apparent that these logicals shrink (to be precise, in PBC) due to the nucleation of an island of identity operators $\id_i$, and their corresponding magic interval eventually becomes mutually-irreducible with respect to that of $(\YL(t), \ZL(t))$ at $t = t_*$. Note that the boundary conditions are, in fact, not crucial for this dynamics: one could always use OBC by adding specific single-qubit Clifford gates on qubits $1$ and $L$ that give the same dynamics as PBC. 

\subsubsection{Conjecture}

Here, we outline a conjecture for an effect that may be potentially relevant to the spreading of magic at intermediate times.
Focusing on $U=V$ for concreteness, we conjecture that the shrinking of nonlocal logical operators might be particularly relevant to such intermediate time spreading regimes. 
By nonlocal logical operators, we mean logicals defined before the evolution under $U_t$ (but after the application of the $T$ gate), which have a diameter of the order of the system size $\mathcal{O}(L)$. 
In addition to the illustrative example from the previous section, we next present several observations that motivate this conjecture.

Note that at very early times $t = {\cal O}(1) \ll L$, the MLMIs must come from local logicals due to the bound $W(t) \leq 2v^{\max}_Wt$; this can be easily verified by considering the backward lightcone from each MLMI, which points towards logicals with diameter of size $\mathcal{O}(t) \ll L$. Then, at intermediate times $t \sim L/2v^{\max}_W$, by lightcone arguments alone, one may not distinguish whether MLMIs come from local logicals that grew or nonlocal logicals that shrank. However, suppose nonlocal logicals were not contributing to MLMIs: in this case, it seems unlikely that the slope of $W(t)$ would change, and in particular, that there would be boundary (OBC vs PBC) effects at these intermediate times. To further motivate the latter, note that local logicals which could have become MLMIs are very unlikely to spread to the boundary because $v^{\max}_W \gtrsim 2v_{\rm B}$, thus suggesting that the operators which contribute to these effects are in fact the nonlocal logicals since these are typically modified by gates near the boundaries starting from $t=0$.

The general lesson suggested by the SDKI-f example is that initially nonlocal logicals may affect the dynamics of the full linear extent of magic at intermediate times by shrinking. While we supplemented this example with certain general observations, it remains a conjecture to what extent the shrinking of nonlocal logicals generally affects magic spreading. Assessing this conjecture may offer insights into the origin of the spreading at intermediate time scales. Such an assessment, as well as alternative arguments for the crossover to slower (OBC) or faster (PBC) spreading regimes, represent interesting open questions, to the best of our knowledge.

\section{Butterfly and entanglement velocities in identity-doped random Clifford circuits}\label{app:rand_cliff_velocity_scales}
In this appendix, we present the derivation of the average butterfly velocity in identity-doped random Clifford circuits at doping rate $p$ [Eq.~\eqref{eq:vPdep}] and give a heuristic argument for the dependence of the entanglement velocity on the butterfly velocity [Eq.~\eqref{eq:vPdep}]. 

As pointed out in Ref.~\cite{JYPC_PRR}, the motion of each end of an operator string in a scrambling Clifford circuit can be described through a biased persistent random walk (to be contrasted with the biased random walk of Haar-random circuits~\cite{nahumPRX17,opSprVedika,cvkPRX18}). Consider the right end of an operator string as an example: at time $t_i$, it has probability $\alpha_+$ or $\alpha_-$ to reverse its direction of movement, depending on whether the end was moving to the right from $t_{i-1}$ to $t_i$, or to the left, respectively. Due to the even-odd staggering pattern of the brickwork circuit, if the right end of an operator string was moving to the right from time $t_{i-1}$ to $t_i$, then the end is located on the first leg of the gate at time $t_i$; conversely, if the right end was moving to the left from time $t_{i-1}$ to $t_{i}$, then it is located on the second leg of the gate at time $t_i$. 

Consider first the case where the right end of the operator string ends on the first leg of the gate: if the gate was chosen to be identity, the right end reverses its direction, and this occurs with probability $p$; conversely, if the gate was chosen to be a random 2-qubit Clifford gate, which occurs with probability $1-p$, the right end may either reverse direction (operator front stays put) or continue along the same direction (operator front grows). It is known that a random 2-qubit Clifford gate maps any Pauli string of length two to any other Pauli string of length two, excluding the double-identity case, uniformly. Out of the 15 possible Pauli strings of length two to be mapped to, 3 correspond to the case $PI\rightarrow P'I$ with $P' \neq I$, where the operator front stays put, and 12 correspond to the case $PI\rightarrow P'Q'$ with $P', Q' \neq I$, where the operator front grows. Therefore, altogether we have
\begin{align}
    \alpha_+=p+\frac{3}{15}(1-p)=p+\frac{1}{5}(1-p).
    \label{eq:alpha_plus}
\end{align}

Analogously, we can calculate the second case where the right end of the operator string ends on the second leg of the gate, yielding
\begin{align}
    \alpha_-=p+\frac{12}{15}(1-p)=p+\frac{4}{5}(1-p).
    \label{eq:alpha_minus}
\end{align}
It is known that for a persistent random walker, $\alpha_+$ and $\alpha_-$ completely determines the drift velocity, which in our case corresponds to the butterfly velocity. Explicitly, using Eq. (95) from Ref.~\cite{JYPC_PRR}, we obtain
\begin{align}
    v_{\rm B}=\frac{\alpha_--\alpha_+}{\alpha_-+\alpha_+}=\frac{3}{5}\left(\frac{1-p}{1+p}\right).
    \label{eq:vB_rand_cliff}
\end{align}
Note that the limiting case of $p=0$ corresponds to applying random Clifford gates, in which case $v_{\rm B}=0.6$, consistent with results from ~\cite{nahumPRX17,cvkPRX18}. Conversely, if $p=1$ we recover $v_{\rm B}=0$ as expected, since there will be no dynamics if all gates are the identity. 

As for the entanglement velocity as a function of $v_{\rm B}$, $v_{\rm E}(v_{\rm B})$, we observe that for all values of $0\le p \le 1$, Eq. (27) from \cite{cvkPRX18} holds approximately
\begin{align}
    v_{\rm E}\approx \frac{\log(1+v_{\rm B})+\log(1-v_{\rm B})}{\log(1+v_{\rm B})-\log(1-v_{\rm B})}.
    \label{eq:ve_as_func_of_vb}
\end{align}
A heuristic argument as to why this relation holds regardless of the rate of identity-doping is that $v_{\rm E}(v_{\rm B})$ depends only on the \textit{shape} of the operator front and not on its \textit{location}~\cite{cvkPRX18}.
Therefore, while $p$ changes the rate at which the operator front propagates, it does not distort the shape of the front, and thus the relation between butterfly and entanglement speeds remains unchanged. 

\section{Connection between shape of the Pauli spectrum and the stabilizer R\'enyi entropy}
\label{app:PS_SRE}

In this appendix, we present in more detail the connection between the stabilizer R\'enyi entropy, a scalar measure, and the shape of the Pauli spectrum in a stabilizer state doped with a single $T$ gate. For concreteness, we use the 2-SRE, $\mathcal{M}_2$ for pure states and $\tilde{\mathcal{M}}_2$ for mixed states, cf. Eq.~\eqref{eq:SRE}.

Consider a pure stabilizer initial state $|\psi\rangle$ on $L$ qubits with density matrix $\rho_0=|\psi\rangle\langle\psi|=\prod_{i=1}^{L}\frac{1+g_i}{2}$, where $\{g_1,g_2,\ldots,g_L\}$ are a generator set of the stabilizer group of $\rho_0$. By expanding the product of sums, one directly observes there are exactly $2^L$ non-trivial Pauli coefficients, each being equal to $2^{-L}$. We can write  $\rho_0=\sum_{i=1}^{4^L}c_i\cdot P_i$, where we expand $\rho_0$ using the $4^L$ Pauli strings $\{P_i\}$ as a basis, with expansion coefficients $c_i$ given by 
\begin{align}\label{eq:c_i_stab}
    c_i=\begin{cases}
        2^{-L},\, & \text{if }\,1\le i\le 2^L,\\
        0\,, \, & \text{if }\, i>2^L.
    \end{cases}    
\end{align}
The Pauli spectrum is thus $\Xi(\rho)=\{\xi_i\}\coloneqq \{c_i^2\cdot2^L\}$, with $\xi_i$ arranged in descending order. Plotting $\xi_i$ against indices $i$ thus yields a step function
\begin{align}\label{eq:xi_i_stab}
    \xi_i=\begin{cases}
        2^{-L},\, & \text{if }\,1\le i\le 2^L,\\
        0\,, \, & \text{if }\, i>2^L.
    \end{cases}    
\end{align}

Now, locally injecting one unit of magic by applying a single $T$ gate on $\rho_0$ results in $\rho = T\rho_0 T^\dagger$ given by Eq.~\eqref{eq:rho}. As mentioned in the main text and detailed in App.~\ref{subsec:evo_T_gate}, one can always find a gauge for the stabilizer generators $\{ g_i \}_{i=1}^L$ such that $T$ commutes with $g_j$ for $j\geq 2$, and does not commute with $g_1$. 
Therefore, from Eq.~\eqref{eq:rho}, one sees that after injecting one unit of magic into a stabilizer state, the step-function in the Pauli spectrum splits into two steps
\begin{align}\label{eq:xi_i_T}
    \xi_i=\begin{cases}
        2^{-L}\,,\, & \text{if }\,1\le i\le 2^{L-1},\\
        2^{-L-1}\,,\, & \text{if }\, 2^{L-1}<i\le 3\cdot 2^{L-1},\\
        0\,, \, & \text{if }\, i>3\cdot 2^{L-1}.
    \end{cases}    
\end{align}

\begin{figure}[tp]
    \centering
    \includegraphics[width=13 cm]{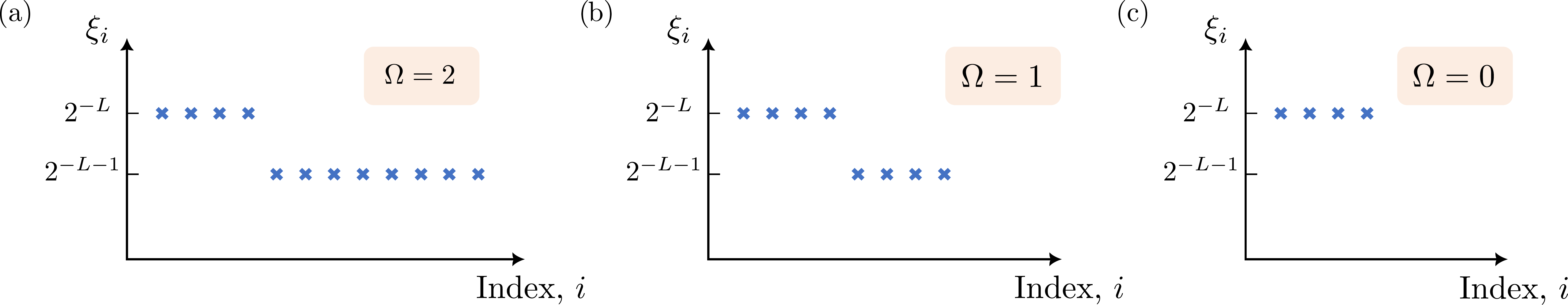}
    \caption{Shape of the Pauli spectrum $\Xi(\rho) = \{ \xi_i \}$ for (sub)system magic given by (a) $\sre(\rho)=\log_2(4/3)$, (b) $\sre(\rho)=\log_2(6/5)$, and (c) $\sre(\rho)=0$, where we only plot the nonzero entries $\xi_i$. Note that the norm of the distribution is irrelevant, and $\sre$ depends only on the ratio $\Omega$.}
    \label{fig:omega}
\end{figure}%

We shall refer to the coefficients with value $\xi_i=2^{-L}$ as ``regular'' coefficients because (mixed) stabilizer states only have such nonzero coefficients.
In contrast, we refer to coefficients with value $\xi_i=2^{-L-1}$ as the ``magical'' coefficients, as they only exist in (mixed) states with finite magic.
Crucially, the ratio of the number of magical coefficients to the number of regular coefficients, defined as $\Omega$, is
\begin{align}
    \Omega(\rho)\coloneqq \frac{\textrm{\# magical coefficients}}{\textrm{\# regular coefficients}}=
    \begin{cases}
        2\,,\, &\text{if there is 1 full unit of magic in }\rho,\\
        0\,,\, &\text{if there is no magic in }\rho.
    \end{cases}
    \label{eq:omega_full_state}
\end{align}
So far we have established the connection between the shape of the Pauli spectrum and the amount of magic in the full state. To observe the spatial extent of magic over time, we consider the state at time $t$,
\begin{align}
    \rho(t) = U(t)T\rho_0 T^\dagger U^\dagger(t)=\left(\frac{1}{2}+\frac{\ZL(t)}{2\sqrt{2}}-\frac{\YL(t)}{2\sqrt{2}}\right)\prod_{i=2}^{L}\frac{1+g_i(t)}{2}. 
\end{align}
Consider a subsystem $A$ and its complement $B = [L]\setminus A$, there are then three possible scenarios, as also summarized in~\autoref{tab:bip} in the main text. We next compute $\Omega(\rho_A(t))$ from Eq.~\eqref{eq:omega_full_state} for each of these cases.
First, if both $\ZL(t)$ and $\YL(t)$ can have their supports entirely in $A$ [\autoref{thm:bip} case (\textit{i})], then $\Omega(\rho_A)=2$, and $\rho_A$ contains one unit of magic.
Second, if only one of $\ZL(t)$ and $\YL(t)$ can have its support entirely in $A$ [\autoref{thm:bip} cases (\textit{iv, v})], while the other one has its support at least partially outside of $A$ and therefore gets killed when tracing out $B$ to obtain $\rho_A$, then $\Omega(\rho_A)=1$ and $\rho_A$ only contains half a unit of magic.
Third, if both $\ZL(t)$ and $\YL(t)$ have their supports at least partially outside of $A$ and therefore both get killed when tracing out $B$ [\autoref{thm:bip} cases (\textit{ii, iii})], then $\Omega(\rho)=0$ and $\rho_A$ contains no magic. Note that when one or more $g_i$'s get killed when tracing out $B$, the norm of the Pauli spectrum $\Xi(\rho_A)$ gets reduced, but its shape remains unchanged, seen from the observation that $\Omega(\rho_A)$ does not depend on the number of (surviving) terms in the product $\prod_{i=2}^{L}[(1+g_i(t))/2]$.

In summary, the amount of magic in the subsystem $\rho_A$ is entirely determined by the shape of the Pauli spectrum $\Xi(\rho_A)$, which is entirely described by the ratio $\Omega(\rho_A)$. Conveniently, for possibly mixed stabilizer states doped with a single $T$ gate, there is a one-to-one correspondence between the values of $\Omega(\rho_A)$ and the SRE $\tilde{\mathcal{M}}_2(\rho_A)$
\begin{align}\label{eq:omega_subsys}
\Omega=\begin{cases}
    2\,,\, &\text{iff } \tilde{\mathcal{M}}_2=\log_2(4/3),\\
    1\,,\, &\text{iff } \tilde{\mathcal{M}}_2=\log_2(6/5),\\
    0\,,\, &\text{iff } \tilde{\mathcal{M}}_2=0.
\end{cases}
\end{align}

\section{Distribution of length of typical MLMIs}
\label{app:distribution}

In this appendix, we present numerical results on the distribution of the lengths of MLMIs. As shown in Fig.~\ref{fig:distrib_typ_MLMI}(a), as time progresses, the typical size of MLMIs  $\ell_{\textrm{typ}}(t)$ grows linearly with time. This is indicated by the peak of the distribution shifting linearly to the right over time. The linear shift is clearly seen in the colormap plot in the inset of Fig.~\ref{fig:distrib_typ_MLMI}(a). This is further corroborated in Fig.~\ref{fig:distrib_typ_MLMI}(b), where from the linear fit, we observe that $\ell_{\textrm{typ}}(t)\approx 2 v_{\rm E} t$. The inset shows that this relation holds consistently throughout various values of $p$. Therefore, we conclude that the typical growth behavior of MLMIs is well represented by the growth behavior of the symmetric-centered MLMI $\ell_{\textrm{typ}}(t)\sim \ell(t)$. 
Finally, note that the distribution sharpens as time progresses, and the peak interval length saturates at $\ell_{\textrm{typ}}\simeq L/2$.

\begin{figure}[hbp]
    \centering
    \includegraphics[width=0.85 \textwidth]{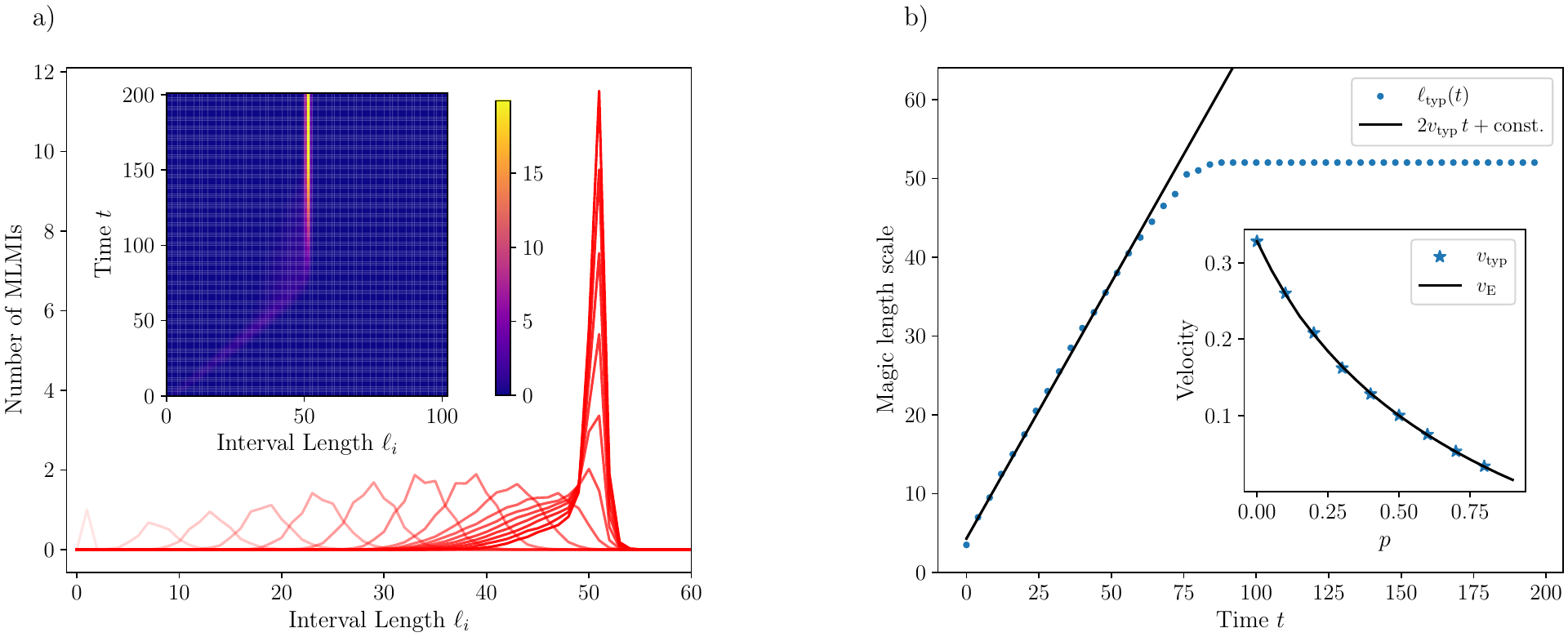}
    \caption{(a) Distribution of minimal linear magic intervals in random Clifford circuits for $L=102$, $t=200$, and $p=0.0$. Data averaged over 2000 circuit realizations.
    Darker red curves indicate later times, with the time range being $1\le t\le 126$.  
    Inset: evolution of the distribution with time; colorbar shows the number of MLMIs. 
    (b) Growth of the typical size of MLMIs $\ell_{\rm typ}$, defined as the peak of the distribution.
    The early-time regime is fitted with a linear function whose slope is defined as $v_{\rm{typ}}$. Inset: Dependence of $v_{\rm{typ}}$ on identity doping rate $p$, which is consistent with $v_{\rm{typ}}(p) \simeq v_{\rm E}(p)$. }
    \label{fig:distrib_typ_MLMI}
\end{figure}%

\section{Numerical results on identity-doped dual-unitary Clifford circuits}
\label{app:du}

In addition to the results on identity-doped random Clifford circuits presented in the main text, in this appendix, we present results on a different class of scrambling Clifford dynamics: identity-doped dual-unitary (DU) Clifford circuits. 
The intrinsic velocity scales, $v_{\rm E}$ and $v_{\rm B}$, in these doped DU circuits have different dependencies on the identity-doping rate $p$ than in random Clifford circuits. In particular, both $v_{\rm E}$ and $v_{\rm B}$ vary over a larger range of values. Therefore, studying these circuits not only helps test the generality of our findings regarding magic spreading but also allows access to additional parameter regimes.

Dual-unitary gates are two-site gates satisfying both the usual temporal unitarity condition as well as a spatial unitarity condition~\cite{bertini2025exactlysolvablemanybodydynamics}.
As pointed out in~\cite{JYPC_PRR,crooks_gates_2020}, out of the four classes of two-site Clifford gates invariant under local-basis rotation, two are dual-unitary: the SWAP and the iSWAP class. The SWAP class are ``poor scramblers"~\cite{sommersCrystallineQC,JYPC_PRR}, in that they do not scramble information but merely transport it. In contrast, the iSWAP class are typically ``good scramblers"~\cite{sommersCrystallineQC,JYPC_PRR}. To erase dependence on single-site basis choice, we choose a representative subclass of gates, the SDKI-r gates, as introduced in~\cite{JYPC_PRR}, which correspond to the self-dual kicked Ising model at the free fermion point ($J=\pi/4$, $g=\pi/4$, $h=0$) plus single-site random Clifford gates. These 2-qubit gates are defined as
\begin{align}\label{eq:SDKIr_def}
    U_{\rm{SDKI-r}}\coloneqq (v_1\otimes v_2){\rm CZ}(H\otimes H){\rm CZ}
\end{align}
where $v_1$ and $v_2$ are single-qubit Clifford gates drawn independently and uniformly at random, ${\rm CZ}$ is the two-qubit control-$Z$ gate, and $H$ is the Hadamard gate. 

\begin{figure}[bp]
    \centering
    \includegraphics[width=7cm]{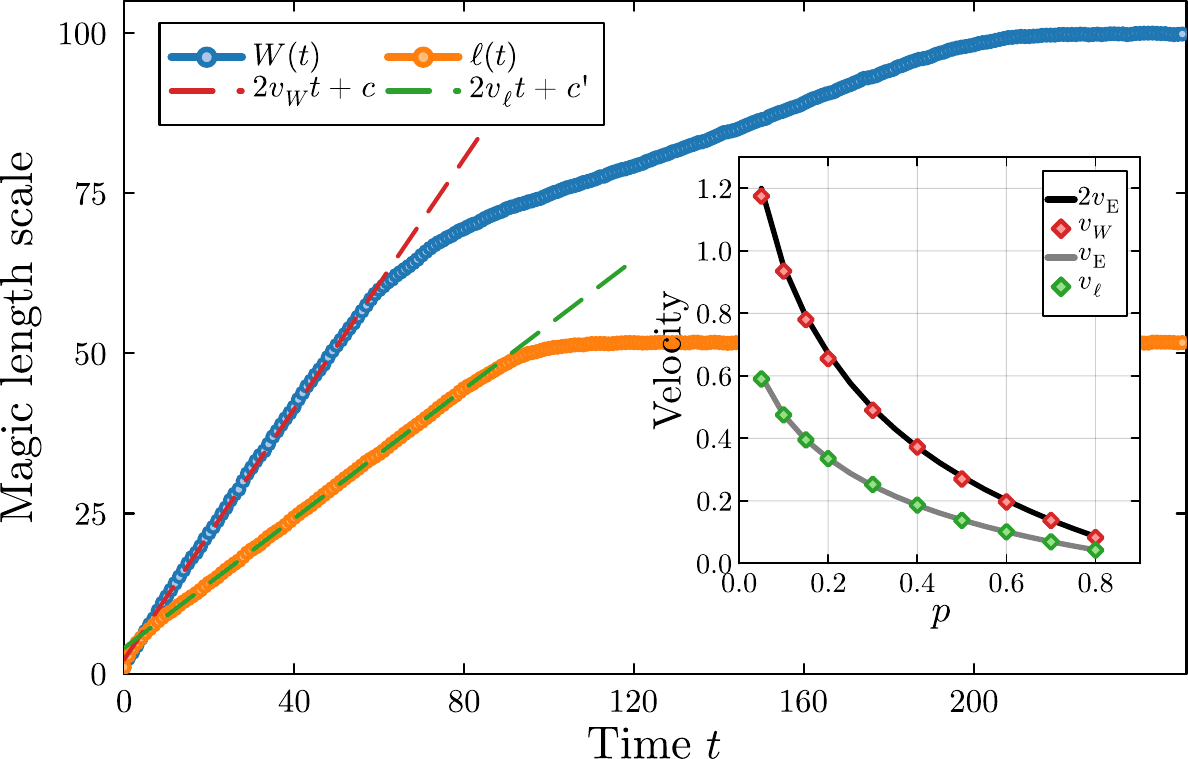}%
    \caption{Growth of magic length scales over time in identity-doped SDKI-r circuits, for $L=102$ and identity-doping rate $p=0.3$. Blue (orange) dots show FLEOM (LML). At early times, both MLS grow ballistically at fitted rates $2v_W$ and $2v_{\rm \ell}$, respectively (dashed lines). Inset: Dependence of $v_W$ and $v_{\rm \ell}$ on $p$ compared to the entanglement velocity $v_{\rm E}(p)$. Data averaged over 2000 circuit realizations, negligible error bars not shown.}
    \label{fig:width_SDKIr}
\end{figure}%
Doping the SDKI-r circuit with identity gates undermines the dual-unitary property of the circuit, thus rendering the dynamics more generic. Nevertheless, the butterfly velocity $v_{\rm{B}}(p)$ in such circuits differs from that in identity-doped random Clifford circuits described by Eq.~\eqref{eq:vPdep} and follows instead \cite{JYPC_PRR}
\begin{align}
    v_{\rm{B}}=\left( 1+\frac{8}{3}\frac{p}{(1-p)} \right)^{-1}.
\end{align}
Meanwhile, the dependence of $v_{\rm{E}}$ on $v_{\rm{B}}$ still holds approximately in the same way as that given in Eq.~\eqref{eq:vPdep}.
Let us emphasize that here, unlike the case of (doped) random Clifford circuits, $v_{\rm E}$ is significantly different from $v_{\rm B}/2$.

Having introduced the identity-doped DU Clifford circuits of interest, we present in Fig.~\ref{fig:width_SDKIr} the spreading of FLEOM and LML in such circuits, as well as the dependence of their velocities on the identity-doping rate. 
Compared to the case of identity-doped random Clifford circuits, the growth behavior of $W(t)$ and $\ell(t)$ is qualitatively the same: they both spread ballistically [Fig.~\ref{fig:width_rand_cliff}]. The only difference is in the values of $v_{\rm{W}}$ and $v_{\rm{\ell}}$. 
We see that relations $v_{\rm{W}}\simeq 2 v_{\rm E}$ and $v_{\rm \ell}\simeq v_{\rm E}$ remain valid, as in the identity-doped random Clifford setup [cf. Fig.~\ref{fig:width_rand_cliff} and inset]. 

To conclude, two useful lessons that can be learned from studying the identity-doped SDKI-r circuits are: (\textit{i}) the relations of $v_{\rm{W}}\simeq 2 v_{\rm E}$ and $v_{\rm \ell}\simeq v_{\rm E}$ seem to generally hold in scrambling random Clifford circuits;
(\textit{ii}) as seen in the inset of Fig.~\ref{fig:width_SDKIr}, at $p=0.05$ for example, $v_W\approx 1.2>1$, which clearly indicates that the spreading of magic can go beyond the geometric light cone due to entanglement build-up (at least according to FLEOM). This latter observation is consistent with our upper bound for the growth of $W(t)$ from Eq.~\eqref{eq:Wupper}.
Lastly, we remark that dynamics at the DU point without doping ($p=0$) is highly-fined tuned and qualitatively different from the dynamics at any $p>0$. While interesting in its own right, we leave such dynamics in DU Clifford circuits for future investigations.

\section{Overlap between the left- and rightmost MLMIs}\label{app:overlap}

Here we provide numerical results regarding the overlap $\omega(t)$ between the left- and rightmost MLMIs.
Fig. \ref{fig:overlap_scale_combined} (a) and (b) show $\omega(t)$ for different values of $L$ and $p$, respectively. The initial growth regime is fitted with an algebraic ansatz of $\sim t^{0.45}$, approximately consistent with the expected behavior due to diffusive broadening of operator fronts, which predicts $\omega\sim\sqrt{t}$. Upon varying system size $L$, the growth exponent remains mostly unchanged, whereas the peak value grows with increasing $L$. Similarly, for different doping rates, the initial growth behavior is again consistent with $\omega\sim t^{0.45}$, independent of $p$. At late times, the left and rightmost MLMIs again overlap minimally on ${\cal O}(1)$ qubits on average. These observations confirm that $W(t)\simeq 2 \ell (t)-\mathcal{O}(\sqrt{t})\simeq 4v_{\rm E} t$ to leading order, as illustrated in Fig.~\ref{fig:FLEOMoverlap_main}.

\begin{figure}[h]
    \centering
    \includegraphics[width=0.7\linewidth]{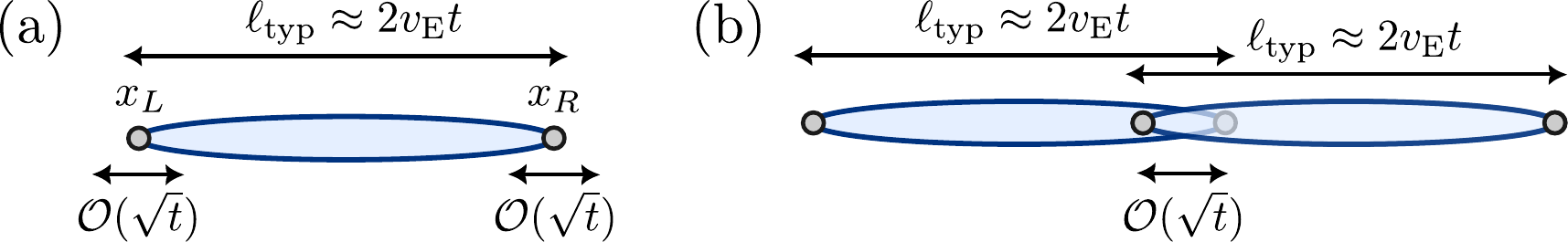}
    \caption{(a) Growth of the typical MLMI length $\ell_{\rm{typ}}$ due to spreading of logical operators $\overline{X}$ and $\overline{Y}$. The leading-order growth is linear in $t$, with a subleading correction of order $\mathcal{O}(t)$ due to diffusive broadening of the (logical) operator fronts. (b) The left- and rightmost MLMIs both grow typically as $\simeq 2v_{\rm E}t$, with an overlap that grows as $\mathcal{O}(\sqrt{t})$ due to diffusive broadening of the (logical) operator fronts, thus leading to $W(t)\simeq 2\ell_{\rm typ}(t)+\mathcal{O}(\sqrt{t})$.  }
    \label{fig:FLEOMoverlap_main}
\end{figure}

\begin{figure}[h]
    \centering
    \includegraphics[width=\textwidth]{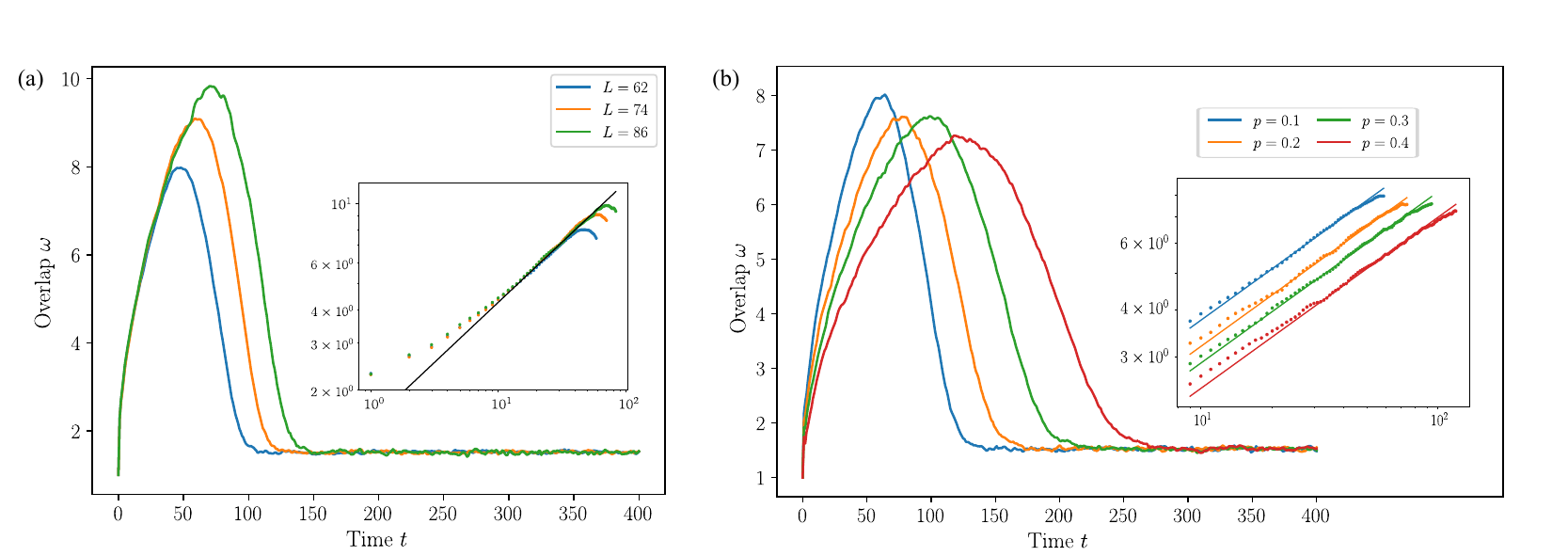}%
    \caption{
    Overlap $\omega (t)$ between the left- and rightmost mutually-independent magic intervals for $(a)$ system sizes $L=62,\, 74,\, 86$, identity-doping rate $p=0$, and $(b)$ $L=62$, $p=0.1,0.2,0.3,0.4$. Results are averaged over 2000 random Clifford circuit realizations. The inset shows the initial growth regime, consistent with $\omega(t) \propto t^{0.45}$. Negligible error bars not shown.}
    \label{fig:overlap_scale_combined}
\end{figure}
\section{Scaling of LML and FLEOM with system size}\label{app:L_scaling}

In this Appendix, we show the scaling of LML and FLEOM with system size $L$. As seen in Fig. \ref{fig:LML_with_L}, for LML, the growth rate is consistently $2v_{\rm E}$, independent of the system size, while the saturation value scales linearly with system size and is approximately $ \ell(t \gg 1) \approx L/2 +1$; for FLEOM, the initial growth rate is consistent with $4v_{\rm E}$, and the time at which the growth slows down due to boundary effect, $t_{*}$, also scales linearly with $L$. The slower growth rate after $t_{*}$ is constant upon varying system sizes, and the saturation value scales linearly with $L$, given by $W(t \gg 1)\approx L$. 
\begin{figure}[h]
    \centering
    \includegraphics[width=\textwidth]{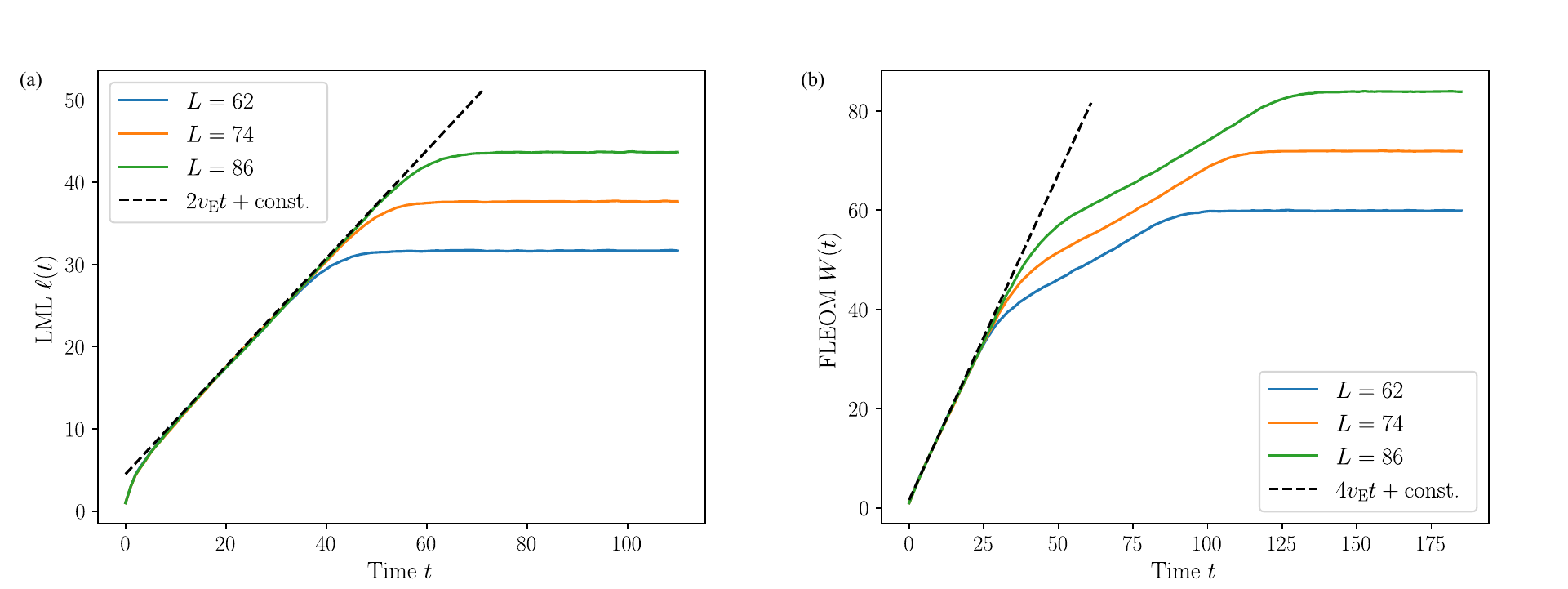}%
    \caption{
    (a) Linear magic length (LML) and (b) full linear extent of magic (FLEOM) for system sizes $L=62,\, 74,\, 86$ and identity-doping rate $p=0$, averaged over 2000 random Clifford circuit realizations. Negligible error bars not shown.
    }
    \label{fig:LML_with_L}
\end{figure}

\end{onecolumngrid}

\begin{thebibliography}{115}%
    \makeatletter
    \providecommand \@ifxundefined [1]{%
     \@ifx{#1\undefined}
    }%
    \providecommand \@ifnum [1]{%
     \ifnum #1\expandafter \@firstoftwo
     \else \expandafter \@secondoftwo
     \fi
    }%
    \providecommand \@ifx [1]{%
     \ifx #1\expandafter \@firstoftwo
     \else \expandafter \@secondoftwo
     \fi
    }%
    \providecommand \natexlab [1]{#1}%
    \providecommand \enquote  [1]{``#1''}%
    \providecommand \bibnamefont  [1]{#1}%
    \providecommand \bibfnamefont [1]{#1}%
    \providecommand \citenamefont [1]{#1}%
    \providecommand \href@noop [0]{\@secondoftwo}%
    \providecommand \href [0]{\begingroup \@sanitize@url \@href}%
    \providecommand \@href[1]{\@@startlink{#1}\@@href}%
    \providecommand \@@href[1]{\endgroup#1\@@endlink}%
    \providecommand \@sanitize@url [0]{\catcode `\\12\catcode `\$12\catcode
      `\&12\catcode `\#12\catcode `\^12\catcode `\_12\catcode `\%12\relax}%
    \providecommand \@@startlink[1]{}%
    \providecommand \@@endlink[0]{}%
    \providecommand \url  [0]{\begingroup\@sanitize@url \@url }%
    \providecommand \@url [1]{\endgroup\@href {#1}{\urlprefix }}%
    \providecommand \urlprefix  [0]{URL }%
    \providecommand \Eprint [0]{\href }%
    \providecommand \doibase [0]{https://doi.org/}%
    \providecommand \selectlanguage [0]{\@gobble}%
    \providecommand \bibinfo  [0]{\@secondoftwo}%
    \providecommand \bibfield  [0]{\@secondoftwo}%
    \providecommand \translation [1]{[#1]}%
    \providecommand \BibitemOpen [0]{}%
    \providecommand \bibitemStop [0]{}%
    \providecommand \bibitemNoStop [0]{.\EOS\space}%
    \providecommand \EOS [0]{\spacefactor3000\relax}%
    \providecommand \BibitemShut  [1]{\csname bibitem#1\endcsname}%
    \let\auto@bib@innerbib\@empty
    %</preamble>
    \bibitem [{\citenamefont {Shor}(1995)}]{shor}%
      \BibitemOpen
      \bibfield  {author} {\bibinfo {author} {\bibfnamefont {P.~W.}\ \bibnamefont
      {Shor}},\ }\href {https://doi.org/10.1103/PhysRevA.52.R2493} {\bibfield
      {journal} {\bibinfo  {journal} {Phys. Rev. A}\ }\textbf {\bibinfo {volume}
      {52}},\ \bibinfo {pages} {R2493} (\bibinfo {year} {1995})}\BibitemShut
      {NoStop}%
    \bibitem [{\citenamefont {Calderbank}\ and\ \citenamefont
      {Shor}(1996)}]{CalderbankShor}%
      \BibitemOpen
      \bibfield  {author} {\bibinfo {author} {\bibfnamefont {A.~R.}\ \bibnamefont
      {Calderbank}}\ and\ \bibinfo {author} {\bibfnamefont {P.~W.}\ \bibnamefont
      {Shor}},\ }\href {https://doi.org/10.1103/PhysRevA.54.1098} {\bibfield
      {journal} {\bibinfo  {journal} {Phys. Rev. A}\ }\textbf {\bibinfo {volume}
      {54}},\ \bibinfo {pages} {1098} (\bibinfo {year} {1996})}\BibitemShut
      {NoStop}%
    \bibitem [{\citenamefont {Steane}(1996)}]{steane}%
      \BibitemOpen
      \bibfield  {author} {\bibinfo {author} {\bibfnamefont {A.~M.}\ \bibnamefont
      {Steane}},\ }\href {https://doi.org/10.1103/PhysRevLett.77.793} {\bibfield
      {journal} {\bibinfo  {journal} {Phys. Rev. Lett.}\ }\textbf {\bibinfo
      {volume} {77}},\ \bibinfo {pages} {793} (\bibinfo {year} {1996})}\BibitemShut
      {NoStop}%
    \bibitem [{\citenamefont {Gottesman}(1997)}]{gottesman1997stabilizer}%
      \BibitemOpen
      \bibfield  {author} {\bibinfo {author} {\bibfnamefont {D.}~\bibnamefont
      {Gottesman}},\ }\href@noop {} {\bibinfo {title} {Stabilizer codes and quantum
      error correction}} (\bibinfo {year} {1997}),\ \Eprint
      {https://arxiv.org/abs/quant-ph/9705052} {arXiv:quant-ph/9705052}
      \BibitemShut {NoStop}%
    \bibitem [{\citenamefont {Bravyi}\ and\ \citenamefont
      {Kitaev}(1998)}]{surface}%
      \BibitemOpen
      \bibfield  {author} {\bibinfo {author} {\bibfnamefont {S.~B.}\ \bibnamefont
      {Bravyi}}\ and\ \bibinfo {author} {\bibfnamefont {A.~Y.}\ \bibnamefont
      {Kitaev}},\ }\href@noop {} {\bibinfo {title} {Quantum codes on a lattice with
      boundary}} (\bibinfo {year} {1998}),\ \Eprint
      {https://arxiv.org/abs/quant-ph/9811052} {arXiv:quant-ph/9811052}
      \BibitemShut {NoStop}%
    \bibitem [{\citenamefont {Gottesman}\ \emph {et~al.}(2001)\citenamefont
      {Gottesman}, \citenamefont {Kitaev},\ and\ \citenamefont {Preskill}}]{gkp}%
      \BibitemOpen
      \bibfield  {author} {\bibinfo {author} {\bibfnamefont {D.}~\bibnamefont
      {Gottesman}}, \bibinfo {author} {\bibfnamefont {A.}~\bibnamefont {Kitaev}},\
      and\ \bibinfo {author} {\bibfnamefont {J.}~\bibnamefont {Preskill}},\ }\href
      {https://doi.org/10.1103/PhysRevA.64.012310} {\bibfield  {journal} {\bibinfo
      {journal} {Phys. Rev. A}\ }\textbf {\bibinfo {volume} {64}},\ \bibinfo
      {pages} {012310} (\bibinfo {year} {2001})}\BibitemShut {NoStop}%
    \bibitem [{\citenamefont {Breuckmann}\ and\ \citenamefont
      {Eberhardt}(2021)}]{qLDPC}%
      \BibitemOpen
      \bibfield  {author} {\bibinfo {author} {\bibfnamefont {N.~P.}\ \bibnamefont
      {Breuckmann}}\ and\ \bibinfo {author} {\bibfnamefont {J.~N.}\ \bibnamefont
      {Eberhardt}},\ }\href {https://doi.org/10.1103/PRXQuantum.2.040101}
      {\bibfield  {journal} {\bibinfo  {journal} {PRX Quantum}\ }\textbf {\bibinfo
      {volume} {2}},\ \bibinfo {pages} {040101} (\bibinfo {year}
      {2021})}\BibitemShut {NoStop}%
    \bibitem [{\citenamefont {Eastin}\ and\ \citenamefont
      {Knill}(2009)}]{EastinKnill}%
      \BibitemOpen
      \bibfield  {author} {\bibinfo {author} {\bibfnamefont {B.}~\bibnamefont
      {Eastin}}\ and\ \bibinfo {author} {\bibfnamefont {E.}~\bibnamefont {Knill}},\
      }\href {https://doi.org/10.1103/PhysRevLett.102.110502} {\bibfield  {journal}
      {\bibinfo  {journal} {Phys. Rev. Lett.}\ }\textbf {\bibinfo {volume} {102}},\
      \bibinfo {pages} {110502} (\bibinfo {year} {2009})}\BibitemShut {NoStop}%
    \bibitem [{\citenamefont {Zeng}\ \emph {et~al.}(2007)\citenamefont {Zeng},
      \citenamefont {Cross},\ and\ \citenamefont
      {Chuang}}]{zeng2007transversalityversusuniversalityadditive}%
      \BibitemOpen
      \bibfield  {author} {\bibinfo {author} {\bibfnamefont {B.}~\bibnamefont
      {Zeng}}, \bibinfo {author} {\bibfnamefont {A.}~\bibnamefont {Cross}},\ and\
      \bibinfo {author} {\bibfnamefont {I.~L.}\ \bibnamefont {Chuang}},\
      }\href@noop {} {\bibinfo {title} {Transversality versus universality for
      additive quantum codes}} (\bibinfo {year} {2007}),\ \Eprint
      {https://arxiv.org/abs/arXiv:0706.1382} {arXiv:0706.1382} \BibitemShut
      {NoStop}%
    \bibitem [{\citenamefont {Bravyi}\ and\ \citenamefont
      {K\"onig}(2013)}]{BravyiKoenig}%
      \BibitemOpen
      \bibfield  {author} {\bibinfo {author} {\bibfnamefont {S.}~\bibnamefont
      {Bravyi}}\ and\ \bibinfo {author} {\bibfnamefont {R.}~\bibnamefont
      {K\"onig}},\ }\href {https://doi.org/10.1103/PhysRevLett.110.170503}
      {\bibfield  {journal} {\bibinfo  {journal} {Phys. Rev. Lett.}\ }\textbf
      {\bibinfo {volume} {110}},\ \bibinfo {pages} {170503} (\bibinfo {year}
      {2013})}\BibitemShut {NoStop}%
    \bibitem [{\citenamefont {Bravyi}\ and\ \citenamefont
      {Kitaev}(2005)}]{bravyi2005universal}%
      \BibitemOpen
      \bibfield  {author} {\bibinfo {author} {\bibfnamefont {S.}~\bibnamefont
      {Bravyi}}\ and\ \bibinfo {author} {\bibfnamefont {A.}~\bibnamefont
      {Kitaev}},\ }\href {https://doi.org/10.1103/PhysRevA.71.022316} {\bibfield
      {journal} {\bibinfo  {journal} {Phys. Rev. A}\ }\textbf {\bibinfo {volume}
      {71}},\ \bibinfo {pages} {022316} (\bibinfo {year} {2005})}\BibitemShut
      {NoStop}%
    \bibitem [{\citenamefont {Feynman}(1982)}]{feynman1982simulating}%
      \BibitemOpen
      \bibfield  {author} {\bibinfo {author} {\bibfnamefont {R.~P.}\ \bibnamefont
      {Feynman}},\ }\href {https://doi.org/https://doi.org/10.1007/BF02650179}
      {\bibfield  {journal} {\bibinfo  {journal} {Int. J. Theor. Phys.}\ }\textbf
      {\bibinfo {volume} {21}},\ \bibinfo {pages} {467} (\bibinfo {year}
      {1982})}\BibitemShut {NoStop}%
    \bibitem [{\citenamefont {Bravyi}\ \emph {et~al.}(2016)\citenamefont {Bravyi},
      \citenamefont {Smith},\ and\ \citenamefont {Smolin}}]{Bravyi_PBC2016}%
      \BibitemOpen
      \bibfield  {author} {\bibinfo {author} {\bibfnamefont {S.}~\bibnamefont
      {Bravyi}}, \bibinfo {author} {\bibfnamefont {G.}~\bibnamefont {Smith}},\ and\
      \bibinfo {author} {\bibfnamefont {J.~A.}\ \bibnamefont {Smolin}},\ }\href
      {http://dx.doi.org/10.1103/PhysRevX.6.021043} {\bibfield  {journal} {\bibinfo
       {journal} {Phys. Rev. X}\ }\textbf {\bibinfo {volume} {6}},\ \bibinfo
      {pages} {021043} (\bibinfo {year} {2016})}\BibitemShut {NoStop}%
    \bibitem [{\citenamefont {Bravyi}\ and\ \citenamefont
      {Gosset}(2016)}]{bravyi2016improved}%
      \BibitemOpen
      \bibfield  {author} {\bibinfo {author} {\bibfnamefont {S.}~\bibnamefont
      {Bravyi}}\ and\ \bibinfo {author} {\bibfnamefont {D.}~\bibnamefont
      {Gosset}},\ }\href {https://doi.org/10.1103/PhysRevLett.116.250501}
      {\bibfield  {journal} {\bibinfo  {journal} {Phys. Rev. Lett.}\ }\textbf
      {\bibinfo {volume} {116}},\ \bibinfo {pages} {250501} (\bibinfo {year}
      {2016})}\BibitemShut {NoStop}%
    \bibitem [{\citenamefont {Bravyi}\ \emph {et~al.}(2019)\citenamefont {Bravyi},
      \citenamefont {Browne}, \citenamefont {Calpin}, \citenamefont {Campbell},
      \citenamefont {Gosset},\ and\ \citenamefont {Howard}}]{magic3}%
      \BibitemOpen
      \bibfield  {author} {\bibinfo {author} {\bibfnamefont {S.}~\bibnamefont
      {Bravyi}}, \bibinfo {author} {\bibfnamefont {D.}~\bibnamefont {Browne}},
      \bibinfo {author} {\bibfnamefont {P.}~\bibnamefont {Calpin}}, \bibinfo
      {author} {\bibfnamefont {E.}~\bibnamefont {Campbell}}, \bibinfo {author}
      {\bibfnamefont {D.}~\bibnamefont {Gosset}},\ and\ \bibinfo {author}
      {\bibfnamefont {M.}~\bibnamefont {Howard}},\ }\href
      {https://doi.org/10.22331/q-2019-09-02-181} {\bibfield  {journal} {\bibinfo
      {journal} {{Quantum}}\ }\textbf {\bibinfo {volume} {3}},\ \bibinfo {pages}
      {181} (\bibinfo {year} {2019})}\BibitemShut {NoStop}%
    \bibitem [{\citenamefont {Schuch}\ \emph {et~al.}(2008)\citenamefont {Schuch},
      \citenamefont {Wolf}, \citenamefont {Verstraete},\ and\ \citenamefont
      {Cirac}}]{schuch2008entropy}%
      \BibitemOpen
      \bibfield  {author} {\bibinfo {author} {\bibfnamefont {N.}~\bibnamefont
      {Schuch}}, \bibinfo {author} {\bibfnamefont {M.~M.}\ \bibnamefont {Wolf}},
      \bibinfo {author} {\bibfnamefont {F.}~\bibnamefont {Verstraete}},\ and\
      \bibinfo {author} {\bibfnamefont {J.~I.}\ \bibnamefont {Cirac}},\ }\href
      {https://doi.org/10.1103/PhysRevLett.100.030504} {\bibfield  {journal}
      {\bibinfo  {journal} {Phys. Rev. Lett.}\ }\textbf {\bibinfo {volume} {100}},\
      \bibinfo {pages} {030504} (\bibinfo {year} {2008})}\BibitemShut {NoStop}%
    \bibitem [{\citenamefont {Vidal}(2003)}]{vidal2003efficient}%
      \BibitemOpen
      \bibfield  {author} {\bibinfo {author} {\bibfnamefont {G.}~\bibnamefont
      {Vidal}},\ }\href {https://doi.org/10.1103/PhysRevLett.91.147902} {\bibfield
      {journal} {\bibinfo  {journal} {Phys. Rev. Lett.}\ }\textbf {\bibinfo
      {volume} {91}},\ \bibinfo {pages} {147902} (\bibinfo {year}
      {2003})}\BibitemShut {NoStop}%
    \bibitem [{\citenamefont {Vidal}(2004)}]{vidal2004efficient}%
      \BibitemOpen
      \bibfield  {author} {\bibinfo {author} {\bibfnamefont {G.}~\bibnamefont
      {Vidal}},\ }\href {https://doi.org/10.1103/PhysRevLett.93.040502} {\bibfield
      {journal} {\bibinfo  {journal} {Phys. Rev. Lett.}\ }\textbf {\bibinfo
      {volume} {93}},\ \bibinfo {pages} {040502} (\bibinfo {year}
      {2004})}\BibitemShut {NoStop}%
    \bibitem [{\citenamefont {Markov}\ and\ \citenamefont
      {Shi}(2008)}]{Markov_2008}%
      \BibitemOpen
      \bibfield  {author} {\bibinfo {author} {\bibfnamefont {I.~L.}\ \bibnamefont
      {Markov}}\ and\ \bibinfo {author} {\bibfnamefont {Y.}~\bibnamefont {Shi}},\
      }\href {https://doi.org/10.1137/050644756} {\bibfield  {journal} {\bibinfo
      {journal} {SIAM Journal on Computing}\ }\textbf {\bibinfo {volume} {38}},\
      \bibinfo {pages} {963–981} (\bibinfo {year} {2008})}\BibitemShut {NoStop}%
    \bibitem [{\citenamefont {Aaronson}\ and\ \citenamefont
      {Arkhipov}(2011)}]{boson_sampling}%
      \BibitemOpen
      \bibfield  {author} {\bibinfo {author} {\bibfnamefont {S.}~\bibnamefont
      {Aaronson}}\ and\ \bibinfo {author} {\bibfnamefont {A.}~\bibnamefont
      {Arkhipov}},\ }in\ \href {https://doi.org/10.1145/1993636.1993682} {\emph
      {\bibinfo {booktitle} {Proceedings of the Forty-Third Annual ACM Symposium on
      Theory of Computing}}},\ \bibinfo {series and number} {STOC '11}\ (\bibinfo
      {year} {2011})\ p.\ \bibinfo {pages} {333–342}\BibitemShut {NoStop}%
    \bibitem [{\citenamefont {Oszmaniec}\ \emph {et~al.}(2022)\citenamefont
      {Oszmaniec}, \citenamefont {Dangniam}, \citenamefont {Morales},\ and\
      \citenamefont {Zimbor\'as}}]{ferm_sampling}%
      \BibitemOpen
      \bibfield  {author} {\bibinfo {author} {\bibfnamefont {M.}~\bibnamefont
      {Oszmaniec}}, \bibinfo {author} {\bibfnamefont {N.}~\bibnamefont {Dangniam}},
      \bibinfo {author} {\bibfnamefont {M.~E.}\ \bibnamefont {Morales}},\ and\
      \bibinfo {author} {\bibfnamefont {Z.}~\bibnamefont {Zimbor\'as}},\ }\href
      {https://doi.org/10.1103/PRXQuantum.3.020328} {\bibfield  {journal} {\bibinfo
       {journal} {PRX Quantum}\ }\textbf {\bibinfo {volume} {3}},\ \bibinfo {pages}
      {020328} (\bibinfo {year} {2022})}\BibitemShut {NoStop}%
    \bibitem [{\citenamefont {Dias}\ and\ \citenamefont
      {K\"onig}(2024{\natexlab{a}})}]{DiasFermionic}%
      \BibitemOpen
      \bibfield  {author} {\bibinfo {author} {\bibfnamefont {B.}~\bibnamefont
      {Dias}}\ and\ \bibinfo {author} {\bibfnamefont {R.}~\bibnamefont {K\"onig}},\
      }\href {https://doi.org/10.22331/q-2024-05-21-1350} {\bibfield  {journal}
      {\bibinfo  {journal} {{Quantum}}\ }\textbf {\bibinfo {volume} {8}},\ \bibinfo
      {pages} {1350} (\bibinfo {year} {2024}{\natexlab{a}})}\BibitemShut {NoStop}%
    \bibitem [{\citenamefont {Dias}\ and\ \citenamefont
      {K\"onig}(2024{\natexlab{b}})}]{DiasBosonic}%
      \BibitemOpen
      \bibfield  {author} {\bibinfo {author} {\bibfnamefont {B.}~\bibnamefont
      {Dias}}\ and\ \bibinfo {author} {\bibfnamefont {R.}~\bibnamefont {K\"onig}},\
      }\href {https://doi.org/10.1103/PhysRevA.110.042402} {\bibfield  {journal}
      {\bibinfo  {journal} {Phys. Rev. A}\ }\textbf {\bibinfo {volume} {110}},\
      \bibinfo {pages} {042402} (\bibinfo {year} {2024}{\natexlab{b}})}\BibitemShut
      {NoStop}%
    \bibitem [{\citenamefont {Zhou}\ \emph {et~al.}(2020)\citenamefont {Zhou},
      \citenamefont {Yang}, \citenamefont {Hamma},\ and\ \citenamefont
      {Chamon}}]{single_T_gate}%
      \BibitemOpen
      \bibfield  {author} {\bibinfo {author} {\bibfnamefont {S.}~\bibnamefont
      {Zhou}}, \bibinfo {author} {\bibfnamefont {Z.-C.}\ \bibnamefont {Yang}},
      \bibinfo {author} {\bibfnamefont {A.}~\bibnamefont {Hamma}},\ and\ \bibinfo
      {author} {\bibfnamefont {C.}~\bibnamefont {Chamon}},\ }\href
      {https://doi.org/10.21468/SciPostPhys.9.6.087} {\bibfield  {journal}
      {\bibinfo  {journal} {SciPost Phys.}\ }\textbf {\bibinfo {volume} {9}},\
      \bibinfo {pages} {087} (\bibinfo {year} {2020})}\BibitemShut {NoStop}%
    \bibitem [{\citenamefont {Leone}\ \emph {et~al.}(2021)\citenamefont {Leone},
      \citenamefont {Oliviero}, \citenamefont {Zhou},\ and\ \citenamefont
      {Hamma}}]{Leone2021quantumchaosis}%
      \BibitemOpen
      \bibfield  {author} {\bibinfo {author} {\bibfnamefont {L.}~\bibnamefont
      {Leone}}, \bibinfo {author} {\bibfnamefont {S.~F.~E.}\ \bibnamefont
      {Oliviero}}, \bibinfo {author} {\bibfnamefont {Y.}~\bibnamefont {Zhou}},\
      and\ \bibinfo {author} {\bibfnamefont {A.}~\bibnamefont {Hamma}},\ }\href
      {https://doi.org/10.22331/q-2021-05-04-453} {\bibfield  {journal} {\bibinfo
      {journal} {{Quantum}}\ }\textbf {\bibinfo {volume} {5}},\ \bibinfo {pages}
      {453} (\bibinfo {year} {2021})}\BibitemShut {NoStop}%
    \bibitem [{\citenamefont {Cotler}\ \emph {et~al.}(2023)\citenamefont {Cotler},
      \citenamefont {Mark}, \citenamefont {Huang}, \citenamefont {Hern\'andez},
      \citenamefont {Choi}, \citenamefont {Shaw}, \citenamefont {Endres},\ and\
      \citenamefont {Choi}}]{dt1}%
      \BibitemOpen
      \bibfield  {author} {\bibinfo {author} {\bibfnamefont {J.~S.}\ \bibnamefont
      {Cotler}}, \bibinfo {author} {\bibfnamefont {D.~K.}\ \bibnamefont {Mark}},
      \bibinfo {author} {\bibfnamefont {H.-Y.}\ \bibnamefont {Huang}}, \bibinfo
      {author} {\bibfnamefont {F.}~\bibnamefont {Hern\'andez}}, \bibinfo {author}
      {\bibfnamefont {J.}~\bibnamefont {Choi}}, \bibinfo {author} {\bibfnamefont
      {A.~L.}\ \bibnamefont {Shaw}}, \bibinfo {author} {\bibfnamefont
      {M.}~\bibnamefont {Endres}},\ and\ \bibinfo {author} {\bibfnamefont
      {S.}~\bibnamefont {Choi}},\ }\href
      {https://doi.org/10.1103/PRXQuantum.4.010311} {\bibfield  {journal} {\bibinfo
       {journal} {PRX Quantum}\ }\textbf {\bibinfo {volume} {4}},\ \bibinfo {pages}
      {010311} (\bibinfo {year} {2023})}\BibitemShut {NoStop}%
    \bibitem [{\citenamefont {Choi~\textit{et al.}}(2023)}]{dt2}%
      \BibitemOpen
      \bibfield  {author} {\bibinfo {author} {\bibfnamefont {J.}~\bibnamefont
      {Choi~\textit{et al.}}},\ }\href {https://doi.org/10.1038/s41586-022-05442-1}
      {\bibfield  {journal} {\bibinfo  {journal} {Nature}\ }\textbf {\bibinfo
      {volume} {613}},\ \bibinfo {pages} {468–473} (\bibinfo {year}
      {2023})}\BibitemShut {NoStop}%
    \bibitem [{\citenamefont {Ho}\ and\ \citenamefont {Choi}(2022)}]{dt9}%
      \BibitemOpen
      \bibfield  {author} {\bibinfo {author} {\bibfnamefont {W.~W.}\ \bibnamefont
      {Ho}}\ and\ \bibinfo {author} {\bibfnamefont {S.}~\bibnamefont {Choi}},\
      }\href {https://doi.org/10.1103/PhysRevLett.128.060601} {\bibfield  {journal}
      {\bibinfo  {journal} {Phys. Rev. Lett.}\ }\textbf {\bibinfo {volume} {128}},\
      \bibinfo {pages} {060601} (\bibinfo {year} {2022})}\BibitemShut {NoStop}%
    \bibitem [{\citenamefont {Claeys}\ and\ \citenamefont {Lamacraft}(2022)}]{dt3}%
      \BibitemOpen
      \bibfield  {author} {\bibinfo {author} {\bibfnamefont {P.~W.}\ \bibnamefont
      {Claeys}}\ and\ \bibinfo {author} {\bibfnamefont {A.}~\bibnamefont
      {Lamacraft}},\ }\href {https://doi.org/10.22331/q-2022-06-15-738} {\bibfield
      {journal} {\bibinfo  {journal} {{Quantum}}\ }\textbf {\bibinfo {volume}
      {6}},\ \bibinfo {pages} {738} (\bibinfo {year} {2022})}\BibitemShut {NoStop}%
    \bibitem [{\citenamefont {Ippoliti}\ and\ \citenamefont {Ho}(2023)}]{dt4}%
      \BibitemOpen
      \bibfield  {author} {\bibinfo {author} {\bibfnamefont {M.}~\bibnamefont
      {Ippoliti}}\ and\ \bibinfo {author} {\bibfnamefont {W.~W.}\ \bibnamefont
      {Ho}},\ }\href {https://doi.org/10.1103/PRXQuantum.4.030322} {\bibfield
      {journal} {\bibinfo  {journal} {PRX Quantum}\ }\textbf {\bibinfo {volume}
      {4}},\ \bibinfo {pages} {030322} (\bibinfo {year} {2023})}\BibitemShut
      {NoStop}%
    \bibitem [{\citenamefont {Liu}\ \emph {et~al.}(2024)\citenamefont {Liu},
      \citenamefont {Huang},\ and\ \citenamefont {Ho}}]{dt5}%
      \BibitemOpen
      \bibfield  {author} {\bibinfo {author} {\bibfnamefont {C.}~\bibnamefont
      {Liu}}, \bibinfo {author} {\bibfnamefont {Q.~C.}\ \bibnamefont {Huang}},\
      and\ \bibinfo {author} {\bibfnamefont {W.~W.}\ \bibnamefont {Ho}},\ }\href
      {https://doi.org/10.1103/PhysRevLett.133.260401} {\bibfield  {journal}
      {\bibinfo  {journal} {Phys. Rev. Lett.}\ }\textbf {\bibinfo {volume} {133}},\
      \bibinfo {pages} {260401} (\bibinfo {year} {2024})}\BibitemShut {NoStop}%
    \bibitem [{\citenamefont {Vairogs}\ and\ \citenamefont {Yan}(2025)}]{dt6}%
      \BibitemOpen
      \bibfield  {author} {\bibinfo {author} {\bibfnamefont {C.}~\bibnamefont
      {Vairogs}}\ and\ \bibinfo {author} {\bibfnamefont {B.}~\bibnamefont {Yan}},\
      }\href {https://doi.org/10.1103/3ttm-vhdt} {\bibfield  {journal} {\bibinfo
      {journal} {Phys. Rev. Res.}\ }\textbf {\bibinfo {volume} {7}},\ \bibinfo
      {pages} {L022069} (\bibinfo {year} {2025})}\BibitemShut {NoStop}%
    \bibitem [{\citenamefont {Fava}\ \emph {et~al.}(2025)\citenamefont {Fava},
      \citenamefont {Kurchan},\ and\ \citenamefont
      {Pappalardi}}]{designs_via_free_probability}%
      \BibitemOpen
      \bibfield  {author} {\bibinfo {author} {\bibfnamefont {M.}~\bibnamefont
      {Fava}}, \bibinfo {author} {\bibfnamefont {J.}~\bibnamefont {Kurchan}},\ and\
      \bibinfo {author} {\bibfnamefont {S.}~\bibnamefont {Pappalardi}},\ }\href
      {https://doi.org/10.1103/PhysRevX.15.011031} {\bibfield  {journal} {\bibinfo
      {journal} {Phys. Rev. X}\ }\textbf {\bibinfo {volume} {15}},\ \bibinfo
      {pages} {011031} (\bibinfo {year} {2025})}\BibitemShut {NoStop}%
    \bibitem [{\citenamefont {Liu}\ \emph {et~al.}(2025)\citenamefont {Liu},
      \citenamefont {Ippoliti},\ and\ \citenamefont {Ho}}]{dt10}%
      \BibitemOpen
      \bibfield  {author} {\bibinfo {author} {\bibfnamefont {C.}~\bibnamefont
      {Liu}}, \bibinfo {author} {\bibfnamefont {M.}~\bibnamefont {Ippoliti}},\ and\
      \bibinfo {author} {\bibfnamefont {W.~W.}\ \bibnamefont {Ho}},\ }\href@noop {}
      {\bibinfo {title} {Coherence-induced deep thermalization transition in random
      permutation quantum dynamics}} (\bibinfo {year} {2025}),\ \Eprint
      {https://arxiv.org/abs/2510.18369} {arXiv:2510.18369} \BibitemShut {NoStop}%
    \bibitem [{\citenamefont {Bejan}\ \emph {et~al.}(2025)\citenamefont {Bejan},
      \citenamefont {B\'eri},\ and\ \citenamefont {McGinley}}]{dt7}%
      \BibitemOpen
      \bibfield  {author} {\bibinfo {author} {\bibfnamefont {M.}~\bibnamefont
      {Bejan}}, \bibinfo {author} {\bibfnamefont {B.}~\bibnamefont {B\'eri}},\ and\
      \bibinfo {author} {\bibfnamefont {M.}~\bibnamefont {McGinley}},\ }\href
      {https://doi.org/10.1103/v8kp-39ry} {\bibfield  {journal} {\bibinfo
      {journal} {Phys. Rev. Lett.}\ }\textbf {\bibinfo {volume} {135}},\ \bibinfo
      {pages} {020401} (\bibinfo {year} {2025})}\BibitemShut {NoStop}%
    \bibitem [{\citenamefont {Lóio}\ \emph {et~al.}(2025)\citenamefont {Lóio},
      \citenamefont {Lami}, \citenamefont {Leone}, \citenamefont {McGinley},
      \citenamefont {Turkeshi},\ and\ \citenamefont {Nardis}}]{dt8}%
      \BibitemOpen
      \bibfield  {author} {\bibinfo {author} {\bibfnamefont {H.}~\bibnamefont
      {Lóio}}, \bibinfo {author} {\bibfnamefont {G.}~\bibnamefont {Lami}},
      \bibinfo {author} {\bibfnamefont {L.}~\bibnamefont {Leone}}, \bibinfo
      {author} {\bibfnamefont {M.}~\bibnamefont {McGinley}}, \bibinfo {author}
      {\bibfnamefont {X.}~\bibnamefont {Turkeshi}},\ and\ \bibinfo {author}
      {\bibfnamefont {J.~D.}\ \bibnamefont {Nardis}},\ }\href@noop {} {\bibinfo
      {title} {Quantum state designs via magic teleportation}} (\bibinfo {year}
      {2025}),\ \Eprint {https://arxiv.org/abs/2510.13950} {arXiv:2510.13950}
      \BibitemShut {NoStop}%
    \bibitem [{\citenamefont {Webb}(2016)}]{cliff3design}%
      \BibitemOpen
      \bibfield  {author} {\bibinfo {author} {\bibfnamefont {Z.}~\bibnamefont
      {Webb}},\ }\href@noop {} {\bibinfo {title} {The clifford group forms a
      unitary 3-design}} (\bibinfo {year} {2016}),\ \Eprint
      {https://arxiv.org/abs/1510.02769} {arXiv:1510.02769} \BibitemShut {NoStop}%
    \bibitem [{\citenamefont {Zhu}\ \emph {et~al.}(2016)\citenamefont {Zhu},
      \citenamefont {Kueng}, \citenamefont {Grassl},\ and\ \citenamefont
      {Gross}}]{cliffordgroupfailsgracefully}%
      \BibitemOpen
      \bibfield  {author} {\bibinfo {author} {\bibfnamefont {H.}~\bibnamefont
      {Zhu}}, \bibinfo {author} {\bibfnamefont {R.}~\bibnamefont {Kueng}}, \bibinfo
      {author} {\bibfnamefont {M.}~\bibnamefont {Grassl}},\ and\ \bibinfo {author}
      {\bibfnamefont {D.}~\bibnamefont {Gross}},\ }\href@noop {} {\bibinfo {title}
      {The clifford group fails gracefully to be a unitary 4-design}} (\bibinfo
      {year} {2016}),\ \Eprint {https://arxiv.org/abs/1609.08172}
      {arXiv:1609.08172} \BibitemShut {NoStop}%
    \bibitem [{\citenamefont {Liu}\ and\ \citenamefont {Winter}(2022)}]{magicMB1}%
      \BibitemOpen
      \bibfield  {author} {\bibinfo {author} {\bibfnamefont {Z.-W.}\ \bibnamefont
      {Liu}}\ and\ \bibinfo {author} {\bibfnamefont {A.}~\bibnamefont {Winter}},\
      }\href {https://doi.org/10.1103/PRXQuantum.3.020333} {\bibfield  {journal}
      {\bibinfo  {journal} {PRX Quantum}\ }\textbf {\bibinfo {volume} {3}},\
      \bibinfo {pages} {020333} (\bibinfo {year} {2022})}\BibitemShut {NoStop}%
    \bibitem [{\citenamefont {Ellison}\ \emph {et~al.}(2021)\citenamefont
      {Ellison}, \citenamefont {Kato}, \citenamefont {Liu},\ and\ \citenamefont
      {Hsieh}}]{magicMB8}%
      \BibitemOpen
      \bibfield  {author} {\bibinfo {author} {\bibfnamefont {T.~D.}\ \bibnamefont
      {Ellison}}, \bibinfo {author} {\bibfnamefont {K.}~\bibnamefont {Kato}},
      \bibinfo {author} {\bibfnamefont {Z.-W.}\ \bibnamefont {Liu}},\ and\ \bibinfo
      {author} {\bibfnamefont {T.~H.}\ \bibnamefont {Hsieh}},\ }\href
      {https://doi.org/10.22331/q-2021-12-28-612} {\bibfield  {journal} {\bibinfo
      {journal} {{Quantum}}\ }\textbf {\bibinfo {volume} {5}},\ \bibinfo {pages}
      {612} (\bibinfo {year} {2021})}\BibitemShut {NoStop}%
    \bibitem [{\citenamefont {Wei}\ and\ \citenamefont
      {Liu}(2025)}]{wei2025longrangenonstabilizernessquantumcodes}%
      \BibitemOpen
      \bibfield  {author} {\bibinfo {author} {\bibfnamefont {F.}~\bibnamefont
      {Wei}}\ and\ \bibinfo {author} {\bibfnamefont {Z.-W.}\ \bibnamefont {Liu}},\
      }\href@noop {} {\bibinfo {title} {Long-range nonstabilizerness from quantum
      codes, orders, and correlations}} (\bibinfo {year} {2025}),\ \Eprint
      {https://arxiv.org/abs/2503.04566} {arXiv:2503.04566} \BibitemShut {NoStop}%
    \bibitem [{\citenamefont {Tarabunga}\ \emph {et~al.}(2023)\citenamefont
      {Tarabunga}, \citenamefont {Tirrito}, \citenamefont {Chanda},\ and\
      \citenamefont {Dalmonte}}]{magicMB2}%
      \BibitemOpen
      \bibfield  {author} {\bibinfo {author} {\bibfnamefont {P.~S.}\ \bibnamefont
      {Tarabunga}}, \bibinfo {author} {\bibfnamefont {E.}~\bibnamefont {Tirrito}},
      \bibinfo {author} {\bibfnamefont {T.}~\bibnamefont {Chanda}},\ and\ \bibinfo
      {author} {\bibfnamefont {M.}~\bibnamefont {Dalmonte}},\ }\href
      {https://doi.org/10.1103/PRXQuantum.4.040317} {\bibfield  {journal} {\bibinfo
       {journal} {PRX Quantum}\ }\textbf {\bibinfo {volume} {4}},\ \bibinfo {pages}
      {040317} (\bibinfo {year} {2023})}\BibitemShut {NoStop}%
    \bibitem [{\citenamefont {Tarabunga}\ and\ \citenamefont
      {Castelnovo}(2024)}]{magicMB4}%
      \BibitemOpen
      \bibfield  {author} {\bibinfo {author} {\bibfnamefont {P.~S.}\ \bibnamefont
      {Tarabunga}}\ and\ \bibinfo {author} {\bibfnamefont {C.}~\bibnamefont
      {Castelnovo}},\ }\href {https://doi.org/10.22331/q-2024-05-14-1347}
      {\bibfield  {journal} {\bibinfo  {journal} {{Quantum}}\ }\textbf {\bibinfo
      {volume} {8}},\ \bibinfo {pages} {1347} (\bibinfo {year} {2024})}\BibitemShut
      {NoStop}%
    \bibitem [{\citenamefont {Collura}\ \emph {et~al.}(2025)\citenamefont
      {Collura}, \citenamefont {Nardis}, \citenamefont {Alba},\ and\ \citenamefont
      {Lami}}]{magicMB6}%
      \BibitemOpen
      \bibfield  {author} {\bibinfo {author} {\bibfnamefont {M.}~\bibnamefont
      {Collura}}, \bibinfo {author} {\bibfnamefont {J.~D.}\ \bibnamefont {Nardis}},
      \bibinfo {author} {\bibfnamefont {V.}~\bibnamefont {Alba}},\ and\ \bibinfo
      {author} {\bibfnamefont {G.}~\bibnamefont {Lami}},\ }\href@noop {} {\bibinfo
      {title} {The non-stabilizerness of fermionic gaussian states}} (\bibinfo
      {year} {2025}),\ \Eprint {https://arxiv.org/abs/2412.05367}
      {arXiv:2412.05367} \BibitemShut {NoStop}%
    \bibitem [{\citenamefont {Korbany}\ \emph {et~al.}(2025)\citenamefont
      {Korbany}, \citenamefont {Gullans},\ and\ \citenamefont {Piroli}}]{LRMandPM}%
      \BibitemOpen
      \bibfield  {author} {\bibinfo {author} {\bibfnamefont {D.~A.}\ \bibnamefont
      {Korbany}}, \bibinfo {author} {\bibfnamefont {M.~J.}\ \bibnamefont
      {Gullans}},\ and\ \bibinfo {author} {\bibfnamefont {L.}~\bibnamefont
      {Piroli}},\ }\href {https://doi.org/10.1103/1hlj-h6t9} {\bibfield  {journal}
      {\bibinfo  {journal} {Phys. Rev. Lett.}\ }\textbf {\bibinfo {volume} {135}},\
      \bibinfo {pages} {160404} (\bibinfo {year} {2025})}\BibitemShut {NoStop}%
    \bibitem [{\citenamefont {Santra}\ \emph
      {et~al.}(2025{\natexlab{a}})\citenamefont {Santra}, \citenamefont
      {Mildenberger}, \citenamefont {Ballini}, \citenamefont {Bottarelli},
      \citenamefont {Wauters},\ and\ \citenamefont
      {Hauke}}]{santra2025quantumresourcesnonabelianlattice}%
      \BibitemOpen
      \bibfield  {author} {\bibinfo {author} {\bibfnamefont {G.~C.}\ \bibnamefont
      {Santra}}, \bibinfo {author} {\bibfnamefont {J.}~\bibnamefont
      {Mildenberger}}, \bibinfo {author} {\bibfnamefont {E.}~\bibnamefont
      {Ballini}}, \bibinfo {author} {\bibfnamefont {A.}~\bibnamefont {Bottarelli}},
      \bibinfo {author} {\bibfnamefont {M.~M.}\ \bibnamefont {Wauters}},\ and\
      \bibinfo {author} {\bibfnamefont {P.}~\bibnamefont {Hauke}},\ }\href@noop {}
      {\bibinfo {title} {Quantum resources in non-abelian lattice gauge theories:
      Nonstabilizerness, multipartite entanglement, and fermionic non-gaussianity}}
      (\bibinfo {year} {2025}{\natexlab{a}}),\ \Eprint
      {https://arxiv.org/abs/2510.07385} {arXiv:2510.07385} \BibitemShut {NoStop}%
    \bibitem [{\citenamefont {Falcão}\ \emph {et~al.}(2025)\citenamefont
      {Falcão}, \citenamefont {Sierant}, \citenamefont {Zakrzewski},\ and\
      \citenamefont {Tirrito}}]{falcão2025magicdynamicsmanybodylocalized}%
      \BibitemOpen
      \bibfield  {author} {\bibinfo {author} {\bibfnamefont {P.~R.~N.}\
      \bibnamefont {Falcão}}, \bibinfo {author} {\bibfnamefont {P.}~\bibnamefont
      {Sierant}}, \bibinfo {author} {\bibfnamefont {J.}~\bibnamefont
      {Zakrzewski}},\ and\ \bibinfo {author} {\bibfnamefont {E.}~\bibnamefont
      {Tirrito}},\ }\href@noop {} {\bibinfo {title} {Magic dynamics in many-body
      localized systems}} (\bibinfo {year} {2025}),\ \Eprint
      {https://arxiv.org/abs/2503.07468} {arXiv:2503.07468} \BibitemShut {NoStop}%
    \bibitem [{\citenamefont {Oliviero}\ \emph {et~al.}(2022)\citenamefont
      {Oliviero}, \citenamefont {Leone},\ and\ \citenamefont
      {Hamma}}]{PhysRevA.106.042426}%
      \BibitemOpen
      \bibfield  {author} {\bibinfo {author} {\bibfnamefont {S.~F.~E.}\
      \bibnamefont {Oliviero}}, \bibinfo {author} {\bibfnamefont {L.}~\bibnamefont
      {Leone}},\ and\ \bibinfo {author} {\bibfnamefont {A.}~\bibnamefont {Hamma}},\
      }\href {https://doi.org/10.1103/PhysRevA.106.042426} {\bibfield  {journal}
      {\bibinfo  {journal} {Phys. Rev. A}\ }\textbf {\bibinfo {volume} {106}},\
      \bibinfo {pages} {042426} (\bibinfo {year} {2022})}\BibitemShut {NoStop}%
    \bibitem [{\citenamefont {White}\ \emph {et~al.}(2021)\citenamefont {White},
      \citenamefont {Cao},\ and\ \citenamefont {Swingle}}]{magicMB3}%
      \BibitemOpen
      \bibfield  {author} {\bibinfo {author} {\bibfnamefont {C.~D.}\ \bibnamefont
      {White}}, \bibinfo {author} {\bibfnamefont {C.}~\bibnamefont {Cao}},\ and\
      \bibinfo {author} {\bibfnamefont {B.}~\bibnamefont {Swingle}},\ }\href
      {https://doi.org/10.1103/PhysRevB.103.075145} {\bibfield  {journal} {\bibinfo
       {journal} {Phys. Rev. B}\ }\textbf {\bibinfo {volume} {103}},\ \bibinfo
      {pages} {075145} (\bibinfo {year} {2021})}\BibitemShut {NoStop}%
    \bibitem [{\citenamefont {Cao}\ \emph {et~al.}(2025{\natexlab{a}})\citenamefont
      {Cao}, \citenamefont {Cheng}, \citenamefont {Hamma}, \citenamefont {Leone},
      \citenamefont {Munizzi},\ and\ \citenamefont
      {Oliviero}}]{cao2025gravitationalbackreactionmagical}%
      \BibitemOpen
      \bibfield  {author} {\bibinfo {author} {\bibfnamefont {C.}~\bibnamefont
      {Cao}}, \bibinfo {author} {\bibfnamefont {G.}~\bibnamefont {Cheng}}, \bibinfo
      {author} {\bibfnamefont {A.}~\bibnamefont {Hamma}}, \bibinfo {author}
      {\bibfnamefont {L.}~\bibnamefont {Leone}}, \bibinfo {author} {\bibfnamefont
      {W.}~\bibnamefont {Munizzi}},\ and\ \bibinfo {author} {\bibfnamefont
      {S.~F.~E.}\ \bibnamefont {Oliviero}},\ }\href@noop {} {\bibinfo {title}
      {Gravitational back-reaction is magical}} (\bibinfo {year}
      {2025}{\natexlab{a}}),\ \Eprint {https://arxiv.org/abs/2403.07056}
      {arXiv:2403.07056} \BibitemShut {NoStop}%
    \bibitem [{\citenamefont {Frau}\ \emph {et~al.}(2025)\citenamefont {Frau},
      \citenamefont {Tarabunga}, \citenamefont {Collura}, \citenamefont {Tirrito},\
      and\ \citenamefont {Dalmonte}}]{magicMB10}%
      \BibitemOpen
      \bibfield  {author} {\bibinfo {author} {\bibfnamefont {M.}~\bibnamefont
      {Frau}}, \bibinfo {author} {\bibfnamefont {P.~S.}\ \bibnamefont {Tarabunga}},
      \bibinfo {author} {\bibfnamefont {M.}~\bibnamefont {Collura}}, \bibinfo
      {author} {\bibfnamefont {E.}~\bibnamefont {Tirrito}},\ and\ \bibinfo {author}
      {\bibfnamefont {M.}~\bibnamefont {Dalmonte}},\ }\href
      {https://doi.org/10.21468/SciPostPhys.18.5.165} {\bibfield  {journal}
      {\bibinfo  {journal} {SciPost Phys.}\ }\textbf {\bibinfo {volume} {18}},\
      \bibinfo {pages} {165} (\bibinfo {year} {2025})}\BibitemShut {NoStop}%
    \bibitem [{\citenamefont {Hoshino}\ \emph {et~al.}(2025)\citenamefont
      {Hoshino}, \citenamefont {Oshikawa},\ and\ \citenamefont
      {Ashida}}]{hoshino2025stabilizerrenyientropyconformal}%
      \BibitemOpen
      \bibfield  {author} {\bibinfo {author} {\bibfnamefont {M.}~\bibnamefont
      {Hoshino}}, \bibinfo {author} {\bibfnamefont {M.}~\bibnamefont {Oshikawa}},\
      and\ \bibinfo {author} {\bibfnamefont {Y.}~\bibnamefont {Ashida}},\
      }\href@noop {} {\bibinfo {title} {Stabilizer r\'enyi entropy and conformal
      field theory}} (\bibinfo {year} {2025}),\ \Eprint
      {https://arxiv.org/abs/2503.13599} {arXiv:2503.13599} \BibitemShut {NoStop}%
    \bibitem [{\citenamefont {Smith}\ \emph {et~al.}(2025)\citenamefont {Smith},
      \citenamefont {Papi\ifmmode~\acute{c}\else \'{c}\fi{}},\ and\ \citenamefont
      {Hallam}}]{magicMB7}%
      \BibitemOpen
      \bibfield  {author} {\bibinfo {author} {\bibfnamefont {R.}~\bibnamefont
      {Smith}}, \bibinfo {author} {\bibfnamefont {Z.}~\bibnamefont
      {Papi\ifmmode~\acute{c}\else \'{c}\fi{}}},\ and\ \bibinfo {author}
      {\bibfnamefont {A.}~\bibnamefont {Hallam}},\ }\href
      {https://doi.org/10.1103/jz4d-vdhj} {\bibfield  {journal} {\bibinfo
      {journal} {Phys. Rev. B}\ }\textbf {\bibinfo {volume} {111}},\ \bibinfo
      {pages} {245148} (\bibinfo {year} {2025})}\BibitemShut {NoStop}%
    \bibitem [{\citenamefont {Hartse}\ \emph {et~al.}(2025)\citenamefont {Hartse},
      \citenamefont {Fidkowski},\ and\ \citenamefont {Mueller}}]{magicMB5}%
      \BibitemOpen
      \bibfield  {author} {\bibinfo {author} {\bibfnamefont {J.}~\bibnamefont
      {Hartse}}, \bibinfo {author} {\bibfnamefont {L.}~\bibnamefont {Fidkowski}},\
      and\ \bibinfo {author} {\bibfnamefont {N.}~\bibnamefont {Mueller}},\ }\href
      {https://doi.org/10.1103/n5hb-l5p5} {\bibfield  {journal} {\bibinfo
      {journal} {Phys. Rev. Lett.}\ }\textbf {\bibinfo {volume} {135}},\ \bibinfo
      {pages} {060402} (\bibinfo {year} {2025})}\BibitemShut {NoStop}%
    \bibitem [{\citenamefont {Sil}\ and\ \citenamefont
      {Roy}(2025)}]{sil2025quantumcomplexityconstrainedmanybody}%
      \BibitemOpen
      \bibfield  {author} {\bibinfo {author} {\bibfnamefont {A.}~\bibnamefont
      {Sil}}\ and\ \bibinfo {author} {\bibfnamefont {S.~S.}\ \bibnamefont {Roy}},\
      }\href@noop {} {\bibinfo {title} {Quantum complexity in constrained many-body
      models: Scars, fragmentation, and chaos}} (\bibinfo {year} {2025}),\ \Eprint
      {https://arxiv.org/abs/2510.16570} {arXiv:2510.16570} \BibitemShut {NoStop}%
    \bibitem [{\citenamefont {Bejan}\ \emph {et~al.}(2024)\citenamefont {Bejan},
      \citenamefont {McLauchlan},\ and\ \citenamefont {B\'eri}}]{bejan2024prxq}%
      \BibitemOpen
      \bibfield  {author} {\bibinfo {author} {\bibfnamefont {M.}~\bibnamefont
      {Bejan}}, \bibinfo {author} {\bibfnamefont {C.}~\bibnamefont {McLauchlan}},\
      and\ \bibinfo {author} {\bibfnamefont {B.}~\bibnamefont {B\'eri}},\ }\href
      {https://doi.org/10.1103/PRXQuantum.5.030332} {\bibfield  {journal} {\bibinfo
       {journal} {PRX Quantum}\ }\textbf {\bibinfo {volume} {5}},\ \bibinfo {pages}
      {030332} (\bibinfo {year} {2024})}\BibitemShut {NoStop}%
    \bibitem [{\citenamefont {Fux}\ \emph {et~al.}(2024)\citenamefont {Fux},
      \citenamefont {Tirrito}, \citenamefont {Dalmonte},\ and\ \citenamefont
      {Fazio}}]{FuxPRR}%
      \BibitemOpen
      \bibfield  {author} {\bibinfo {author} {\bibfnamefont {G.~E.}\ \bibnamefont
      {Fux}}, \bibinfo {author} {\bibfnamefont {E.}~\bibnamefont {Tirrito}},
      \bibinfo {author} {\bibfnamefont {M.}~\bibnamefont {Dalmonte}},\ and\
      \bibinfo {author} {\bibfnamefont {R.}~\bibnamefont {Fazio}},\ }\href
      {https://doi.org/10.1103/PhysRevResearch.6.L042030} {\bibfield  {journal}
      {\bibinfo  {journal} {Phys. Rev. Res.}\ }\textbf {\bibinfo {volume} {6}},\
      \bibinfo {pages} {L042030} (\bibinfo {year} {2024})}\BibitemShut {NoStop}%
    \bibitem [{\citenamefont {Tirrito}\ \emph
      {et~al.}(2025{\natexlab{a}})\citenamefont {Tirrito}, \citenamefont {Lumia},
      \citenamefont {Paviglianiti}, \citenamefont {Lami}, \citenamefont {Silva},
      \citenamefont {Turkeshi},\ and\ \citenamefont
      {Collura}}]{tirrito2025magicphasetransitionsmonitored}%
      \BibitemOpen
      \bibfield  {author} {\bibinfo {author} {\bibfnamefont {E.}~\bibnamefont
      {Tirrito}}, \bibinfo {author} {\bibfnamefont {L.}~\bibnamefont {Lumia}},
      \bibinfo {author} {\bibfnamefont {A.}~\bibnamefont {Paviglianiti}}, \bibinfo
      {author} {\bibfnamefont {G.}~\bibnamefont {Lami}}, \bibinfo {author}
      {\bibfnamefont {A.}~\bibnamefont {Silva}}, \bibinfo {author} {\bibfnamefont
      {X.}~\bibnamefont {Turkeshi}},\ and\ \bibinfo {author} {\bibfnamefont
      {M.}~\bibnamefont {Collura}},\ }\href@noop {} {\bibinfo {title} {Magic phase
      transitions in monitored gaussian fermions}} (\bibinfo {year}
      {2025}{\natexlab{a}}),\ \Eprint {https://arxiv.org/abs/2507.07179}
      {arXiv:2507.07179} \BibitemShut {NoStop}%
    \bibitem [{\citenamefont {Trigueros}\ and\ \citenamefont
      {Guzmán}(2025)}]{trigueros2025nonstabilizernesserrorresiliencenoisy}%
      \BibitemOpen
      \bibfield  {author} {\bibinfo {author} {\bibfnamefont {F.~B.}\ \bibnamefont
      {Trigueros}}\ and\ \bibinfo {author} {\bibfnamefont {J.~A.~M.}\ \bibnamefont
      {Guzmán}},\ }\href@noop {} {\bibinfo {title} {Nonstabilizerness and error
      resilience in noisy quantum circuits}} (\bibinfo {year} {2025}),\ \Eprint
      {https://arxiv.org/abs/2506.18976} {arXiv:2506.18976} \BibitemShut {NoStop}%
    \bibitem [{\citenamefont {Scocco}\ \emph {et~al.}(2025)\citenamefont {Scocco},
      \citenamefont {Mok}, \citenamefont {Aolita}, \citenamefont {Collura},\ and\
      \citenamefont {Haug}}]{scocco2025risefallnonstabilizernessrandom}%
      \BibitemOpen
      \bibfield  {author} {\bibinfo {author} {\bibfnamefont {A.}~\bibnamefont
      {Scocco}}, \bibinfo {author} {\bibfnamefont {W.-K.}\ \bibnamefont {Mok}},
      \bibinfo {author} {\bibfnamefont {L.}~\bibnamefont {Aolita}}, \bibinfo
      {author} {\bibfnamefont {M.}~\bibnamefont {Collura}},\ and\ \bibinfo {author}
      {\bibfnamefont {T.}~\bibnamefont {Haug}},\ }\href@noop {} {\bibinfo {title}
      {Rise and fall of nonstabilizerness via random measurements}} (\bibinfo
      {year} {2025}),\ \Eprint {https://arxiv.org/abs/2507.11619}
      {arXiv:2507.11619} \BibitemShut {NoStop}%
    \bibitem [{\citenamefont {Tarabunga}\ and\ \citenamefont
      {Tirrito}(2025)}]{TarabungaTirrito2025}%
      \BibitemOpen
      \bibfield  {author} {\bibinfo {author} {\bibfnamefont {P.~S.}\ \bibnamefont
      {Tarabunga}}\ and\ \bibinfo {author} {\bibfnamefont {E.}~\bibnamefont
      {Tirrito}},\ }\href {https://doi.org/10.1038/s41534-025-01104-y} {\bibfield
      {journal} {\bibinfo  {journal} {npj Quantum Information}\ }\textbf {\bibinfo
      {volume} {11}},\ \bibinfo {pages} {166} (\bibinfo {year} {2025})}\BibitemShut
      {NoStop}%
    \bibitem [{\citenamefont {Gu}\ \emph {et~al.}(2025)\citenamefont {Gu},
      \citenamefont {Oliviero},\ and\ \citenamefont {Leone}}]{GuPRXQ}%
      \BibitemOpen
      \bibfield  {author} {\bibinfo {author} {\bibfnamefont {A.}~\bibnamefont
      {Gu}}, \bibinfo {author} {\bibfnamefont {S.~F.}\ \bibnamefont {Oliviero}},\
      and\ \bibinfo {author} {\bibfnamefont {L.}~\bibnamefont {Leone}},\ }\href
      {https://doi.org/10.1103/PRXQuantum.6.020324} {\bibfield  {journal} {\bibinfo
       {journal} {PRX Quantum}\ }\textbf {\bibinfo {volume} {6}},\ \bibinfo {pages}
      {020324} (\bibinfo {year} {2025})}\BibitemShut {NoStop}%
    \bibitem [{\citenamefont {Wang}\ \emph {et~al.}(2025)\citenamefont {Wang},
      \citenamefont {Yang}, \citenamefont {Zhou},\ and\ \citenamefont
      {Chen}}]{wang2025magictransitionmonitoredfree}%
      \BibitemOpen
      \bibfield  {author} {\bibinfo {author} {\bibfnamefont {C.}~\bibnamefont
      {Wang}}, \bibinfo {author} {\bibfnamefont {Z.-C.}\ \bibnamefont {Yang}},
      \bibinfo {author} {\bibfnamefont {T.}~\bibnamefont {Zhou}},\ and\ \bibinfo
      {author} {\bibfnamefont {X.}~\bibnamefont {Chen}},\ }\href@noop {} {\bibinfo
      {title} {Magic transition in monitored free fermion dynamics}} (\bibinfo
      {year} {2025}),\ \Eprint {https://arxiv.org/abs/2507.10688}
      {arXiv:2507.10688} \BibitemShut {NoStop}%
    \bibitem [{\citenamefont {Zhang}\ and\ \citenamefont
      {Gu}(2024)}]{zhang2024quantummagicdynamicsrandom}%
      \BibitemOpen
      \bibfield  {author} {\bibinfo {author} {\bibfnamefont {Y.}~\bibnamefont
      {Zhang}}\ and\ \bibinfo {author} {\bibfnamefont {Y.}~\bibnamefont {Gu}},\
      }\href@noop {} {\bibinfo {title} {Quantum magic dynamics in random circuits}}
      (\bibinfo {year} {2024}),\ \Eprint {https://arxiv.org/abs/2410.21128}
      {arXiv:2410.21128} \BibitemShut {NoStop}%
    \bibitem [{\citenamefont {Zhang}\ \emph {et~al.}(2025)\citenamefont {Zhang},
      \citenamefont {Vijay}, \citenamefont {Gu},\ and\ \citenamefont
      {Bao}}]{zhang2025designsmagicaugmentedcliffordcircuits}%
      \BibitemOpen
      \bibfield  {author} {\bibinfo {author} {\bibfnamefont {Y.}~\bibnamefont
      {Zhang}}, \bibinfo {author} {\bibfnamefont {S.}~\bibnamefont {Vijay}},
      \bibinfo {author} {\bibfnamefont {Y.}~\bibnamefont {Gu}},\ and\ \bibinfo
      {author} {\bibfnamefont {Y.}~\bibnamefont {Bao}},\ }\href@noop {} {\bibinfo
      {title} {Designs from magic-augmented clifford circuits}} (\bibinfo {year}
      {2025}),\ \Eprint {https://arxiv.org/abs/2507.02828} {arXiv:2507.02828}
      \BibitemShut {NoStop}%
    \bibitem [{\citenamefont {Hou}\ \emph {et~al.}(2025)\citenamefont {Hou},
      \citenamefont {Cao},\ and\ \citenamefont
      {Yang}}]{hou2025stabilizerentanglementenhancesmagic}%
      \BibitemOpen
      \bibfield  {author} {\bibinfo {author} {\bibfnamefont {Z.-Y.}\ \bibnamefont
      {Hou}}, \bibinfo {author} {\bibfnamefont {C.}~\bibnamefont {Cao}},\ and\
      \bibinfo {author} {\bibfnamefont {Z.-C.}\ \bibnamefont {Yang}},\ }\href@noop
      {} {\bibinfo {title} {Stabilizer entanglement enhances magic injection}}
      (\bibinfo {year} {2025}),\ \Eprint {https://arxiv.org/abs/2503.20873}
      {arXiv:2503.20873} \BibitemShut {NoStop}%
    \bibitem [{\citenamefont {Russomanno}\ \emph {et~al.}(2025)\citenamefont
      {Russomanno}, \citenamefont {Passarelli}, \citenamefont {Rossini},\ and\
      \citenamefont {Lucignano}}]{njgn-fksh}%
      \BibitemOpen
      \bibfield  {author} {\bibinfo {author} {\bibfnamefont {A.}~\bibnamefont
      {Russomanno}}, \bibinfo {author} {\bibfnamefont {G.}~\bibnamefont
      {Passarelli}}, \bibinfo {author} {\bibfnamefont {D.}~\bibnamefont
      {Rossini}},\ and\ \bibinfo {author} {\bibfnamefont {P.}~\bibnamefont
      {Lucignano}},\ }\href {https://doi.org/10.1103/njgn-fksh} {\bibfield
      {journal} {\bibinfo  {journal} {Phys. Rev. B}\ }\textbf {\bibinfo {volume}
      {112}},\ \bibinfo {pages} {064312} (\bibinfo {year} {2025})}\BibitemShut
      {NoStop}%
    \bibitem [{\citenamefont {Passarelli}\ \emph {et~al.}(2025)\citenamefont
      {Passarelli}, \citenamefont {Lucignano}, \citenamefont {Rossini},\ and\
      \citenamefont {Russomanno}}]{Passarelli2025chaosmagicin}%
      \BibitemOpen
      \bibfield  {author} {\bibinfo {author} {\bibfnamefont {G.}~\bibnamefont
      {Passarelli}}, \bibinfo {author} {\bibfnamefont {P.}~\bibnamefont
      {Lucignano}}, \bibinfo {author} {\bibfnamefont {D.}~\bibnamefont {Rossini}},\
      and\ \bibinfo {author} {\bibfnamefont {A.}~\bibnamefont {Russomanno}},\
      }\href {https://doi.org/10.22331/q-2025-03-05-1653} {\bibfield  {journal}
      {\bibinfo  {journal} {{Quantum}}\ }\textbf {\bibinfo {volume} {9}},\ \bibinfo
      {pages} {1653} (\bibinfo {year} {2025})}\BibitemShut {NoStop}%
    \bibitem [{\citenamefont {Santra}\ \emph
      {et~al.}(2025{\natexlab{b}})\citenamefont {Santra}, \citenamefont {Windey},
      \citenamefont {Bandyopadhyay}, \citenamefont {Legramandi},\ and\
      \citenamefont {Hauke}}]{santra2025complexitytransitionschaoticquantum}%
      \BibitemOpen
      \bibfield  {author} {\bibinfo {author} {\bibfnamefont {G.~C.}\ \bibnamefont
      {Santra}}, \bibinfo {author} {\bibfnamefont {A.}~\bibnamefont {Windey}},
      \bibinfo {author} {\bibfnamefont {S.}~\bibnamefont {Bandyopadhyay}}, \bibinfo
      {author} {\bibfnamefont {A.}~\bibnamefont {Legramandi}},\ and\ \bibinfo
      {author} {\bibfnamefont {P.}~\bibnamefont {Hauke}},\ }\href@noop {} {\bibinfo
      {title} {Complexity transitions in chaotic quantum systems:
      Nonstabilizerness, entanglement, and fractal dimension in syk and random
      matrix models}} (\bibinfo {year} {2025}{\natexlab{b}}),\ \Eprint
      {https://arxiv.org/abs/2505.09707} {arXiv:2505.09707} \BibitemShut {NoStop}%
    \bibitem [{\citenamefont {Rakovszky}\ \emph {et~al.}(2019)\citenamefont
      {Rakovszky}, \citenamefont {Gopalakrishnan}, \citenamefont {Parameswaran},\
      and\ \citenamefont {Pollmann}}]{PhysRevB.100.125115}%
      \BibitemOpen
      \bibfield  {author} {\bibinfo {author} {\bibfnamefont {T.}~\bibnamefont
      {Rakovszky}}, \bibinfo {author} {\bibfnamefont {S.}~\bibnamefont
      {Gopalakrishnan}}, \bibinfo {author} {\bibfnamefont {S.~A.}\ \bibnamefont
      {Parameswaran}},\ and\ \bibinfo {author} {\bibfnamefont {F.}~\bibnamefont
      {Pollmann}},\ }\href {https://doi.org/10.1103/PhysRevB.100.125115} {\bibfield
       {journal} {\bibinfo  {journal} {Phys. Rev. B}\ }\textbf {\bibinfo {volume}
      {100}},\ \bibinfo {pages} {125115} (\bibinfo {year} {2019})}\BibitemShut
      {NoStop}%
    \bibitem [{\citenamefont {Varikuti}\ \emph {et~al.}(2025)\citenamefont
      {Varikuti}, \citenamefont {Bandyopadhyay},\ and\ \citenamefont
      {Hauke}}]{varikuti2025impactcliffordoperationsnonstabilizing}%
      \BibitemOpen
      \bibfield  {author} {\bibinfo {author} {\bibfnamefont {N.~D.}\ \bibnamefont
      {Varikuti}}, \bibinfo {author} {\bibfnamefont {S.}~\bibnamefont
      {Bandyopadhyay}},\ and\ \bibinfo {author} {\bibfnamefont {P.}~\bibnamefont
      {Hauke}},\ }\href@noop {} {\bibinfo {title} {Impact of clifford operations on
      non-stabilizing power and quantum chaos}} (\bibinfo {year} {2025}),\ \Eprint
      {https://arxiv.org/abs/2505.14793} {arXiv:2505.14793} \BibitemShut {NoStop}%
    \bibitem [{\citenamefont {Mezei}\ and\ \citenamefont
      {Stanford}(2017)}]{Mezei_2017}%
      \BibitemOpen
      \bibfield  {author} {\bibinfo {author} {\bibfnamefont {M.}~\bibnamefont
      {Mezei}}\ and\ \bibinfo {author} {\bibfnamefont {D.}~\bibnamefont
      {Stanford}},\ }\href {https://doi.org/10.1007/jhep05(2017)065} {\bibfield
      {journal} {\bibinfo  {journal} {J. High Energ. Phys.}\ }\textbf {\bibinfo
      {volume} {2017}}\bibinfo  {number} { (5)}}\BibitemShut {NoStop}%
    \bibitem [{\citenamefont {Nahum}\ \emph {et~al.}(2017)\citenamefont {Nahum},
      \citenamefont {Ruhman}, \citenamefont {Vijay},\ and\ \citenamefont
      {Haah}}]{nahumPRX17}%
      \BibitemOpen
    \bibfield  {number} {  }\bibfield  {author} {\bibinfo {author} {\bibfnamefont
      {A.}~\bibnamefont {Nahum}}, \bibinfo {author} {\bibfnamefont
      {J.}~\bibnamefont {Ruhman}}, \bibinfo {author} {\bibfnamefont
      {S.}~\bibnamefont {Vijay}},\ and\ \bibinfo {author} {\bibfnamefont
      {J.}~\bibnamefont {Haah}},\ }\href
      {https://doi.org/10.1103/PhysRevX.7.031016} {\bibfield  {journal} {\bibinfo
      {journal} {Phys. Rev. X}\ }\textbf {\bibinfo {volume} {7}},\ \bibinfo {pages}
      {031016} (\bibinfo {year} {2017})}\BibitemShut {NoStop}%
    \bibitem [{\citenamefont {von Keyserlingk}\ \emph {et~al.}(2018)\citenamefont
      {von Keyserlingk}, \citenamefont {Rakovszky}, \citenamefont {Pollmann},\ and\
      \citenamefont {Sondhi}}]{cvkPRX18}%
      \BibitemOpen
      \bibfield  {author} {\bibinfo {author} {\bibfnamefont {C.~W.}\ \bibnamefont
      {von Keyserlingk}}, \bibinfo {author} {\bibfnamefont {T.}~\bibnamefont
      {Rakovszky}}, \bibinfo {author} {\bibfnamefont {F.}~\bibnamefont
      {Pollmann}},\ and\ \bibinfo {author} {\bibfnamefont {S.~L.}\ \bibnamefont
      {Sondhi}},\ }\href {https://doi.org/10.1103/PhysRevX.8.021013} {\bibfield
      {journal} {\bibinfo  {journal} {Phys. Rev. X}\ }\textbf {\bibinfo {volume}
      {8}},\ \bibinfo {pages} {021013} (\bibinfo {year} {2018})}\BibitemShut
      {NoStop}%
    \bibitem [{\citenamefont {Bertini}\ \emph {et~al.}(2019)\citenamefont
      {Bertini}, \citenamefont {Kos},\ and\ \citenamefont {Prosen}}]{bertiniPRX}%
      \BibitemOpen
      \bibfield  {author} {\bibinfo {author} {\bibfnamefont {B.}~\bibnamefont
      {Bertini}}, \bibinfo {author} {\bibfnamefont {P.}~\bibnamefont {Kos}},\ and\
      \bibinfo {author} {\bibfnamefont {T.}~\bibnamefont {Prosen}},\ }\href
      {https://doi.org/10.1103/PhysRevX.9.021033} {\bibfield  {journal} {\bibinfo
      {journal} {Phys. Rev. X}\ }\textbf {\bibinfo {volume} {9}},\ \bibinfo {pages}
      {021033} (\bibinfo {year} {2019})}\BibitemShut {NoStop}%
    \bibitem [{\citenamefont {Turkeshi}\ \emph {et~al.}(2025)\citenamefont
      {Turkeshi}, \citenamefont {Tirrito},\ and\ \citenamefont
      {Sierant}}]{Turkeshi_2025}%
      \BibitemOpen
      \bibfield  {author} {\bibinfo {author} {\bibfnamefont {X.}~\bibnamefont
      {Turkeshi}}, \bibinfo {author} {\bibfnamefont {E.}~\bibnamefont {Tirrito}},\
      and\ \bibinfo {author} {\bibfnamefont {P.}~\bibnamefont {Sierant}},\ }\href
      {http://dx.doi.org/10.1038/s41467-025-57704-x} {\bibfield  {journal}
      {\bibinfo  {journal} {Nat. Comm.}\ }\textbf {\bibinfo {volume} {16}},\
      \bibinfo {pages} {2575} (\bibinfo {year} {2025})}\BibitemShut {NoStop}%
    \bibitem [{\citenamefont {Tirrito}\ \emph
      {et~al.}(2025{\natexlab{b}})\citenamefont {Tirrito}, \citenamefont
      {Turkeshi},\ and\ \citenamefont
      {Sierant}}]{tirrito2025anticoncentrationnonstabilizernessspreadingergodic}%
      \BibitemOpen
      \bibfield  {author} {\bibinfo {author} {\bibfnamefont {E.}~\bibnamefont
      {Tirrito}}, \bibinfo {author} {\bibfnamefont {X.}~\bibnamefont {Turkeshi}},\
      and\ \bibinfo {author} {\bibfnamefont {P.}~\bibnamefont {Sierant}},\
      }\href@noop {} {\bibinfo {title} {Anticoncentration and nonstabilizerness
      spreading under ergodic quantum dynamics}} (\bibinfo {year}
      {2025}{\natexlab{b}}),\ \Eprint {https://arxiv.org/abs/2412.10229}
      {arXiv:2412.10229} \BibitemShut {NoStop}%
    \bibitem [{\citenamefont {Tirrito}\ \emph
      {et~al.}(2025{\natexlab{c}})\citenamefont {Tirrito}, \citenamefont
      {Tarabunga}, \citenamefont {Bhakuni}, \citenamefont {Dalmonte}, \citenamefont
      {Sierant},\ and\ \citenamefont
      {Turkeshi}}]{tirrito2025universalspreadingnonstabilizernessquantum}%
      \BibitemOpen
      \bibfield  {author} {\bibinfo {author} {\bibfnamefont {E.}~\bibnamefont
      {Tirrito}}, \bibinfo {author} {\bibfnamefont {P.~S.}\ \bibnamefont
      {Tarabunga}}, \bibinfo {author} {\bibfnamefont {D.~S.}\ \bibnamefont
      {Bhakuni}}, \bibinfo {author} {\bibfnamefont {M.}~\bibnamefont {Dalmonte}},
      \bibinfo {author} {\bibfnamefont {P.}~\bibnamefont {Sierant}},\ and\ \bibinfo
      {author} {\bibfnamefont {X.}~\bibnamefont {Turkeshi}},\ }\href@noop {}
      {\bibinfo {title} {Universal spreading of nonstabilizerness and quantum
      transport}} (\bibinfo {year} {2025}{\natexlab{c}}),\ \Eprint
      {https://arxiv.org/abs/2506.12133} {arXiv:2506.12133} \BibitemShut {NoStop}%
    \bibitem [{\citenamefont {Nielsen}\ and\ \citenamefont
      {Chuang}(2010)}]{Nielsen2010}%
      \BibitemOpen
      \bibfield  {author} {\bibinfo {author} {\bibfnamefont {M.~A.}\ \bibnamefont
      {Nielsen}}\ and\ \bibinfo {author} {\bibfnamefont {I.~L.}\ \bibnamefont
      {Chuang}},\ }\href {https://doi.org/10.1017/CBO9780511976667} {\emph
      {\bibinfo {title} {Quantum Computation and Quantum Information: 10th
      Anniversary Edition}}}\ (\bibinfo  {publisher} {CUP},\ \bibinfo {year}
      {2010})\BibitemShut {NoStop}%
    \bibitem [{\citenamefont {Yao}\ and\ \citenamefont {Claeys}(2024)}]{JYPC_PRR}%
      \BibitemOpen
      \bibfield  {author} {\bibinfo {author} {\bibfnamefont {J.}~\bibnamefont
      {Yao}}\ and\ \bibinfo {author} {\bibfnamefont {P.~W.}\ \bibnamefont
      {Claeys}},\ }\href {https://doi.org/10.1103/PhysRevResearch.6.043077}
      {\bibfield  {journal} {\bibinfo  {journal} {Phys. Rev. Res.}\ }\textbf
      {\bibinfo {volume} {6}},\ \bibinfo {pages} {043077} (\bibinfo {year}
      {2024})}\BibitemShut {NoStop}%
    \bibitem [{SM()}]{SM}%
      \BibitemOpen
      \href@noop {} {}\bibinfo {note} {See the Supplemental Material for the proofs
      of Theorem 1 (including a BMG algorithm), equivalent and extended operational
      definitions of magic resources, and locality bounds on the LML; a discussion
      of the interplay between entanglement growth and operator spreading, and a
      review of their corresponding velocities in identity-doped Clifford circuits;
      a connection between the shape of the Pauli spectrum and the SRE; additional
      numerical results on the typical MLMI length, identity-doped dual-unitary
      Clifford circuits, the overlap of MLMIs, and system size independence of the
      spreading rates. Contains Refs.~\cite{GHZ_trio, Audenaert_2005,
      bejan2024prxq, cvkPRX18, nahumPRX17,JYPC_PRR,
      bertini2025exactlysolvablemanybodydynamics,sommersCrystallineQC,crooks_gates_2020,
      opSprVedika}.}\BibitemShut {Stop}%
    \bibitem [{\citenamefont {Gottesman}(1998)}]{gottesman1998heisenberg}%
      \BibitemOpen
      \bibfield  {author} {\bibinfo {author} {\bibfnamefont {D.}~\bibnamefont
      {Gottesman}},\ }\href@noop {} {\bibinfo {title} {The heisenberg
      representation of quantum computers}} (\bibinfo {year} {1998}),\ \Eprint
      {https://arxiv.org/abs/quant-ph/9807006} {arXiv:quant-ph/9807006}
      \BibitemShut {NoStop}%
    \bibitem [{\citenamefont {Leone}\ \emph {et~al.}(2022)\citenamefont {Leone},
      \citenamefont {Oliviero},\ and\ \citenamefont {Hamma}}]{sre}%
      \BibitemOpen
      \bibfield  {author} {\bibinfo {author} {\bibfnamefont {L.}~\bibnamefont
      {Leone}}, \bibinfo {author} {\bibfnamefont {S.~F.~E.}\ \bibnamefont
      {Oliviero}},\ and\ \bibinfo {author} {\bibfnamefont {A.}~\bibnamefont
      {Hamma}},\ }\href {https://doi.org/10.1103/PhysRevLett.128.050402} {\bibfield
       {journal} {\bibinfo  {journal} {Phys. Rev. Lett.}\ }\textbf {\bibinfo
      {volume} {128}},\ \bibinfo {pages} {050402} (\bibinfo {year}
      {2022})}\BibitemShut {NoStop}%
    \bibitem [{\citenamefont {Howard}\ and\ \citenamefont {Campbell}(2017)}]{RoM}%
      \BibitemOpen
      \bibfield  {author} {\bibinfo {author} {\bibfnamefont {M.}~\bibnamefont
      {Howard}}\ and\ \bibinfo {author} {\bibfnamefont {E.}~\bibnamefont
      {Campbell}},\ }\href {https://doi.org/10.1103/PhysRevLett.118.090501}
      {\bibfield  {journal} {\bibinfo  {journal} {Phys. Rev. Lett.}\ }\textbf
      {\bibinfo {volume} {118}},\ \bibinfo {pages} {090501} (\bibinfo {year}
      {2017})}\BibitemShut {NoStop}%
    \bibitem [{Note1()}]{Note1}%
      \BibitemOpen
      \bibinfo {note} {The Pauli spectrum is defined in terms of a basis of
      $L$-qubit Pauli operators. Such a basis is formed by all Hermitian Pauli
      strings $P$ with phase $+1$.}\BibitemShut {Stop}%
    \bibitem [{Note2()}]{Note2}%
      \BibitemOpen
      \bibinfo {note} {This notion of magic is relevant to active quantum error
      correction and monitored quantum dynamics, where a key question related to
      the complexity of classically simulating the dynamics is which measurements
      reduce the amount of magic in a state~\cite {bejan2024prxq}.}\BibitemShut
      {Stop}%
    \bibitem [{\citenamefont {Koç}\ and\ \citenamefont {Arachchige}(1991)}]{GF2}%
      \BibitemOpen
      \bibfield  {author} {\bibinfo {author} {\bibfnamefont {C.~K.}\ \bibnamefont
      {Koç}}\ and\ \bibinfo {author} {\bibfnamefont {S.~N.}\ \bibnamefont
      {Arachchige}},\ }\href {https://doi.org/10.1016/0743-7315(91)90115-P}
      {\bibfield  {journal} {\bibinfo  {journal} {J. Parallel Distrib. Comput.
      \textbf{13}, 118-122}\ } (\bibinfo {year} {1991})}\BibitemShut {NoStop}%
    \bibitem [{\citenamefont {Bravyi}\ and\ \citenamefont
      {Terhal}(2009)}]{BravyiTerhal2009}%
      \BibitemOpen
      \bibfield  {author} {\bibinfo {author} {\bibfnamefont {S.}~\bibnamefont
      {Bravyi}}\ and\ \bibinfo {author} {\bibfnamefont {B.}~\bibnamefont
      {Terhal}},\ }\href {https://doi.org/10.1088/1367-2630/11/4/043029} {\bibfield
       {journal} {\bibinfo  {journal} {New J. Phys.}\ }\textbf {\bibinfo {volume}
      {11}},\ \bibinfo {pages} {043029} (\bibinfo {year} {2009})}\BibitemShut
      {NoStop}%
    \bibitem [{\citenamefont {Gullans}\ and\ \citenamefont
      {Huse}(2020)}]{gullansPRXpurif}%
      \BibitemOpen
      \bibfield  {author} {\bibinfo {author} {\bibfnamefont {M.~J.}\ \bibnamefont
      {Gullans}}\ and\ \bibinfo {author} {\bibfnamefont {D.~A.}\ \bibnamefont
      {Huse}},\ }\href {https://doi.org/10.1103/PhysRevX.10.041020} {\bibfield
      {journal} {\bibinfo  {journal} {Phys. Rev. X}\ }\textbf {\bibinfo {volume}
      {10}},\ \bibinfo {pages} {041020} (\bibinfo {year} {2020})}\BibitemShut
      {NoStop}%
    \bibitem [{Note3()}]{Note3}%
      \BibitemOpen
      \bibinfo {note} {Focusing on contiguous systems also naturally leads to
      interesting connections between magic length scales for
      projectively-destroyable magic resource and the linear and contiguous code
      distances of QECC~\cite {BravyiTerhal2009, gullansPRXpurif}.}\BibitemShut
      {Stop}%
    \bibitem [{Note4()}]{Note4}%
      \BibitemOpen
      \bibinfo {note} {Here, by generators, we mean both the logical and stabilizer
      generators of, for example, $\rho $ from Eq.~\protect \eqref
      {eq:rho}.}\BibitemShut {Stop}%
    \bibitem [{Note5()}]{Note5}%
      \BibitemOpen
      \bibinfo {note} {We can relabel the logicals $\protect \bar {X}, \protect
      \bar {Y},$ and $\protect \bar {Z}$ as needed, provided they follow the
      \(\protect \mathfrak {su}(2)\) anti-commutation rules; the prefactors $a_\pm
      $ depend on the non-Clifford gate used to produce $\mathinner {|{\psi
      }\rangle }$, but for suitably chosen $\protect \bar {Z}$ and $\protect \bar
      {Y}$, they satisfy $|a_\pm | \protect \neq 0$, ensuring that $\protect \tilde
      {\protect \cal M}_2(\rho _A) > 0$ and ${\protect \cal R}(\rho _A)>1$ for a
      subsystem \(A\) where at least one logical can be reduced. The specific
      values of $|a_\pm |$ set the values of ``full'' [case ($i$)] and ``partial''
      [cases ($iv, v$)] NS. See also~\cite {SM}.}\BibitemShut {Stop}%
    \bibitem [{Note6()}]{Note6}%
      \BibitemOpen
      \bibinfo {note} {For intermediate times, the data from Fig.~\ref
      {fig:width_rand_cliff} is consistent with a growth rate of FLEOM given by
      $\partial _t W \sim 4v_{\protect \rm E}/3$. We attribute the distinct
      intermediate-time growth of $W(t)$ to a combination of effects due to
      boundary conditions and shrinking of large logicals, as detailed in~\cite
      {SM}.}\BibitemShut {Stop}%
    \bibitem [{\citenamefont {Nahum}\ \emph {et~al.}(2018)\citenamefont {Nahum},
      \citenamefont {Vijay},\ and\ \citenamefont {Haah}}]{nahumPRX18}%
      \BibitemOpen
      \bibfield  {author} {\bibinfo {author} {\bibfnamefont {A.}~\bibnamefont
      {Nahum}}, \bibinfo {author} {\bibfnamefont {S.}~\bibnamefont {Vijay}},\ and\
      \bibinfo {author} {\bibfnamefont {J.}~\bibnamefont {Haah}},\ }\href
      {https://doi.org/10.1103/PhysRevX.8.021014} {\bibfield  {journal} {\bibinfo
      {journal} {Phys. Rev. X}\ }\textbf {\bibinfo {volume} {8}},\ \bibinfo {pages}
      {021014} (\bibinfo {year} {2018})}\BibitemShut {NoStop}%
    \bibitem [{\citenamefont {Cerf}\ and\ \citenamefont
      {Cleve}(1997)}]{qSingleton}%
      \BibitemOpen
      \bibfield  {author} {\bibinfo {author} {\bibfnamefont {N.~J.}\ \bibnamefont
      {Cerf}}\ and\ \bibinfo {author} {\bibfnamefont {R.}~\bibnamefont {Cleve}},\
      }\href {https://doi.org/10.1103/PhysRevA.56.1721} {\bibfield  {journal}
      {\bibinfo  {journal} {Phys. Rev. A}\ }\textbf {\bibinfo {volume} {56}},\
      \bibinfo {pages} {1721} (\bibinfo {year} {1997})}\BibitemShut {NoStop}%
    \bibitem [{\citenamefont {Schumacher}\ and\ \citenamefont
      {Nielsen}(1996)}]{schumacher_quantum_1996}%
      \BibitemOpen
      \bibfield  {author} {\bibinfo {author} {\bibfnamefont {B.}~\bibnamefont
      {Schumacher}}\ and\ \bibinfo {author} {\bibfnamefont {M.~A.}\ \bibnamefont
      {Nielsen}},\ }\href {https://doi.org/10.1103/PhysRevA.54.2629} {\bibfield
      {journal} {\bibinfo  {journal} {Phys. Rev. A}\ }\textbf {\bibinfo {volume}
      {54}},\ \bibinfo {pages} {2629} (\bibinfo {year} {1996})}\BibitemShut
      {NoStop}%
    \bibitem [{\citenamefont {Lloyd}(1997)}]{lloyd_capacity_1997}%
      \BibitemOpen
      \bibfield  {author} {\bibinfo {author} {\bibfnamefont {S.}~\bibnamefont
      {Lloyd}},\ }\href {https://doi.org/10.1103/PhysRevA.55.1613} {\bibfield
      {journal} {\bibinfo  {journal} {Phys. Rev. A}\ }\textbf {\bibinfo {volume}
      {55}},\ \bibinfo {pages} {1613} (\bibinfo {year} {1997})}\BibitemShut
      {NoStop}%
    \bibitem [{\citenamefont {Bennett}\ \emph {et~al.}(1997)\citenamefont
      {Bennett}, \citenamefont {DiVincenzo},\ and\ \citenamefont
      {Smolin}}]{bennett_capacities_1997}%
      \BibitemOpen
      \bibfield  {author} {\bibinfo {author} {\bibfnamefont {C.~H.}\ \bibnamefont
      {Bennett}}, \bibinfo {author} {\bibfnamefont {D.~P.}\ \bibnamefont
      {DiVincenzo}},\ and\ \bibinfo {author} {\bibfnamefont {J.~A.}\ \bibnamefont
      {Smolin}},\ }\href {https://doi.org/10.1103/PhysRevLett.78.3217} {\bibfield
      {journal} {\bibinfo  {journal} {Phys. Rev. Lett.}\ }\textbf {\bibinfo
      {volume} {78}},\ \bibinfo {pages} {3217} (\bibinfo {year}
      {1997})}\BibitemShut {NoStop}%
    \bibitem [{\citenamefont {Aziz}\ \emph {et~al.}(2025)\citenamefont {Aziz},
      \citenamefont {Pan}, \citenamefont {Gullans},\ and\ \citenamefont
      {Pixley}}]{aziz2025classicalsimulationslowmagic}%
      \BibitemOpen
      \bibfield  {author} {\bibinfo {author} {\bibfnamefont {K.}~\bibnamefont
      {Aziz}}, \bibinfo {author} {\bibfnamefont {H.}~\bibnamefont {Pan}}, \bibinfo
      {author} {\bibfnamefont {M.~J.}\ \bibnamefont {Gullans}},\ and\ \bibinfo
      {author} {\bibfnamefont {J.~H.}\ \bibnamefont {Pixley}},\ }\href@noop {}
      {\bibinfo {title} {Classical simulations of low magic quantum dynamics}}
      (\bibinfo {year} {2025}),\ \Eprint {https://arxiv.org/abs/2508.20252}
      {arXiv:2508.20252} \BibitemShut {NoStop}%
    \bibitem [{\citenamefont {Cao}\ \emph {et~al.}(2025{\natexlab{b}})\citenamefont
      {Cao}, \citenamefont {Cheng},\ and\ \citenamefont
      {Zhou}}]{cao2025efficientalgorithmcomputeentanglement}%
      \BibitemOpen
      \bibfield  {author} {\bibinfo {author} {\bibfnamefont {C.}~\bibnamefont
      {Cao}}, \bibinfo {author} {\bibfnamefont {G.}~\bibnamefont {Cheng}},\ and\
      \bibinfo {author} {\bibfnamefont {T.}~\bibnamefont {Zhou}},\ }\href@noop {}
      {\bibinfo {title} {An efficient algorithm to compute entanglement in states
      with low magic}} (\bibinfo {year} {2025}{\natexlab{b}}),\ \Eprint
      {https://arxiv.org/abs/2510.06318} {arXiv:2510.06318} \BibitemShut {NoStop}%
    \bibitem [{\citenamefont {Fisher}\ \emph {et~al.}(2023)\citenamefont {Fisher},
      \citenamefont {Khemani}, \citenamefont {Nahum},\ and\ \citenamefont
      {Vijay}}]{fisher2023rqc}%
      \BibitemOpen
      \bibfield  {author} {\bibinfo {author} {\bibfnamefont {M.~P.~A.}\
      \bibnamefont {Fisher}}, \bibinfo {author} {\bibfnamefont {V.}~\bibnamefont
      {Khemani}}, \bibinfo {author} {\bibfnamefont {A.}~\bibnamefont {Nahum}},\
      and\ \bibinfo {author} {\bibfnamefont {S.}~\bibnamefont {Vijay}},\ }\href
      {https://doi.org/10.1146/annurev-conmatphys-031720-030658} {\bibfield
      {journal} {\bibinfo  {journal} {Annu. Rev. Condens. Matt. Phys.}\ }\textbf
      {\bibinfo {volume} {14}},\ \bibinfo {pages} {335} (\bibinfo {year}
      {2023})}\BibitemShut {NoStop}%
    \bibitem [{\citenamefont {Montañà~López}\ and\ \citenamefont
      {Kos}(2024)}]{Monta_L_pez_2024}%
      \BibitemOpen
      \bibfield  {author} {\bibinfo {author} {\bibfnamefont {J.~A.}\ \bibnamefont
      {Montañà~López}}\ and\ \bibinfo {author} {\bibfnamefont {P.}~\bibnamefont
      {Kos}},\ }\href {https://doi.org/10.1088/1751-8121/ad85b0} {\bibfield
      {journal} {\bibinfo  {journal} {J. Phys. A: Math. Theor.}\ }\textbf {\bibinfo
      {volume} {57}},\ \bibinfo {pages} {475301} (\bibinfo {year}
      {2024})}\BibitemShut {NoStop}%
    \bibitem [{\citenamefont {Sierant}\ \emph {et~al.}(2025)\citenamefont
      {Sierant}, \citenamefont {Stornati},\ and\ \citenamefont
      {Turkeshi}}]{sierant2025fermionicmagicresourcesquantum}%
      \BibitemOpen
      \bibfield  {author} {\bibinfo {author} {\bibfnamefont {P.}~\bibnamefont
      {Sierant}}, \bibinfo {author} {\bibfnamefont {P.}~\bibnamefont {Stornati}},\
      and\ \bibinfo {author} {\bibfnamefont {X.}~\bibnamefont {Turkeshi}},\
      }\href@noop {} {\bibinfo {title} {Fermionic magic resources of quantum
      many-body systems}} (\bibinfo {year} {2025}),\ \Eprint
      {https://arxiv.org/abs/2506.00116} {arXiv:2506.00116} \BibitemShut {NoStop}%
    \bibitem [{\citenamefont {Aditya}\ \emph {et~al.}(2025)\citenamefont {Aditya},
      \citenamefont {Summer}, \citenamefont {Sierant},\ and\ \citenamefont
      {Turkeshi}}]{aditya2025mpembaeffectsquantumcomplexity}%
      \BibitemOpen
      \bibfield  {author} {\bibinfo {author} {\bibfnamefont {S.}~\bibnamefont
      {Aditya}}, \bibinfo {author} {\bibfnamefont {A.}~\bibnamefont {Summer}},
      \bibinfo {author} {\bibfnamefont {P.}~\bibnamefont {Sierant}},\ and\ \bibinfo
      {author} {\bibfnamefont {X.}~\bibnamefont {Turkeshi}},\ }\href@noop {}
      {\bibinfo {title} {Mpemba effects in quantum complexity}} (\bibinfo {year}
      {2025}),\ \Eprint {https://arxiv.org/abs/2509.22176} {arXiv:2509.22176}
      \BibitemShut {NoStop}%
    \bibitem [{\citenamefont {Cotler}\ \emph {et~al.}(2019)\citenamefont {Cotler},
      \citenamefont {Han}, \citenamefont {Qi},\ and\ \citenamefont
      {Yang}}]{Cotler_2019}%
      \BibitemOpen
      \bibfield  {author} {\bibinfo {author} {\bibfnamefont {J.}~\bibnamefont
      {Cotler}}, \bibinfo {author} {\bibfnamefont {X.}~\bibnamefont {Han}},
      \bibinfo {author} {\bibfnamefont {X.-L.}\ \bibnamefont {Qi}},\ and\ \bibinfo
      {author} {\bibfnamefont {Z.}~\bibnamefont {Yang}},\ }\href
      {https://doi.org/10.1007/jhep07(2019)042} {\bibfield  {journal} {\bibinfo
      {journal} {J. High Energ. Phys.}\ }\textbf {\bibinfo {volume} {2019}}\bibinfo
       {number} { (7)}}\BibitemShut {NoStop}%
    \bibitem [{\citenamefont {Maity}\ and\ \citenamefont
      {Hamazaki}(2025)}]{maity2025localspreadingstabilizerrenyi}%
      \BibitemOpen
    \bibfield  {number} {  }\bibfield  {author} {\bibinfo {author} {\bibfnamefont
      {S.}~\bibnamefont {Maity}}\ and\ \bibinfo {author} {\bibfnamefont
      {R.}~\bibnamefont {Hamazaki}},\ }\href@noop {} {\bibinfo {title} {Local
      spreading of stabilizer r\'enyi entropy in a brickwork random clifford
      circuit}} (\bibinfo {year} {2025}),\ \Eprint
      {https://arxiv.org/abs/2511.07769} {arXiv:2511.07769} \BibitemShut {NoStop}%
    \bibitem [{\citenamefont {Lee}\ and\ \citenamefont
      {Yoshida}(2024)}]{lee2024randomlymonitoredquantumcodes}%
      \BibitemOpen
      \bibfield  {author} {\bibinfo {author} {\bibfnamefont {D.}~\bibnamefont
      {Lee}}\ and\ \bibinfo {author} {\bibfnamefont {B.}~\bibnamefont {Yoshida}},\
      }\href@noop {} {\bibinfo {title} {Randomly monitored quantum codes}}
      (\bibinfo {year} {2024}),\ \Eprint {https://arxiv.org/abs/2402.00145}
      {arXiv:2402.00145} \BibitemShut {NoStop}%
    \bibitem [{\citenamefont {Bravyi}\ \emph {et~al.}(2006)\citenamefont {Bravyi},
      \citenamefont {Fattal},\ and\ \citenamefont {Gottesman}}]{GHZ_trio}%
      \BibitemOpen
      \bibfield  {author} {\bibinfo {author} {\bibfnamefont {S.}~\bibnamefont
      {Bravyi}}, \bibinfo {author} {\bibfnamefont {D.}~\bibnamefont {Fattal}},\
      and\ \bibinfo {author} {\bibfnamefont {D.}~\bibnamefont {Gottesman}},\ }\href
      {https://doi.org/10.1063/1.2203431} {\bibfield  {journal} {\bibinfo
      {journal} {J. Math. Phys.}\ }\textbf {\bibinfo {volume} {47}},\ \bibinfo
      {pages} {062106} (\bibinfo {year} {2006})}\BibitemShut {NoStop}%
    \bibitem [{\citenamefont {Aaronson}\ and\ \citenamefont
      {Gottesman}(2004)}]{aaronsonPRA2004}%
      \BibitemOpen
      \bibfield  {author} {\bibinfo {author} {\bibfnamefont {S.}~\bibnamefont
      {Aaronson}}\ and\ \bibinfo {author} {\bibfnamefont {D.}~\bibnamefont
      {Gottesman}},\ }\href {https://doi.org/10.1103/PhysRevA.70.052328} {\bibfield
       {journal} {\bibinfo  {journal} {Phys. Rev. A}\ }\textbf {\bibinfo {volume}
      {70}},\ \bibinfo {pages} {052328} (\bibinfo {year} {2004})}\BibitemShut
      {NoStop}%
    \bibitem [{\citenamefont {Audenaert}\ and\ \citenamefont
      {Plenio}(2005)}]{Audenaert_2005}%
      \BibitemOpen
      \bibfield  {author} {\bibinfo {author} {\bibfnamefont {K.~M.~R.}\
      \bibnamefont {Audenaert}}\ and\ \bibinfo {author} {\bibfnamefont {M.~B.}\
      \bibnamefont {Plenio}},\ }\href {https://doi.org/10.1088/1367-2630/7/1/170}
      {\bibfield  {journal} {\bibinfo  {journal} {New J. Phys.}\ }\textbf {\bibinfo
      {volume} {7}},\ \bibinfo {pages} {170–170} (\bibinfo {year}
      {2005})}\BibitemShut {NoStop}%
    \bibitem [{\citenamefont {Beverland}\ \emph {et~al.}(2020)\citenamefont
      {Beverland}, \citenamefont {Campbell}, \citenamefont {Howard},\ and\
      \citenamefont {Kliuchnikov}}]{Beverland_2020}%
      \BibitemOpen
      \bibfield  {author} {\bibinfo {author} {\bibfnamefont {M.}~\bibnamefont
      {Beverland}}, \bibinfo {author} {\bibfnamefont {E.}~\bibnamefont {Campbell}},
      \bibinfo {author} {\bibfnamefont {M.}~\bibnamefont {Howard}},\ and\ \bibinfo
      {author} {\bibfnamefont {V.}~\bibnamefont {Kliuchnikov}},\ }\href
      {https://doi.org/10.1088/2058-9565/ab8963} {\bibfield  {journal} {\bibinfo
      {journal} {Quantum Sci. Technol.}\ }\textbf {\bibinfo {volume} {5}},\
      \bibinfo {pages} {035009} (\bibinfo {year} {2020})}\BibitemShut {NoStop}%
    \bibitem [{\citenamefont {Sommers}\ \emph {et~al.}(2023)\citenamefont
      {Sommers}, \citenamefont {Huse},\ and\ \citenamefont
      {Gullans}}]{sommersCrystallineQC}%
      \BibitemOpen
      \bibfield  {author} {\bibinfo {author} {\bibfnamefont {G.~M.}\ \bibnamefont
      {Sommers}}, \bibinfo {author} {\bibfnamefont {D.~A.}\ \bibnamefont {Huse}},\
      and\ \bibinfo {author} {\bibfnamefont {M.~J.}\ \bibnamefont {Gullans}},\
      }\href {https://doi.org/10.1103/PRXQuantum.4.030313} {\bibfield  {journal}
      {\bibinfo  {journal} {PRX Quantum}\ }\textbf {\bibinfo {volume} {4}},\
      \bibinfo {pages} {030313} (\bibinfo {year} {2023})}\BibitemShut {NoStop}%
    \bibitem [{\citenamefont {Khemani}\ \emph {et~al.}(2018)\citenamefont
      {Khemani}, \citenamefont {Vishwanath},\ and\ \citenamefont
      {Huse}}]{opSprVedika}%
      \BibitemOpen
      \bibfield  {author} {\bibinfo {author} {\bibfnamefont {V.}~\bibnamefont
      {Khemani}}, \bibinfo {author} {\bibfnamefont {A.}~\bibnamefont
      {Vishwanath}},\ and\ \bibinfo {author} {\bibfnamefont {D.~A.}\ \bibnamefont
      {Huse}},\ }\href {https://doi.org/10.1103/PhysRevX.8.031057} {\bibfield
      {journal} {\bibinfo  {journal} {Phys. Rev. X}\ }\textbf {\bibinfo {volume}
      {8}},\ \bibinfo {pages} {031057} (\bibinfo {year} {2018})}\BibitemShut
      {NoStop}%
    \bibitem [{\citenamefont {Bertini}\ \emph {et~al.}(2025)\citenamefont
      {Bertini}, \citenamefont {Claeys},\ and\ \citenamefont
      {Prosen}}]{bertini2025exactlysolvablemanybodydynamics}%
      \BibitemOpen
      \bibfield  {author} {\bibinfo {author} {\bibfnamefont {B.}~\bibnamefont
      {Bertini}}, \bibinfo {author} {\bibfnamefont {P.~W.}\ \bibnamefont
      {Claeys}},\ and\ \bibinfo {author} {\bibfnamefont {T.}~\bibnamefont
      {Prosen}},\ }\href@noop {} {\bibinfo {title} {Exactly solvable many-body
      dynamics from space-time duality}} (\bibinfo {year} {2025}),\ \Eprint
      {https://arxiv.org/abs/2505.11489} {arXiv:2505.11489} \BibitemShut {NoStop}%
    \bibitem [{\citenamefont {Crooks}(2020)}]{crooks_gates_2020}%
      \BibitemOpen
      \bibfield  {author} {\bibinfo {author} {\bibfnamefont {G.~E.}\ \bibnamefont
      {Crooks}},\ }\href {https://threeplusone.com/pubs/on_gates.pdf} {\bibfield
      {journal} {\bibinfo  {journal} {Gates states and circuits}\ } (\bibinfo
      {year} {2020})}\BibitemShut {NoStop}%
    \end{thebibliography}
\end{document}